\documentclass[pdflscape,lscape,iop,apj,numberedappendix,appendixfloats,stfloats,url,microtype]{emulateapj}

\usepackage{paralist,amsmath,amssymb}
\usepackage{longtable,rotating}
\usepackage{float}
\usepackage{morefloats}
\usepackage{placeins}

\usepackage{color,soul}
\graphicspath{{./}}

\shorttitle{1-10 GHz spectropolarimetry of discrete radio sources}
\shortauthors{Anderson et al.}

\begin{document}




\title{A study of broadband Faraday rotation and polarization behaviour over 1.3--10 GH\MakeLowercase{z} in 36 discrete radio sources}


\author{
C. S. Anderson\altaffilmark{1,2,}$^{\dagger}$,
B. M. Gaensler\altaffilmark{1,3},
I. J. Feain\altaffilmark{1,4},
}

\altaffiltext{$\dagger$}{{\bf c.anderson@physics.usyd.edu.au}}
\altaffiltext{1}{Sydney Institute for Astronomy (SIfA), School of Physics, University of Sydney, NSW 2006, Australia}
\altaffiltext{2}{CSIRO Astronomy \& Space Science, Epping, NSW 1710, Australia}
\altaffiltext{3}{Dunlap Institute for Astronomy \& Astrophysics, University of Toronto, Ontario, Canada}
\altaffiltext{4}{Central Clinical School, School of Medicine, University of Sydney, NSW 2006, Australia}


\begin{abstract}

We present a broadband polarization analysis of 36 discrete polarized radio sources over a very broad, densely-sampled frequency band. Our sample was selected on the basis of polarization behaviour apparent in narrowband archival data at 1.4 GHz: half the sample show complicated frequency-dependent polarization behaviour (i.e. Faraday complexity) at these frequencies, while half show comparatively simple behaviour (i.e. they appear \emph{Faraday simple}). We re-observed the sample using the Australia Telescope Compact Array (ATCA) in full polarization, with 6 GHz of densely sampled frequency coverage spanning 1.3 to 10 GHz. We have devised a general polarization modelling technique that allows us to identify multiple polarized emission components in a source, and to characterize their properties. We detect Faraday complex behaviour in almost every source in our sample. Several sources exhibit particularly remarkable polarization behaviour. By comparing our new and archival data, we have identified temporal variability in the broadband integrated polarization spectra of some sources. In a number of cases, the characteristics of the polarized emission components, including the range of Faraday depths over which they emit, their temporal variability, spectral index, and the linear extent of the source, allow us to argue that the spectropolarimetric data encodes information about the magnetoionic environment of active galactic nuclei themselves. Furthermore, the data place direct constraints on the geometry and magnetoionic structure of this material. We discuss the consequences of restricted frequency bands on the detection and interpretation of polarization structures, and implications for upcoming spectropolarimetric surveys.

\end{abstract}





\section{Introduction}\label{sec-intro}

Radio-loud Active Galactic Nuclei (AGN) play a key role in driving the evolution and ecology of the Universe by feeding energy and material into their broader environment. However, we still lack a clear picture of the exact physical processes involved, and indeed, of the detailed physical structure of AGN, jets, and radio lobes more generally. Arguably, the main impediment to an improved understanding of these objects is the observational challenge of spatially resolving the structures of interest, which are small in physical extent and lie at great distance. Very Long Baseline Interferometry (VLBI) observations represent one powerful method of approaching this problem, but at the current time, still provides limited survey capability and limited instantaneous frequency coverage, and key structures of interest still remain near or below the resolution limit (e.g. Taylor \& Zavala 2010). 

Often, the unresolved issues surrounding AGN fuelling and feedback have, at their heart, a set of fundamental questions concerning the structure and dynamics of magnetized plasma associated with AGN. To address these questions, an alternate and complementary approach to VLBI-based studies is available: the frequency-dependant signal encoded in linearly polarized emission by Faraday rotation can be used to probe magnetoionized structure towards radio sources on spatial scales below the observing resolution (e.g. Gardner \& Whiteoak 1966, Burn 1966, Gaensler et al. 2015, Heald et al. 2015). 

The way in which this information is encoded can be described mathematically as follows: First, the linear polarization state of radio emission is described by a complex vector $\boldsymbol{P}$. It is related to the Stokes parameters $Q$ \& $U$, the polarization angle $\psi$, the fractional polarization $p$ and the total intensity $I$ as:
 
 \begin{equation}
\boldsymbol{P} = Q + iU = pIe^{2i\psi}
\label{eq:ComplexPolVec}
 \end{equation}

After being emitted at a distance $L$, linearly polarized radiation will be Faraday rotated by magnetized plasma along the line of sight (LOS) to an observer by an amount equal to

  \begin{equation}
\Delta\psi= \phi\lambda^2
\label{eq:rotation}
 \end{equation}

 where $\psi$ is the polarization angle, $\lambda$ is the observing wavelength, and $\phi$ is the Faraday depth, given by  

 \begin{equation}
\text{$\phi$} = 0.812 \int_{L}^{\text{telescope}} n_e\boldsymbol{B}.\text{d}\boldsymbol{s}~\text{rad m}^{-2}
\label{eq:FaradayDepth}
 \end{equation}
 
and, in turn, $n_e$ [cm$^{-3}$] \& $\boldsymbol{B}$ [$\mu$G] are the thermal electron density and magnetic field along the LOS respectively.

The net observable polarization $\boldsymbol{P}(\lambda^2)$ is obtained by summing the polarized emission emerging from all possible Faraday depths within the synthesized beam of the telescope:

 \begin{equation}
\boldsymbol{P}(\lambda^2) = \int_{-\infty}^{\infty} \boldsymbol{F}(\phi) e^{2i\phi\lambda^2} d\phi
\label{eq:SumPol}
 \end{equation}
 
The function $\boldsymbol{F}(\phi)$ (the so-called Faraday Dispersion Function, henceforth FDF) specifies the distribution of polarized emission as a function of Faraday depth along the LOS, and possesses units of Jy m$^2$ rad$^{-3}$ for a source which is extended both in the plane of the sky and in Faraday depth. Based on the form of this function, two broad types of $\boldsymbol{P}(\lambda^2)$ behaviour can result. When $\boldsymbol{F}(\phi)$ is a $\delta$-function (i.e. polarized emission emanates from a single Faraday depth), the polarization behaviour is characterized by constant $p(\lambda^2)$, linear $\psi(\lambda^2)$ (modulo $\pi$ radians), and a quantity known as the rotation measure (RM) --- the gradient of $\psi(\lambda^2)$. The RM parameterizes the magnitude of the Faraday rotation effect, and can be related to magnetized structure along the LOS through a relation similar to Eqn. \ref{eq:FaradayDepth}. Ultimately this \emph{Faraday simple} behaviour is an idealization, but one which individual polarized emission components from a source can closely approximate (e.g. O'Sullivan et al. 2012). Conversely, when polarized emission occurs over a range of Faraday depths, or if multiple polarized emission components are emitted from different Faraday depths, we refer to a source as being \emph{Faraday complex}. In this case, Stokes $Q$/$I(\lambda^2)$, $U$/$I(\lambda^2)$, $p(\lambda^2)$ \& $\psi(\lambda^2)$ can all show complicated behaviours which reveal magnetoionized material along the LOS, and can be modelled to reveal the structure and properties of this material on scales smaller than the resolution of the observing instrument.

A number of authors have exploited Faraday complex behaviour to study magnetoionized structures towards unresolved radio sources. Early work by Slysh (1965) and Goldstein \& Reid (1984) established the existence of Faraday complex behaviour in discrete radio sources using narrowband polarization data. Observations over broad (but sparsely sampled) frequency bands ($\gtrsim10$ GHz) appear to reveal the existence and properties of magnetized material in the vicinity of radio sources themselves (e.g. Conway et al. 1974, Farnes et al. 2014, Pasetto et al. 2015). Other studies have reached similar conclusions by studying narrower, $\sim$GHz-wide bands with high sampling density (Law et al. 2011, O'Sullivan et al. 2012). In general though, it remains unclear exactly which structures in radio sources are being studied, or indeed, whether other structures along the LOS also make a substantial contribution to the observed polarization behaviour. For example, Anderson et al. (2015) recently analysed $\sim$160 polarized sources over 1.3--2 GHz and found evidence for a significant Galactic contribution to the Faraday complexity of background sources.


As a tool for studying the physics of unresolved magnetoionic structure in the radio source population, broadband spectropolarimetry will come of age in the approaching era of wide field, broadband surveys such as the Polarization Sky Survey of the Universe's Magnetism (POSSUM; Gaensler et al. 2009) and the VLA Sky Survey (VLASS; VSSG 2015). However, even these surveys will have bandwidths which are narrow compared to what can be achieved with contemporary targeted observations of a (much) smaller sample. Currently, we lack a detailed survey of the near-continuously sampled, multi-GHz-bandwidth polarization properties of Faraday complex radio sources. It remains a largely open question what the broadband polarization spectra of AGN look like. The study we present here aims to address this issue by presenting a spectropolarimetric survey of confirmed Faraday complex sources over a very broad 1.3--10 GHz band. Specifically, we seek to address the following questions: To what extent and over what wavelength ranges do deviations from Faraday simple behaviour occur? What are the characteristic Faraday depths and dispersions of the emission components responsible for these deviations? Is it possible to model, in a simple way, polarization data over fractional bandwidths approaching 200\%? To what extent can the properties of individual polarized emission components be distinguished and characterized? Where along the LOS are the emitting/rotating/Faraday dispersing medium(s) located? In particular, are the Faraday effects generated in the sources  themselves (thus providing a method for their study) or in the foreground? What are the implications and opportunities for upcoming broadband spectropolarimetric surveys? Finally, we note what our study is not, which is an in-depth investigation of the physics of individual sources. This would require complementary multiwavelength data which is not presently available for most objects in our sample.

Our paper is set out as follows. We describe the construction of a sample to address our questions in Section \ref{sec-sample}, our observations and their calibration in Section \ref{sec-obscal}, and our spectropolarimetric analysis in Section \ref{sec-specpol}, including imaging, modelling and fitting the data. Ancillary non-spectropolarimetric analysis is described in Section \ref{sec-extras}. We present our results in Section \ref{sec-results} and our discussion in Section \ref{sec-discussion}. A summary of our work and our conclusions are provided in Section \ref{sec-conclusion}.

\section{Sample construction}\label{sec-sample}

\begin{figure*}
\includegraphics[width=0.95\textwidth]{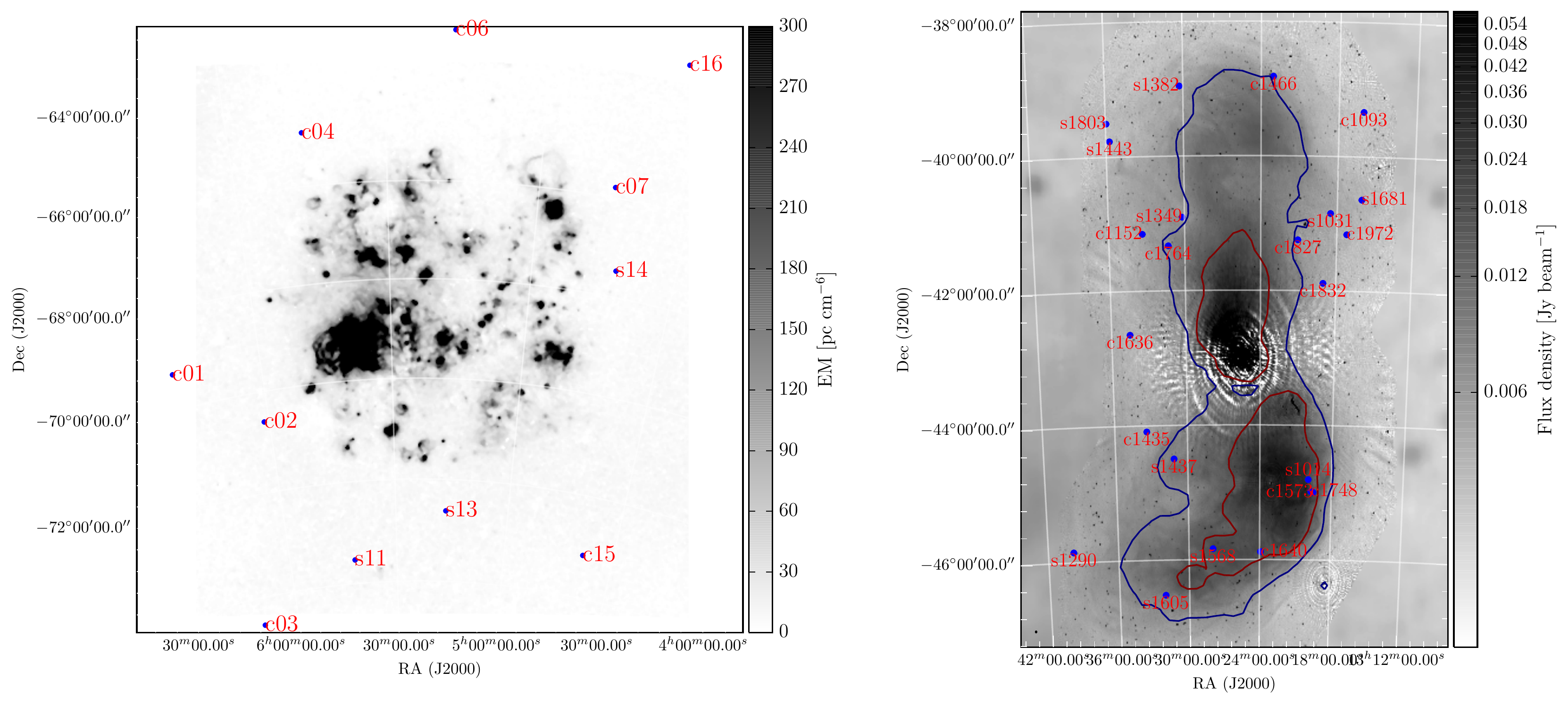}
\caption{The positions of sources in our sample in relation to the LMC (left panel) and Centaurus A radio lobes (right panel). The grayscale for the LMC shows emission measure (see Gaesnler et al. 2005), while the grayscale for the Centaurus radio lobes is a combined ATCA / Parkes radio image at 1.4 GHz (see Feain et al. 2011). The contour levels for the Centaurus radio lobes have been chosen to approximately delineate the extent of the bright middle and faint outer lobes.}
\label{fig:C2570_sourcepoz}
\end{figure*}

The targets for our study were selected from amongst two archival polarization data sets (detailed in Table \ref{table:observations}), which were observed and processed by several independent groups. Kim et al. (1998) conducted a 130 square degree survey of the Large Magellanic Cloud (LMC), during which polarization data was recorded commensally. Gaensler et al. (2005) reprocessed this data and imaged it in full polarization between 1.328 \& 1.432 GHz. Feain et al. (2009) observed a 34 square degree region around the radio galaxy Centaurus A, imaging the data in full polarization between 1.296 \& 1.480 GHz. For the combined total of 572 sources robustly detected in linear polarization in these data, Gaensler et al. (2005) and Feain et al. (2009) extracted Stokes $I$, $Q$, $U$ \& $V$ flux densities at 8 MHz intervals through the respective bands. We applied RM synthesis (Brentjens \& DeBruyn 2005) \& {\sc rmclean} (Heald et al. 2009) to these data, identifying Faraday complex sources using the second moment ($\sigma_{\boldsymbol{F}_{M}}$) of the {\sc rmclean} component distribution $\boldsymbol{F}_M$ (Brown et al. 2011, Anderson et al. 2015). This provides a measure of the spread of {\sc rmclean} components in $\phi$ space, and is given by:

\begin{eqnarray}
\sigma_{\boldsymbol{F}_{M}}&=&K^{-1}\sum\limits_{i=1}^{n}(\phi_i-\mu_\phi)^2 |\boldsymbol{F}_M(\phi_i)|
\end{eqnarray}

where the normalization constant K is given by
\begin{eqnarray}
K=\sum\limits_{i=0}^{n}|\boldsymbol{F}_M(\phi_i)|
\end{eqnarray}

and $\mu_\phi$, the first moment of the distribution, is given by
\begin{eqnarray}
\mu_\phi=K^{-1}\sum\limits_{i=0}^{n}\phi_i|\boldsymbol{F}_M(\phi_i)|
\end{eqnarray}

where $n$ is the number of channels $i$ possessing non-zero values in the {\sc rmclean} model and $\phi_i$ is the Faraday depth of channel $i$. We identified sources for which $\sigma_{\boldsymbol{F}_{M}}>10$ rad m$^{-2}$ as candidates for our Faraday complex sample (e.g. see Anderson et al. 2015), and sources with $\sigma_{\boldsymbol{F}_{M}}=0$ rad m$^{-2}$ (i.e. Faraday simple sources) as a control sample. In addition, we only selected sources where the chance of foreground contamination from the LMC and Centaurus A radio lobes was minimal. In the former case, we selected only sources which fell outside the boundaries of the LMC, and in regions with an emission measure consistent with zero (i.e. with a low column density of free electrons --- see the lefthand panel of Fig. \ref{fig:C2570_sourcepoz}). In the latter case, we generally selected sources which sat outside the Centaurus A radio lobes, though we chose to include several sources which do not meet this criteria (see the righthand panel of Fig. \ref{fig:C2570_sourcepoz}). We note that our choice of array configurations and $uv$ weighting scheme (see Section \ref{sec-extract}) results in the diffuse emission from the lobes being completely resolved out. As such, the only way that the lobes could affect the level of Faraday complexity in our polarization data is if RM variations within the lobes act as a depolarising foreground screen. However, Feain et al. (2009) showed that the lobes do not appear to show RM structure on angular scales below $0\fdg2$, which all of our sources are considerably smaller than.

In total, 7 (13) complex and 7 (13) simple sources were selected from the LMC (Centaurus) datasets, with band-averaged, undebiased (see Simmons \& Stewart 1985) polarized signal-to-noise (S/N) ratios of between 4 \& 350. In this paper, we use source designations that begin with either `cena' or `lmc' to specify the data set that each source was selected from, followed by a `c' or `s'  to specify whether the source was selected because it was Faraday complex or simple, and which end with a sequential numerical designation --- e.g. lmc\_c04. In addition, we have added our two phase calibrators to the sample (named 0515-674 and 1315-46 in this work; see Section \ref{sec-obscal}). All of these sources are listed in column 1 of Table \ref{tab:SourceDat} along with their corresponding RA and Dec. (columns 2 \& 3 respectively).

\begin{deluxetable*}{lclccccc} 
\tabletypesize{\footnotesize} 
\tablecolumns{8} 
\tablewidth{0pt} 
\tablecaption{Summary of observational data used in this work} 
\tablehead{ 
\colhead{(1)} & \colhead{(2)} & \colhead{(3)} & \colhead{(4)} & \colhead{(5)} & \colhead{(6)} & \colhead{(7)} & \colhead{(8)} \\ 
\colhead{Observations} & \colhead{$\nu$ span} & \colhead{Epoch} & \colhead{ATCA config.} & \colhead{Beam} & \colhead{$\delta\phi$} & \colhead{$\phi_\text{Max-scale}$} & \colhead{$|\phi_\text{Max}|$}\\ 
\colhead{} & \colhead{[GHz]} & \colhead{} & \colhead{} & \colhead{["]} & \colhead{[rad m$^{-2}$]} & \colhead{[rad m$^{-2}$]} & \colhead{[rad m$^{-2}$]}
} 
\startdata 
 LMC (Archival) & 1.328--1.432 & 1994 Oct 26 -- Nov 9 &750D & 60$\times$60 & 490 & 72 & 3200  \\
 " & " & 1995 Feb 23 -- Mar 11 & 750A & " & " & " & " \\
 " & " & 1995 Jun 2--7 & 750B & " & " & " & " \\
 " & " & 1995 Oct 15--31 & 750B & " & " & " & " \\
 " & " & 1996 Jan 27 -- Feb 8 & 750C & " & " & " & " \\
 \\
 Cen A (Archival) & 1.296--1.480 & 2006 Dec 20 -- 2007 Jan 7 & 750A & 66$\times$33 & 280 & 77 & 3300  \\
 " & " & 2007 Feb 23 -- Mar 13 & 750D & " & " & " & " \\
 " & " & 2007 Dec 7 -- 30 & 750C & " & " & " & " \\
 " & " & 2008 Feb 16 -- Mar 13 & 750B & " & " & " & " \\
 \\
 This paper & 1.350--9.900 & 2012 Feb 10--12 & 6A & 15$\times$15--1$\times$1 & 70 & 3400 & 7900  \\
 " & " & 2012 Jun 22--24 & 6D & " & " & " & " \\
 " & " & 2012 Aug 17--19 & 6A & " & " & " & "
\enddata 
\tablecomments{For the current observations, the quoted range in synthesized beam size (column 5) corresponds to the change between 1.3 \& 9.9 GHz. $\delta\phi$ (column 6), max-scale (column 7) and $|\phi_{max}|$ (column 8) are the resolution in $\phi$-space, the maximum detectable emission scale in $\phi$-space and the maximum detectable Faraday depth of emission respectively. Note that the current observations have $\phi_\text{Max-scale}\gg\delta\phi$, while the opposite is true for the archival observations.} 
\label{table:observations}
\end{deluxetable*}

\begin{deluxetable*}{l l l c c c l c l c } 
\tabletypesize{\footnotesize} 
\setlength{\tabcolsep}{0.015in} 
\tablewidth{0pt} 
\tablecaption{Selected observational properties of our sample sources} 
\\ 
\tablehead{
\colhead{(1)} & \colhead{(2)} & \colhead{(3)} & \colhead{(4)} & \colhead{(5)} & \colhead{(6)} & \colhead{(7)} & \colhead{(8)} & \colhead{(9)} & \colhead{(10)} 
\\ 
\colhead{Source} & \colhead{RA} & \colhead{Decl.} & \colhead{Self calibrated}& \colhead{Maximum} & \colhead{$\nu_{\text{res}}$} & \colhead{$I_{1.4}$} & \colhead{$I_{\text{model}}$} & \colhead{$\langle\alpha_{1.3}^{10}\rangle$} & \colhead{AT20G HARC} 
\\ 
\colhead{} & \colhead{} & \colhead{} & \colhead{bands} & \colhead{angular size} & \colhead{} & \colhead{} & \colhead{polynomial deg.} & \colhead{} & \colhead{} 
\\ 
\colhead{} & \colhead{} & \colhead{}  & \colhead{} & \colhead{["]} & \colhead{[GHz]} & \colhead{[Jy]} & \colhead{} & \colhead{} & \colhead{} 
}
\startdata 
lmc\_c16 &  04:35:41.95 & -62:58:44.27 & 16, 6, 3 & 4 & $\circ$&0.3707(8)&5&-0.849(7)& - \\ 
lmc\_c15 &  04:38:17.75 & -73:09:56.40 & 16, 6, 3 & 5 & $\circ$ &0.1224(7)&4&-0.915(6)& - \\
lmc\_s14 &  04:42:25.27 & -67:28:05.12 & 16, 6 & 20 & $\circ$ &0.0178(3)&4&-0.96(3)& - \\
lmc\_c07 &  04:45:13.38 & -65:47:10.65 & 16, 6, 3 & - & $\cdot$ &0.1208(3)&5&-0.324(5)& 0.95, 0.95 \\
lmc\_s13 &  05:16:36.84 & -72:37:10.86 & 16, 6, 3 & - & $\cdot$ &0.2021(5)&7&+0.357(9)& 0.99 \\
0515-674 & 05:15:37.56 & -67:21:27.78 & 16, 6, 3 & - & $\cdot$ & 1.5443(2) &4&-0.763(2) & 0.98, 0.98 \\
lmc\_c06 &  05:17:14.75 & -62:51:13.94 & 16, 6 & 17 & $\circ$ &0.0324(6)&4&-0.68(2)& - \\
lmc\_s11 &  05:41:50.69 & -73:32:14.01 & 16, 6, 3 & 5 & $\circ$ &0.987(2)&7&-0.421(4)& 0.87 \\
lmc\_c04 &  05:45:54.59 & -64:53:28.20 & 16, 6, 3 & - & $\cdot$ &0.445(1)&3&-0.552(3)& 0.60, 0.72\\
lmc\_c02 &  06:01:11.54 & -70:36:07.20 & 16, 6, 3 & - & $\cdot$ &0.2996(6)&7&+0.135(8)& -\\ 
lmc\_c03 &  06:10:12.90 & -74:32:05.95 & 16, 6, 3 & 35 & $\circ$ &0.733(2)&7&-0.833(6)& -\\
lmc\_c01 &  06:20:02.55 & -69:16:52.72 & 16, 6, 3 & 4 &$\circ$&0.1014(5)&4&-0.685(5)& - \\
cen\_s1681 &  13:16:05.50 & -40:38:24.00 & 16, 6 & 4 & $\circ$ &0.0184(3)&4&-1.05(2)& - \\
cen\_c1093 & 13:16:06.06 & -39:20:06.00 & 16, 6, 3 & - & $\cdot$ &0.0200(3)&3&-0.664(9)& - \\
cen\_c1972 &  13:17:13.00 & -41:09:34.00 & 16, 6, 3 & 2 & 6.3&0.167(1)&3&-1.416(9)& - \\
0315-46 & 13:18:30.06 & -46:20:34.90 & 16, 6, 3 & - & $\cdot$ & 2.197(4)&4&-1.004(3)& - \\
cen\_s1031 &  13:18:31.40 & -40:50:55.00 & 16, 6 & - & $\cdot$ &0.0174(5)&2&-0.83(1)& - \\
cen\_c1832 &  13:19:01.40 & -41:53:18.00 & 16, 6 & 33 &$\circ$&0.0384(4)&3&-0.880(8)& - \\
cen\_c1573 &  13:19:31.30 & -44:59:43.00 & 16, 6, 3 & 7 & $\circ$ &0.107(1)&5&-1.17(8) & - \\
cen\_c1748 &  13:19:43.00 & -44:59:04.00 & 16, 6, 3 & - & $\cdot$ &0.0235(6)&3&-0.91(2)&- \\
cen\_s1014 &  13:19:56.60 & -44:48:22.00 & 16 & 85 & $\circ$ & 0.042(2) &2& -2.3(2) & - \\
cen\_c1827 &  13:21:04.00 & -41:14:51.00 & 16, 6, 3 & 38 & $\circ$ &0.285(2)&7&-1.33(2)& - \\  
cen\_c1466 &  13:23:04.00 & -38:49:05.00 & 16, 6, 3 & - & $\cdot$ &0.2925(8)&7&-0.16(1)& 0.99\\
cen\_c1640 &  13:23:55.00 & -45:53:02.00 & 16, 6 & - & $\cdot$ &0.0108(4)&2&-0.47(2)& - \\
cen\_s1568 &  13:27:57.30 & -45:50:19.00 & 16, 6, 3 & - & $\cdot$ &0.0096(4)&4&+0.31(1)& - \\
cen\_s1349 &  13:30:11.00 & -40:54:36.00 & 16, 6, 3 & - & $\cdot$ &0.177(1)&4&-0.542(8)& - \\
cen\_s1382 &  13:30:16.30 & -38:57:32.00 &16 & - & $\cdot$ &0.0072(2)&1&-0.45(6)& - \\
cen\_s1437 &  13:31:08.50 & -44:29:54.00 & 16, 6, 3 & - & $\cdot$ &0.0496(3)&4&-0.702(7)& - \\
cen\_c1764 &  13:31:18.50 & -41:19:56.00 & 16, 6 & 50 & $\circ$ &0.392(4)&3&-1.45(2)& - \\
cen\_s1605 &  13:32:02.30 & -46:31:17.00 & 16, 6 & 3 &2.4&0.0098(3)&3&-0.64(2)& - \\
cen\_c1152 &  13:33:20.60 & -41:09:12.00 &16 & - & $\cdot$ &0.0143(5)&2&-0.79(6)& - \\
cen\_c1435 &  13:33:21.40 & -44:05:22.00 & 16, 6 & 90 & $\circ$ &0.179(3)&5&-1.37(2)& - \\
cen\_c1636 &  13:34:31.20 & -42:38:50.00 & 16, 6 & 5 & $\circ$ &0.0101(2)&5&-1.37(3)& - \\
cen\_s1443 &  13:35:43.40 & -39:45:59.00 & 16, 6 & - & $\cdot$ &0.0089(4)&2&-0.70(2)& - \\
cen\_s1803 &  13:35:56.50 & -39:30:06.00 & 16, 6, 3 & 13 & $\circ$ &0.0943(8)&4&-1.25(2)&- \\
cen\_s1290 &  13:39:50.00 & -45:50:59.00 & 16, 6, 3 & - & $\cdot$ &0.0370(3)&4&-0.245(8)& - \\
\enddata 
\tablecomments{In column 5, the maximum angular size was determined from images at 1.4 GHz for extended sources, and from images at the highest frequency at which the source remained visible in cases where the source resolved into multiple separated sub-components towards higher frequencies. In column 6, a `$\cdot$' symbol indicates that a source is unresolved over 1.3--10 GHz, by which we mean its peak-to-total flux density ratio is at least 0.95 over the full range. A `$\circ$' symbol indicates that a source is resolved at all frequencies between 1.3 \& 10 GHz (i.e. a peak-to-total flux density ratio of $<0.95$), while a number indicates the frequency in GHz above which the source becomes resolved. These designations apply to observations conducted with the ATCA array configurations listed in column 4 of Table \ref{table:observations} and imaged using robust $=0$ weighting. $I_{1.4}$ (column 7) is the value of the Stokes I model at 1.4 GHz. $\langle\alpha_{1.3}^{10}\rangle$ (column 9) is the mean spectral index over the 1.3--10 GHz band. The uncertainties on the AT20G high angular resolution catalogue (HARC) visibility ratios listed in column 10 are all $\sim\pm0.03$.} 
\label{tab:SourceDat} 
\end{deluxetable*}

\section{Observations and calibration}\label{sec-obscal}

We observed the 40 sample sources using the Australia Telescope Compact Array (ATCA; Wilson et al. 2011) over 1.1--3.1 GHz, 4.5--6.5 GHz and 8.0--10.0 GHz in full polarization with nominal 1 MHz channel resolution. At each end of each of these bands, 100 channels were flagged due to diminished sensitivity in these regions, resulting in an effective frequency coverage of 1.2--3.0 GHz, 4.6--6.4 GHz and 8.1--9.9 GHz. Henceforth, we refer to these as the 16 cm, 6 cm and 3 cm bands respectively. 

\begin{figure}
\includegraphics[width=0.475\textwidth]{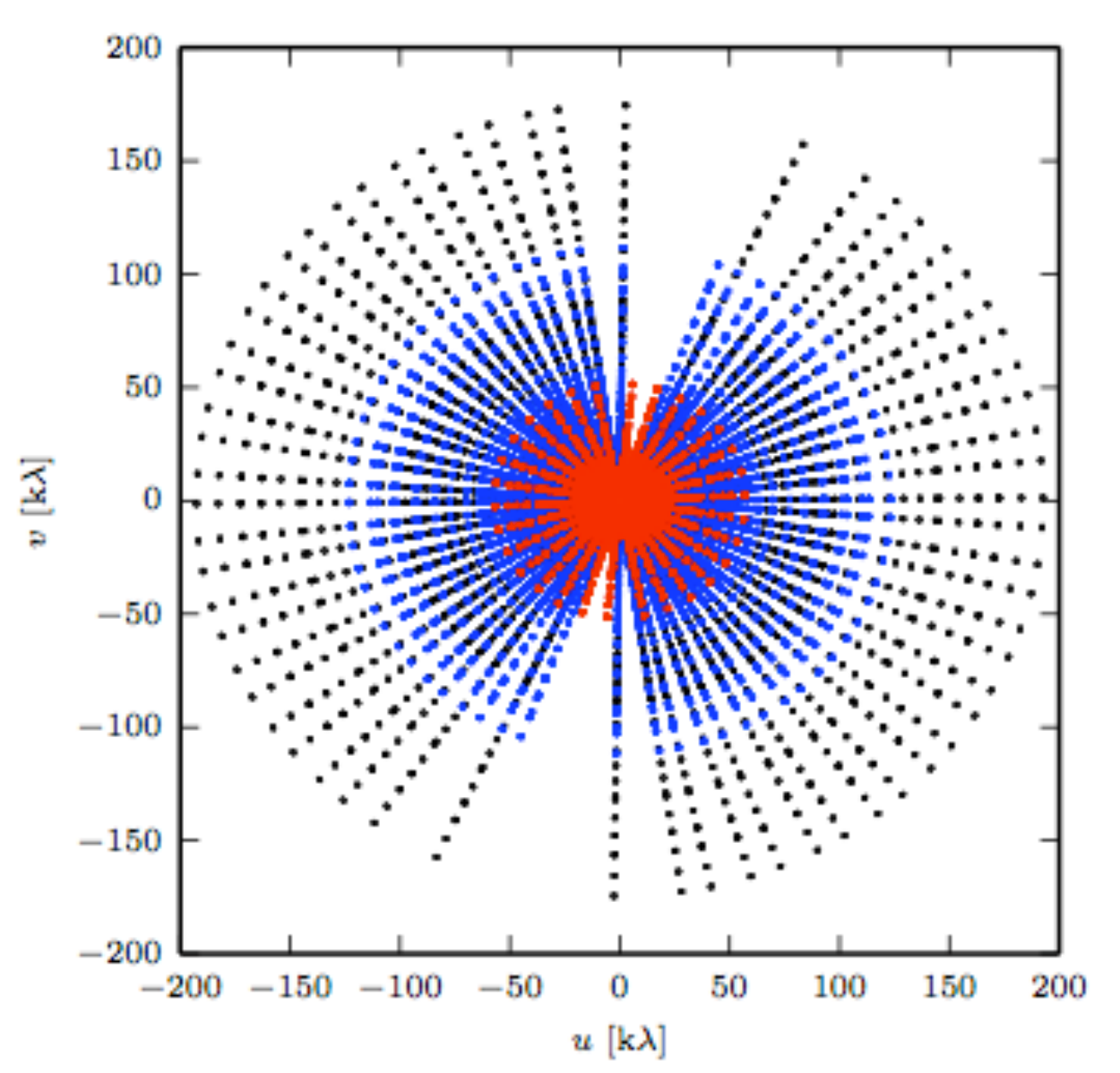}
\caption{The frequency-averaged conjugate $uv$ coverage of our observations for a typical source (lmc\_s11). The red, blue and black data points correspond to the 16, 6 \& 3 cm bands respectively. Each data point represents an average $uv$ coordinate, corresponding to the mean frequency of the lower, middle or upper third of frequencies in each band for a single visit of the telescope to the source.}
\label{fig:C2570_uvcoverage}
\end{figure}

Each source was observed on axis over a full 12 hour synthesis, receiving a total integration time of 23 minutes split over 15 $uv$ cuts for the 16 cm observations, and 50 minutes split over 33 $uv$ cuts for the 6 \& 3 cm band observations. We used 6 km array configurations to achieve high spatial resolution and to maximize overlap in $uv$ coverage between bands (see Fig. \ref{fig:C2570_uvcoverage}) for sources that became resolved towards higher frequencies. We provide images of these sources in Appendix \ref{sec-appendA}. The minimum $uv$ scale in the 16 cm, 6 cm \& 3 cm bands is 1.4 k$\lambda$, 4.8 k$\lambda$ \& 8 k$\lambda$ respectively, corresponding to a maximum angular scale sensitivity of approximately 130", 40" \& 22". Important details of our observations are summarized in Table \ref{table:observations}. 

The data were reduced and calibrated using standard procedures for cm-band ATCA data. Radio-frequency interference (RFI) was flagged iteratively throughout the calibration process using the {\sc sumthreshold} algorithm (Offringa et al. 2010). RFI was so severe below 1.3 GHz that all data below this frequency were discarded. Daily observations of either PKS B1934-638 or PKS B0823-500 were used to calibrate the bandpass response and primary flux scale. The time-dependent complex antenna gains and on-axis polarization leakage were calibrated using PKS B0515-674 and MRC B1315-460 for the LMC and Centaurus sources respectively, observed half-hourly for the 16 cm band observations and every 20 minutes for the 6 \& 3 cm band observations. Independent calibration solutions were derived at 128 MHz intervals in each band then interpolated to account for the frequency dependence of the complex gain and leakage (Schnitzeler et al. 2011). We have included these two calibrators as sources in our sample; their properties are discussed alongside those of the other sample sources in the relevant sections of our paper.

We phase and amplitude self-calibrated the data for all sources in our sample for bands where adequate signal was present. To handle the large fractional bandwidths involved, we split the data for self calibration into sixteen 128 MHz sub bands for the 16 cm data, eight 256 MHz sub bands for the 6 cm data, and four 512 MHz sub bands for the 3 cm data, then derived separate self-calibration solutions in each sub band. The calibration solutions were interpolated and applied continuously across each of the 2 GHz bands. For sources and bands in which good phase and amplitude self-calibration solutions were found, we compared the data extracted from images made both before and after the procedure. We found that for the pre-self-calibration images, the source flux densities were systematically lower (by up to 20\%), and the noise levels higher by a factor of up to 10, than for the post-self-calibrated images. This effect was particularly evident for the 3 cm band data. As such, we chose to discard data for sources in bands where both phase and amplitude self-calibration was unsuccessful. We list the frequency bands in which self calibration was successful for each source in column 4 of Table \ref{tab:SourceDat}.
 
For the archival data extracted from mosaiced images, the estimated maximum polarization leakage in the mosaiced images is $\sim$0.1\% of Stokes I (e.g. Feain et al. 2009). For the current on-axis observations, the frequency-dependent leakage is no greater than $\approx0.05$\% of Stokes I (e.g. Schnitzeler et al. 2011), and as low as 0.002\% of Stokes I when averaged over the full available bandwidth (see results for 0515-674 presented in Section \ref{sec-results-bestfit}).

\section{Spectropolarimetric  analysis}\label{sec-specpol}
  
\subsection{Imaging and polarized signal extraction}\label{sec-extract}

We imaged each source in Stokes $I$, $Q$ \& $U$ at 20 MHz intervals through the 16 cm \& 6 cm bands, and at a substantially coarser resolution of 200 MHz in the 3 cm band after verifying that no components with extreme Faraday depth were present in the calibrated $uv$ data. We used robust $=0$ weighting (Briggs 1995) to moderately down-weight data in sparsely populated regions of the $uv$ plane. The higher spatial resolution of the new vs. archival data (Table \ref{table:observations}) meant that several sources are resolved at all frequencies, some become resolved above some frequency $\nu_{\text{res}}$, while the rest are fully unresolved. We convey this information by recording either `$\circ$', $\nu_{\text{res}}$ or `$\cdot$' (respectively) in column 5 of Table \ref{tab:SourceDat}. For the sources which became resolved above a frequency $\nu_{\text{res}}$, we applied a taper to the $uv$ data above this frequency to match the resolution of the observations at $\nu_{\text{res}}$. For the fully resolved sources, we tapered the $uv$ data to match the scale-size sensitivity at 1.3 GHz. Of the 40 sources originally selected for our new observations, six subsequently proved so faint and/or heavily resolved that we discarded them from the study. Including the two calibrators then, our final sample consists of the 36 sources listed in Table \ref{tab:SourceDat}.

For our spectropolarimetric analysis, we extracted the integrated frequency dependent Stokes $I$, $Q$ \& $U$ flux densities for each source. These quantities were measured at the location of the peak pixel in Stokes I for the fully unresolved and partially resolved sources (in the latter case, after convolving the images at $\nu>\nu_{\text{res}}$ to match the resolution at $\nu_{\text{res}}$). To extract the integrated polarization of the fully resolved sources, we applied a mask to the source at its 10\% flux density radio contour and extracted the integrated flux densities from the unmasked region. The uncertainty in each Stokes parameter was estimated via direct measurement of the RMS noise adjacent to the source, modified appropriately for the number of independent beams sampled in the unmasked regions for `$\circ$' sources. 

Our spectropolarimetric analysis was conducted on the fractional Stokes parameters $q(\nu)$ \& $u(\nu)$, which were obtained by fitting a polynomial to log($I$) vs. log($\nu$), transforming this model into linear space then dividing it out of Stokes $Q$ \& $U$. The 1.4 GHz Stokes I flux density predicted by the model, the degree of the polynomial fit required, and the spectral index of the model at 1.4 GHz are presented in columns 6, 7 \& 8 of Table \ref{tab:SourceDat}  respectively.

\subsection{Modelling Stokes q($\lambda^2$) \& u($\lambda^2$)}\label{sec-modelling}

A key goal of our work is to identify and characterize Faraday thick components. This has traditionally been achieved by fitting $\boldsymbol{P}(\lambda^2)$ with a depolarization model derived for a specific physical arrangement of magnetoionic material (e.g. O'Sullivan et al. 2012). However, this approach involves making assumptions about the structure of the magnetoionic material under investigation. Moreover, we found that existing models were incapable of reproducing the detailed polarization behaviour of several of our high S/N sources. 

Instead, we modelled our data in a general way that allowed us to quantify the number of dominant emission components in each source, their width and shape in $\phi$ space, and their initial polarization angle. We did this by constructing model FDFs from elementary basis functions. For each model, $n$ polarized emission components $\boldsymbol{p}(\phi)$ (which could be Faraday thick or thin) are combined in $\phi$ space as:
 
 \begin{equation}
\boldsymbol{F}_\phi = \boldsymbol{p}_{1}(\phi) + \boldsymbol{p}_{2}(\phi)+...+\boldsymbol{p}_{n}(\phi)
\label{eq:linearcomblsq}
 \end{equation}
 
then transformed into $\lambda^2$ space (see Section \ref{sec-fitting}) using

 \begin{equation}
\boldsymbol{P}(\lambda^2) = \sum\limits_{\phi} \boldsymbol{F}_\phi e^{2i\phi\lambda^2}
\label{eq:genP}
 \end{equation} 
 
 The goodness of fit of the $\boldsymbol{P}(\lambda^2)$ model to the measured polarization data can then be assessed. 
 
 We modelled Faraday-thin FDF components as complex-valued $\delta$ functions of modulus $p$ with polarization angle $\psi_0$ at a Faraday depth of $\phi$. For Faraday thick components, we used a parametric function capable of generating a variety of $\boldsymbol{P}(\lambda^2)$ behaviours --- a Super-Gaussian (e.g. Decker 1994) which takes on complex values, defined by: 
 
\begin{equation}
\boldsymbol{p}(\phi) = -\frac{A}{\sqrt{2\pi}\sigma_\phi} \text{exp}\bigg(2i\psi_0+\frac{-|\phi-\phi_{\text{peak}}|^N}{2\sigma_\phi^N}\bigg)
\label{eq:supergauss}
\end{equation}

where $A$ is an amplitude parameter, $\psi_0$ is the emitted polarization angle, $\phi_{\text{peak}}$ is the mean Faraday depth of the emission component, $N$ is a shape parameter, and $\sigma_\phi$ is a width parameter which is related to the traditional FWHM of the Gaussian $\sigma$ parameter as:

\begin{equation}
\sigma \approx \sigma_\phi(\pi/2)^{2/N-1}
\label{eq:supergaussb}
\end{equation}

\begin{figure}
\centering
\includegraphics[width=0.450\textwidth]{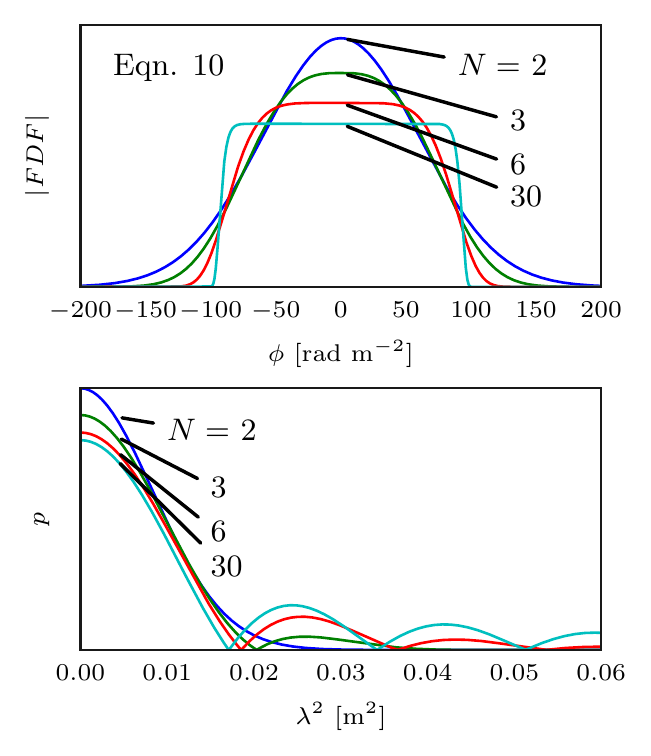}
\caption{\emph{Top:} The modulus (in arbitrary units) of the complex-valued Super-Gaussian function defined by Eqn. \ref{eq:supergauss}, evaluated for $\phi_{\text{peak}}=0$, $\sigma_\phi=60$, $\psi_0=0$ \& for multiple values of $N$ as indicated by the annotations. The curves have been colored to assist in telling them apart. $N=2$ corresponds to a Gaussian function; as $N$ increases, the Gaussian becomes progressively `squared' until achieving a close approximation to a top hat function at $\sim30$. \emph{Bottom:} $|\boldsymbol{P}(\lambda^2)|$ behaviour (in arbitrary units) resulting from the FDF components in the top panel. Note the Gaussian roll off for $N=2$, and the different locations of the nulls \& local maxima for other values of $N$.}
\label{fig:sgdemo}
\end{figure}

With reasonably few parameters, the Super-Gaussian function provides a generic Faraday thick FDF component capable of modelling emission from an assortment of magnetoionic arrangements. For example, when $N=2$, Eqn. \ref{eq:supergauss} reduces to a normal Gaussian function and can be used to model depolarization from a large number of independent magnetoionized cells covering the source in a foreground Faraday rotating medium (i.e. the Burn foreground screen; Burn 1966). As $N$ increases, the Gaussian becomes progressively `squared' (see Fig. \ref{fig:sgdemo}) until at $N\approx30$, $\boldsymbol{p}(\phi)$ becomes a reasonable approximation to a top hat function, which can be used to model a region of mixed emitting and Faraday rotating plasma (i.e. the ``Burn slab"; Burn 1966), linear RM gradients or specific types of foreground depolarization screen (Schnitzeler et al. 2015). Other values of N provide the flexibility to approximate the polarization behaviours expected from non-uniform gradients in the RM of a foreground Faraday rotating medium or an illuminating background source, multiple layers of 2D turbulent foregrounds (Schnitzeler et al. 2015), among other possible structures. While a lone Super-Gaussian cannot describe magnetoionic structures that have a skewed distribution in $\phi$ space, different parametric curves could in principle be used for this purpose. In general, we found that this was not necessary to achieve reasonably good fits to our data (but see Section \ref{sec-results-bestfit}).

We point out that our method avoids the deconvolution ambiguities that afflict the so-called open-ended spectropolarimetric analysis algorithms (see Sun et al. 2015), while at the same time, it avoids making significant a-priori assumptions about the magnetoionic structure of the source under study. It therefore combines some of the benefits of open-ended methods with those of QU-fitting, though with the drawback that using Monte Carlo methods (see Section \ref{sec-fitting} below), the fitting process can be quite time-consuming.

In this work, we refer to FDF models which contain a specific combination of thick and/or thin components as a `model type'. We denote the model type with character strings in which each `S' (`T') indicates the presence of a thin (thick) component (the characters S and T were chosen in reference to `simple' and `thick' respectively). For example, a model consisting of two thin components and a thick component is denoted by the string `SST'. We refer to a specific realization of a model type --- i.e. when all parameters in the model have specified values --- as a `polarization model' or simply `a model'.

\subsection{Fitting the models and selecting the best-fit model(s)}\label{sec-fitting}

We fit our models using Monte Carlo methods. The posterior probability $\text{Pr}(\boldsymbol{\theta}|\boldsymbol{D},\text{M})$ for a set of model parameters ($\boldsymbol{\theta}$) given some data ($\boldsymbol{D}$) and a model type (M) can be calculated using BayesÕ theorem:

\begin{equation}
\text{Pr}(\boldsymbol{\theta}|\boldsymbol{D},\text{M}) = \frac{\text{Pr}(\boldsymbol{D}|\boldsymbol{\theta},\text{M}) \times \text{Pr}(\boldsymbol{\theta}|\text{M})}{\text{Pr}(\boldsymbol{D}|\text{M})} 
\label{eq:bayes}
\end{equation}
 
 The likelihood $\text{Pr}(\boldsymbol{D}|\boldsymbol{\theta},\text{M})=\mathcal{L}$ for polarization data ($q_i$,$u_i$) and a model (q$_{mod,i}$,u$_{mod,i}$) can be quantified as:

\begin{equation}
\mathcal{L} = \prod\limits_{i=1}^n \frac{1}{\pi\sigma_{q_{i}}\sigma_{u_{i}}}\text{exp}\Bigg(-\frac{(q_i-q_{mod,i})^2}{2\sigma_{q_{i}}^2}-\frac{(u_i-u_{mod,i})^2}{2\sigma_{u_{i}}^2}\Bigg)
\label{eq:likelihood}
\end{equation}

For $\text{Pr}(\boldsymbol{\theta}|\text{M})$ --- the prior belief that the model parameters should take on a set of values --- we used a uniform PDF in ranges where these values were physically acceptable: [$-\pi/2$,$\pi/2$) rad for $\psi_0$ (the initial polarization angle), [0,0.7] for $p$ (the fractional polarized amplitude), [-3000,3000] rad m$^{-2}$ for $\phi_{\text{peak}}$ and [0,1500] rad m$^{-2}$ for $\sigma_\phi$. The PDF was set to zero outside these ranges. Note that the prior ranges for  $\phi_{\text{peak}}$ and $\sigma_\phi$ do not probe the full range of parameter space allowed by our data (see Table \ref{table:observations}). We imposed these limits to aid the computational tractability of our fitting procedure.

Finally, since we are only interested in relative probabilities for the purpose of model comparison, we set the normalization factor $\text{Pr}(\boldsymbol{D}|\text{M})=1$.

We used the affine-invariant MCMC sampler {\sc Emcee} (Foreman-Mackay et al. 2013) to sample $\text{Pr}(\boldsymbol{\theta}|\boldsymbol{D},\text{M})$, deeming the parameter values which maximized $\text{Pr}(\boldsymbol{\theta}|\boldsymbol{D},\text{M})$ to specify the best fit capable of being achieved with a given model type. The 1$\sigma$ uncertainties for each fitted parameter were measured directly from the posterior probability distribution (marginalized over all other model parameters). These values are only meaningful when a good fit to the data has been achieved. 

Initially, we fit models of type S, T, SS, ST, TT \& SSS to every source in our sample. Several sources required more complicated model types to achieve a good fit (lmc\_c03, lmc\_c04 \& lmc\_s11; described in Section \ref{sec-results-bestfit}). We constructed these models manually on a case-by-case basis, by adding additional S \& T components to the models until a good fit was achieved. For the best fit obtained for each model type, we quantified the absolute goodness of fit using the reduced chi-squared ($\tilde{\chi}^2$) statistic. 

We compared the relative merit of each fitted model type using the Akaike information criterion (AIC; Akaike 1974). Formally, for a model $M$ the AIC$_M$ is calculated as 

\begin{equation}
\text{AIC}_M = 2k-2\text{ln}(\mathcal{L}_{\text{max}})
\label{eq:AIC}
\end{equation}

where $k$ = no. of fitted model parameters, and $\mathcal{L}_{\text{max}}$ is the maximum of the likelihood distribution for model M. For two models M$_1$ \& M$_2$ with AIC values of AIC$_1$ \& AIC$_2$ respectively, M$_2$ is then exp((AIC$_1-$AIC$_2$)/2) times as likely as M$_1$ to minimize the information lost by over or under fitting the data. We consider a model to be substantially favoured by the AIC when AIC$_1+ 10 <$ AIC$_2$ (i.e. $>99$\% confidence). We refer to the model with the lowest AIC value for a given source as the overall best fit model.
 
\section{Ancillary analysis}\label{sec-extras}

\subsection{RM synthesis \& {\sc rmclean}}\label{sec-extras-RMS}

RM synthesis (Brentjens \& DeBruyn 2005) is a technique for directly calculating $\boldsymbol{P}(\phi)$ (also called the Faraday dispersion spectrum, or FDS) from observations of $\boldsymbol{P}(\lambda^2)$ using a Fourier transform (see also Burn 1966). A procedure called {\sc rmclean} (Heald et al. 2009) is usually then applied to the FDS to remove artefacts caused by imperfect sampling of $\lambda^2$. We applied RM synthesis \& {\sc rmclean} to each source in our sample so that the resulting FDS could be directly compared to our FDF models. 

\subsection{Morphological constraints}\label{sec-extras-AT20G}
We crossmatched our sources with the AT20G high angular resolution catalogue (Chhetri et al. 2013) to provide morphological constraints on $\sim 0.15$" angular scales. This catalogue contains ratios of scalar-averaged visibilities measured on 6km and 4.5km baselines for $\sim 5000$ sources observed by ATCA at 20 GHz. We record the visibility ratios for matched sources in column 9 of Table \ref{tab:SourceDat}.

%
 
\section{Results}\label{sec-results}

\subsection{Spectropolarimetric data}\label{sec-results-observed}

\begin{figure*}[htpb]
\centering
\includegraphics[width=0.95\textwidth]{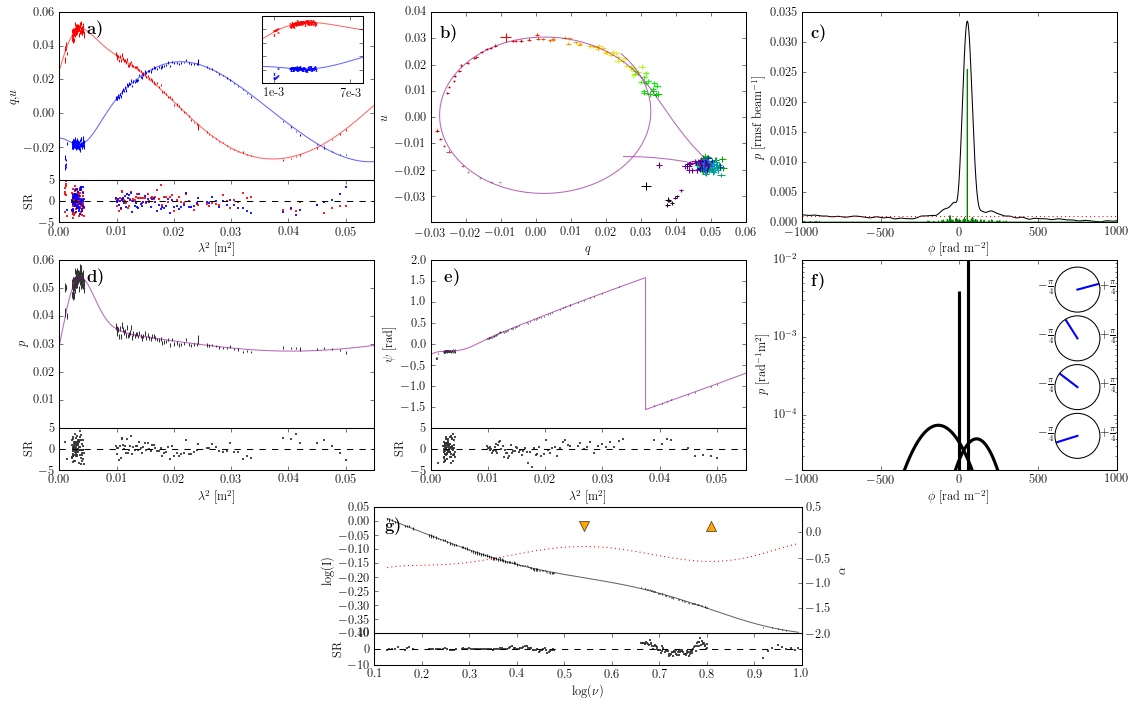}
\caption{Data and models for lmc\_s11 in both linear polarization and total intensity. Panel (a): Stokes $q$ (red) \& $u$ (blue) data + overall best fit polarization model (see Table \ref{tab:fitgoodness}). The inset panel zooms in on detail at short wavelengths, and is only included for sources where significant structure is present in the model / data at these wavelengths. Panel (b): Stokes $u$ vs. $q$ + overall best fit model (purple line). The error bar color changes linearly from blue to red with decreasing $\nu$.  Panel (c): $|$FDS$|$ calculated using RM synthesis and {\sc rmclean} (black line), {\sc rmclean} components (green lines) and {\sc rmclean} cutoff (red dotted line). Panel (d): $p(\lambda^2)$ (black points) + overall best fit polarization model (purple line). Panel (e): $\psi(\lambda^2)$ (black points) + overall best fit polarization model (purple line). Panel (f): Model $|$FDF$|$ (black lines). Note the logarithmic y-axis which differs from panel (c). The inset polar axes display $\psi_0$ for each component in the model FDF, moving from the topmost axis to the bottommost as the $\phi_{\text{peak}}$ value of each component increases. Note that they are half-angle axes, running from $-\pi/2$ to $+\pi/2$ with 0 located at the 12 o'clock position. Panel (g): Stokes I data + polynomial model fit (black line; see Table \ref{tab:fitgoodness} for details) with associated spectral index (red dotted line; secondary axis). Orange `down' (`up') arrows mark values of log($\nu$) where the rate of change of $\alpha$ changes from positive to negative (or vice versa), moving from low to high frequencies. For panels (a), (d), (e) \& (g), the sub panel immediately below the main panel plots the standardized residuals (SR) for each data point (see main text).}
\label{fig:lmc_s11}
\end{figure*}

\begin{figure*}[htpb]
\centering
\includegraphics[width=0.95\textwidth]{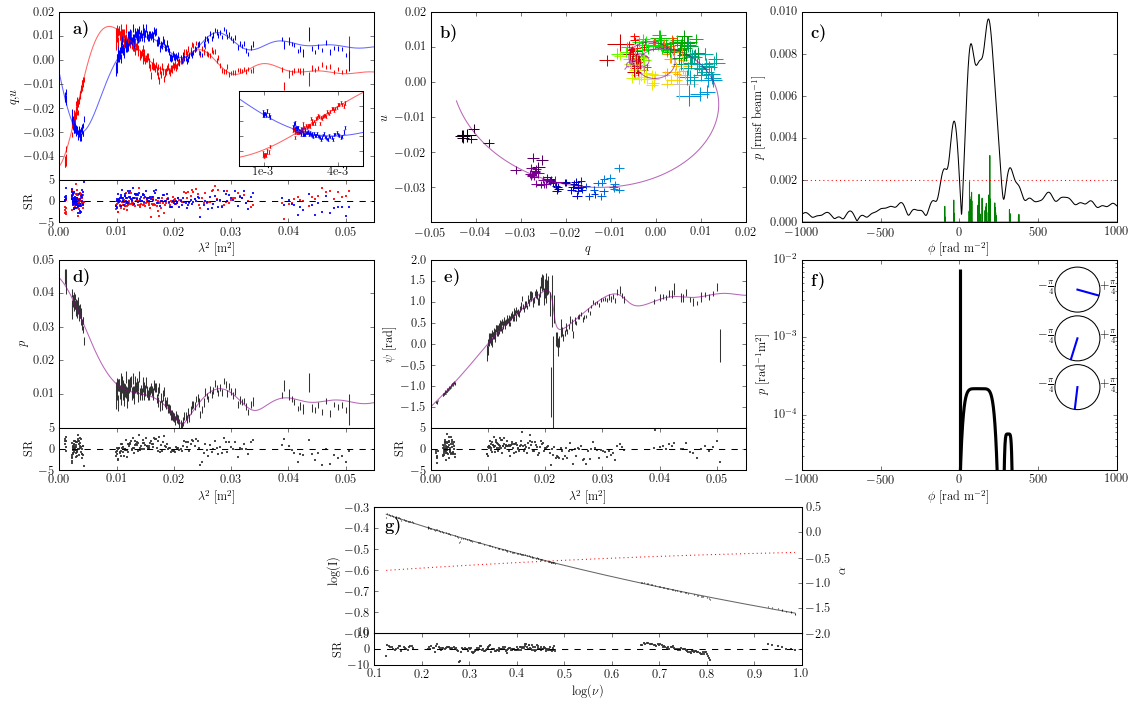}
\caption{As for Fig. \ref{fig:lmc_s11}. Source: lmc\_c04.}
\label{fig:lmc_c04}
\end{figure*}

\begin{figure*}[htpb]
\centering
\includegraphics[width=0.95\textwidth]{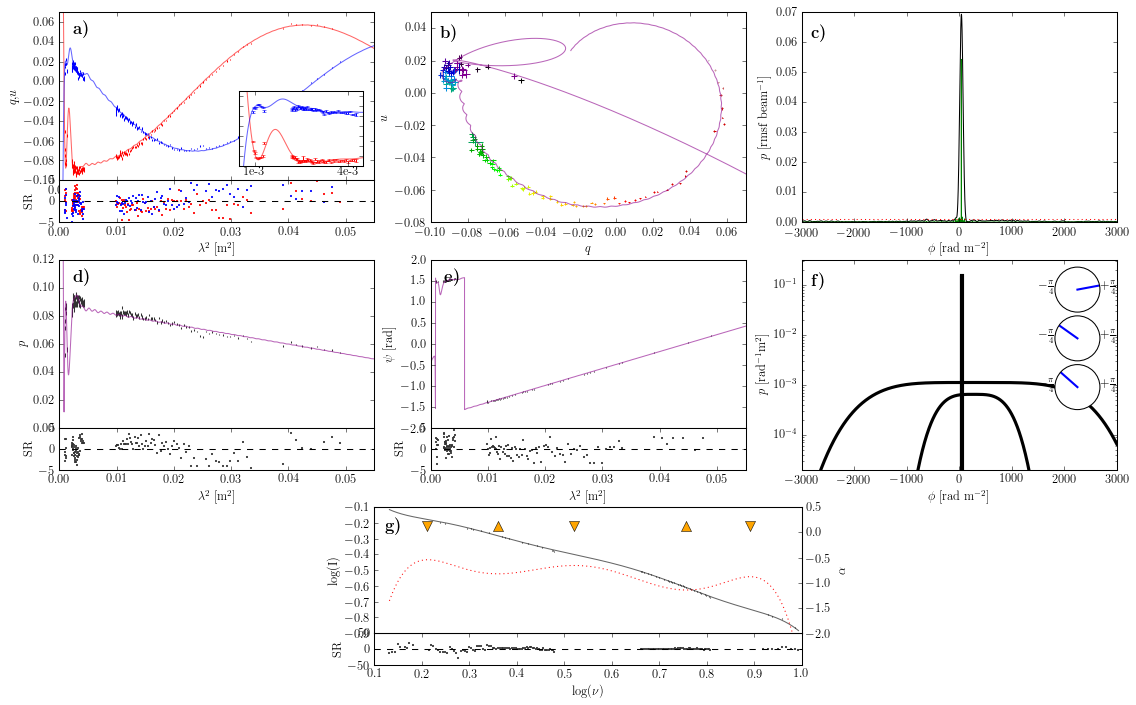}
\caption{As for Fig. \ref{fig:lmc_s11}. Source: lmc\_c03.}
\label{fig:lmc_c03}
\end{figure*}

\begin{figure*}[htpb]
\centering
\includegraphics[width=0.95\textwidth]{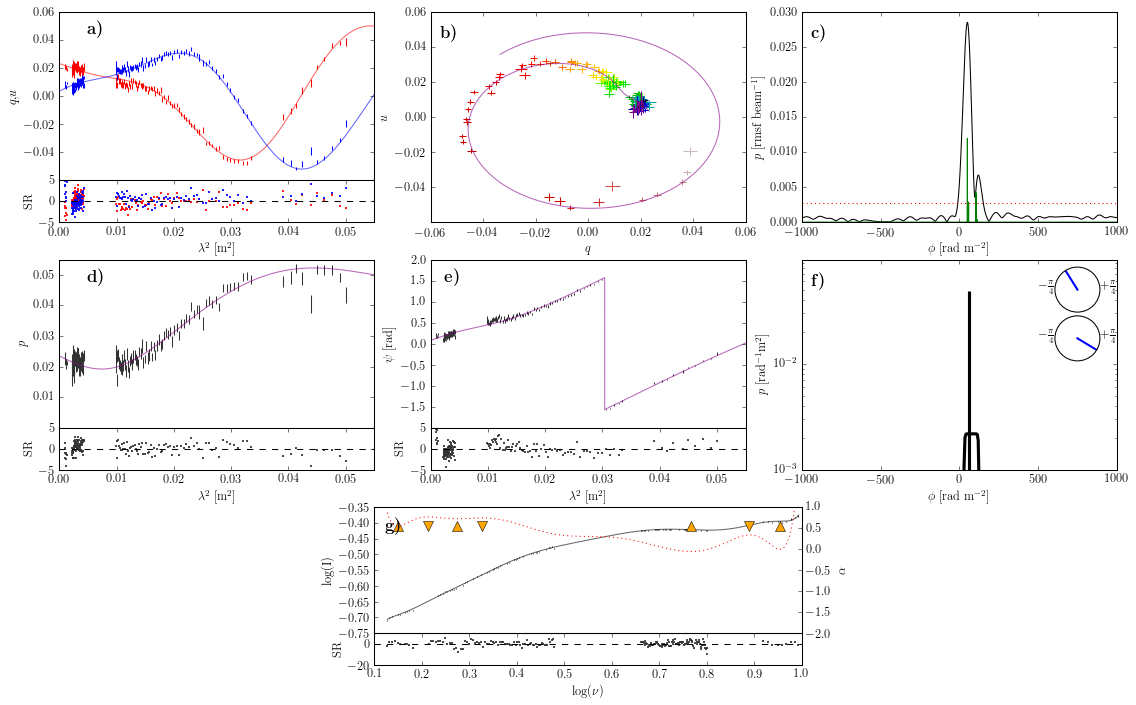}
\caption{As for Fig. \ref{fig:lmc_s11}. Source: lmc\_s13.}
\label{fig:lmc_s13}
\end{figure*}

\begin{figure*}[htpb]
\centering
\includegraphics[width=0.95\textwidth]{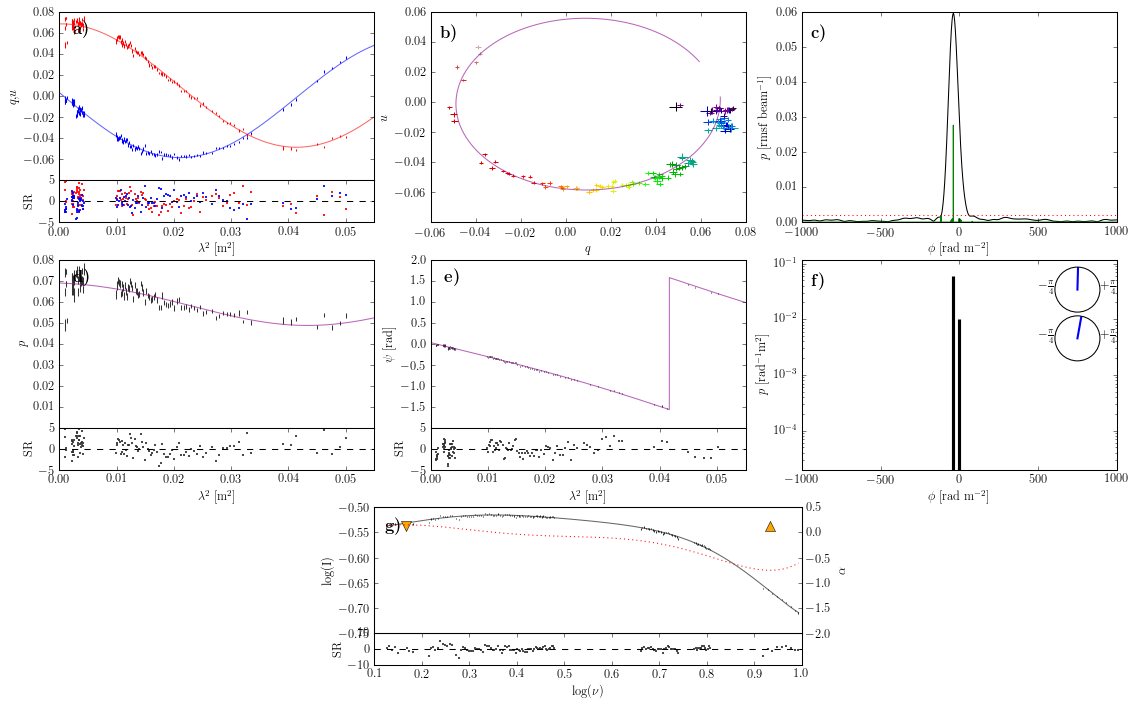}
\caption{As for Fig. \ref{fig:lmc_s11}. Source: cen\_c1466.}
\label{fig:cena_c1466}
\end{figure*}

\begin{figure*}[htpb]
\centering
\includegraphics[width=0.95\textwidth]{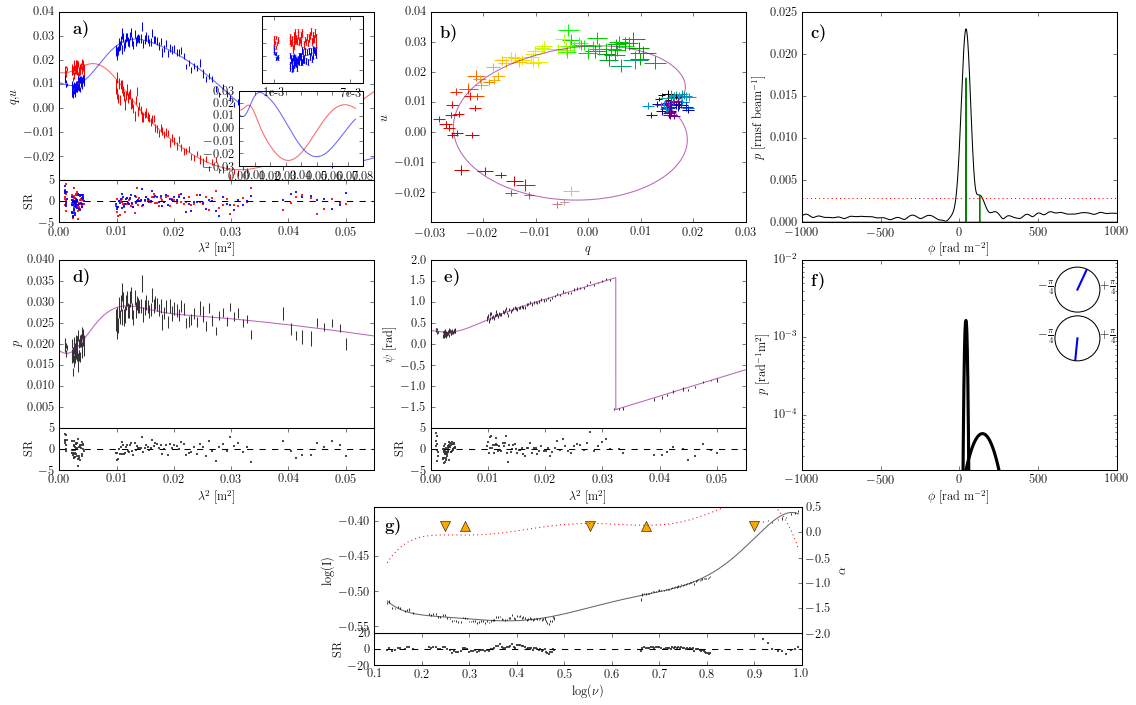}
\caption{As for Fig. \ref{fig:lmc_s11}. Source: lmc\_c02.}
\label{fig:lmc_c02}
\end{figure*}

\begin{figure*}[htpb]
\centering
\includegraphics[width=0.95\textwidth]{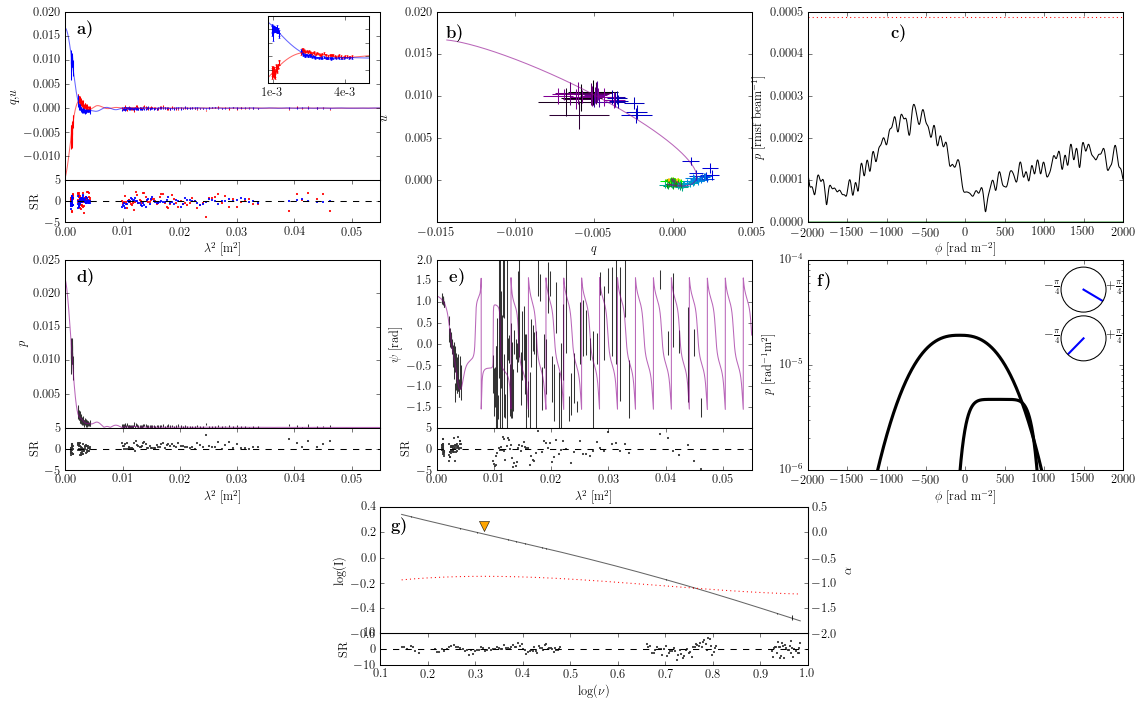}
\caption{As for Fig. \ref{fig:lmc_s11}. Source: 1315-46 (Centaurus A dataset phase calibrator).}
\label{fig:1315-46}
\end{figure*}

\begin{figure*}[htpb]
\centering
\includegraphics[width=0.95\textwidth]{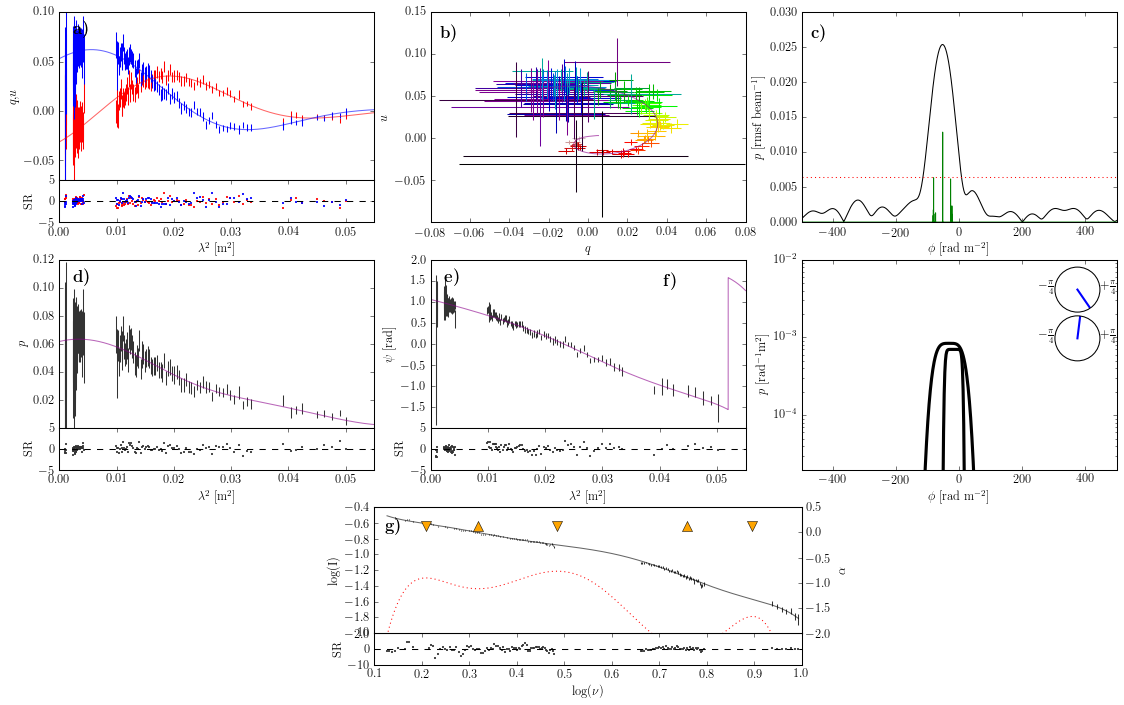}
\caption{As for Fig. \ref{fig:lmc_s11}. Source: cen\_c1827.}
\label{fig:cena_c1827}
\end{figure*}

\begin{figure*}[htpb]
\centering
\includegraphics[width=0.95\textwidth]{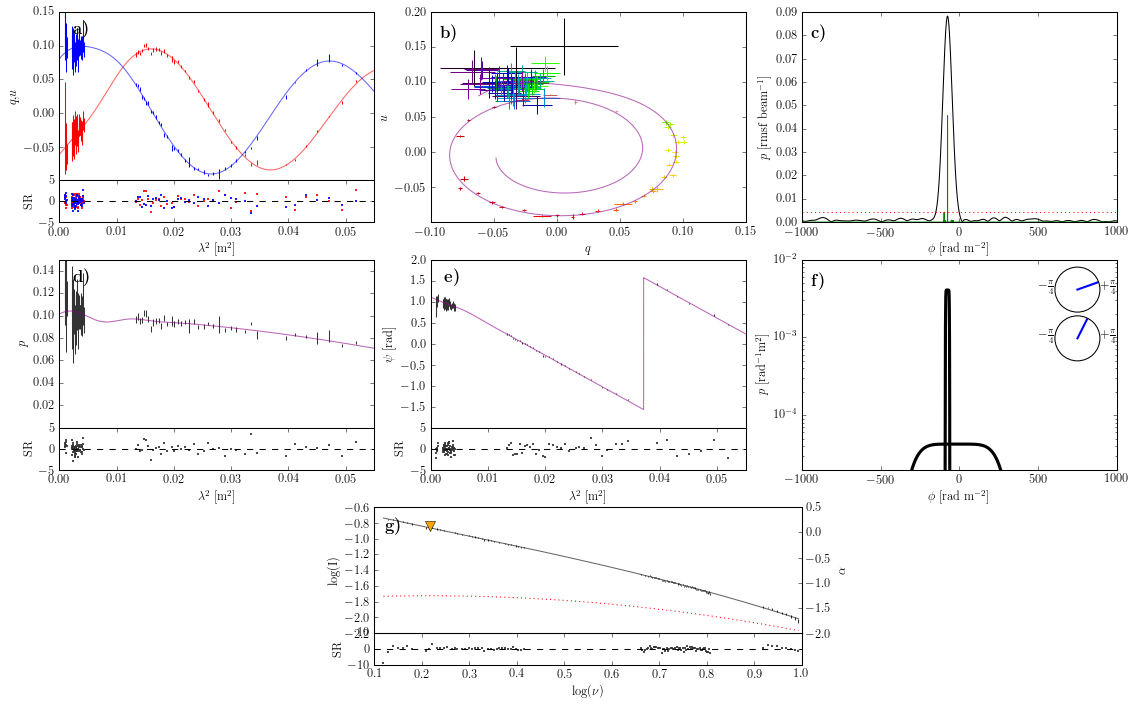}
\caption{As for Fig. \ref{fig:lmc_s11}. Source: cen\_c1972.}
\label{fig:cena_c1972}
\end{figure*}

\begin{figure*}[htpb]
\centering
\includegraphics[width=0.95\textwidth]{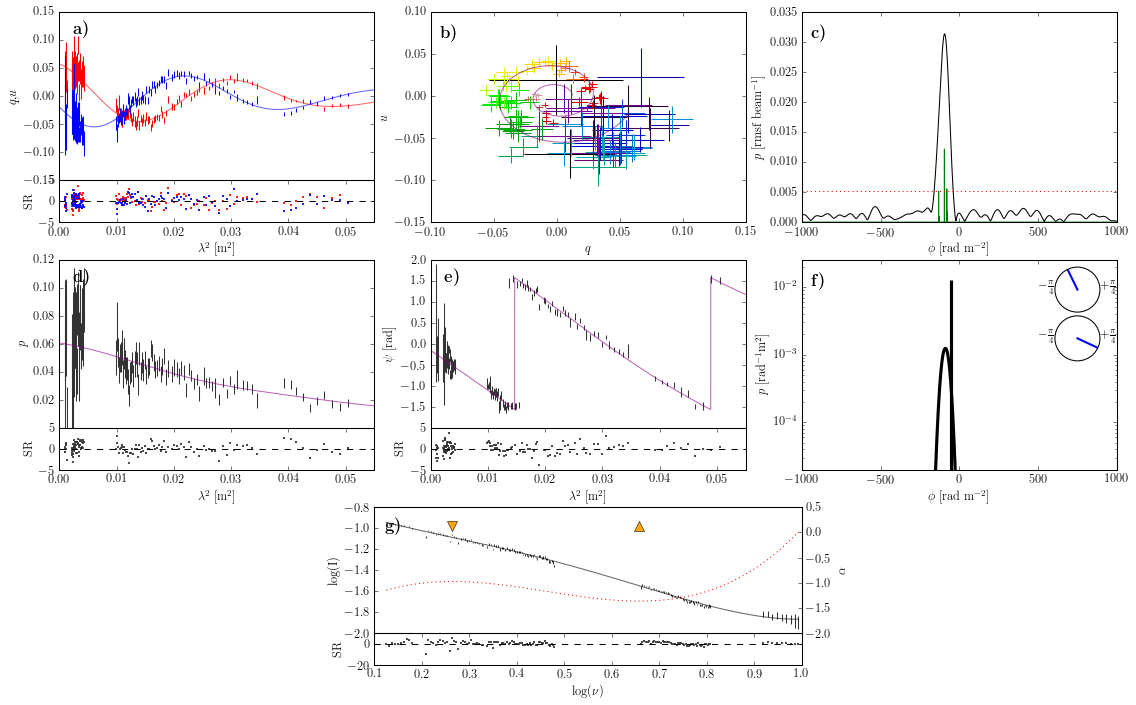}
\caption{As for Fig. \ref{fig:lmc_s11}. Source: cen\_c1573.}
\label{fig:cena_c1573}
\end{figure*}

\begin{figure*}[htpb]
\centering
\includegraphics[width=0.95\textwidth]{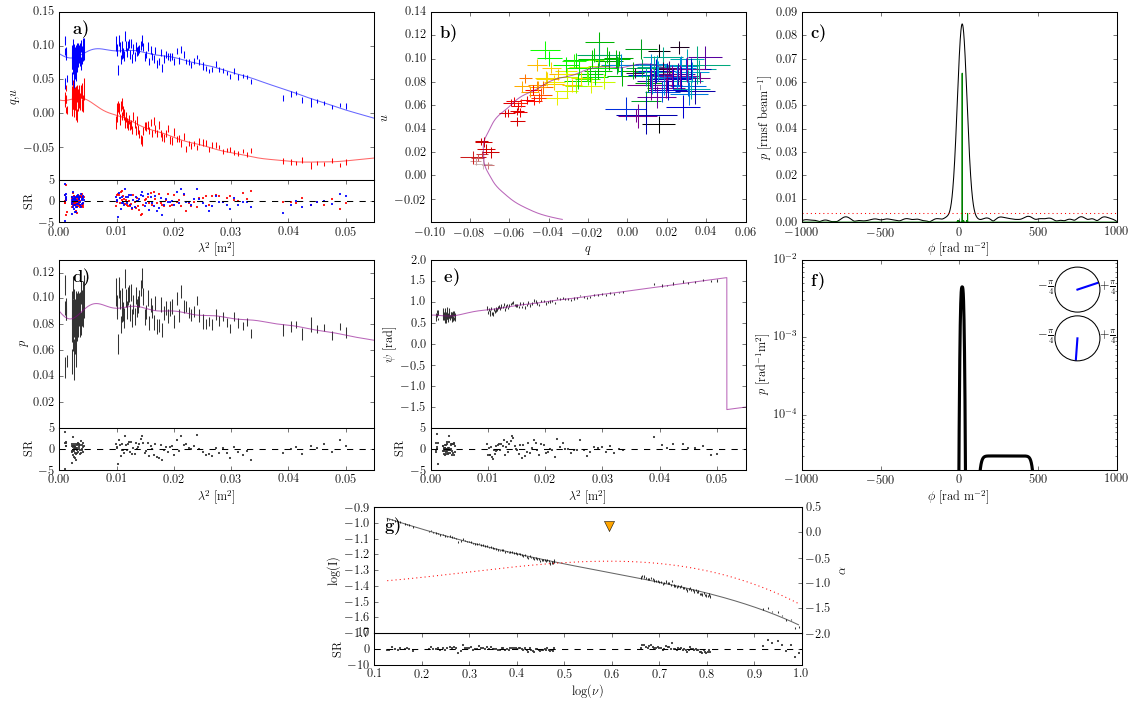}
\caption{As for Fig. \ref{fig:lmc_s11}. Source: lmc\_c01.}
\label{fig:lmc_c01}
\end{figure*}

\begin{figure*}[htpb]
\centering
\includegraphics[width=0.95\textwidth]{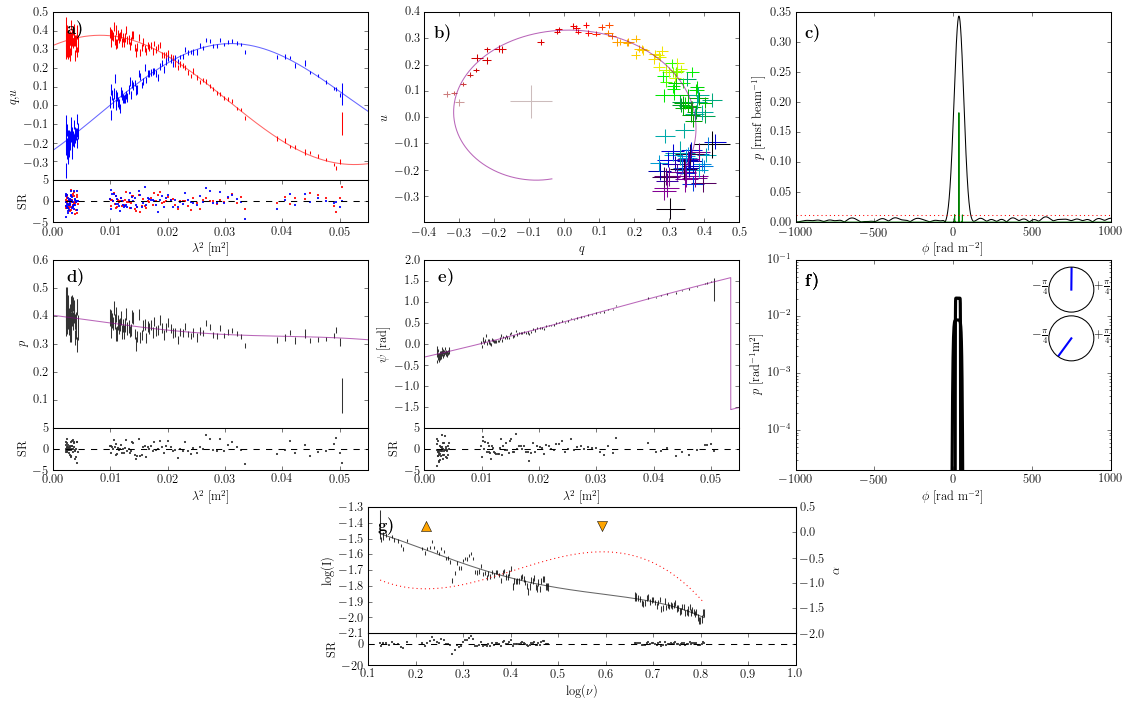}
\caption{As for Fig. \ref{fig:lmc_s11}. Source: lmc\_c06.}
\label{fig:lmc_c06}
\end{figure*}

\begin{figure*}[htpb]
\centering
\includegraphics[width=0.95\textwidth]{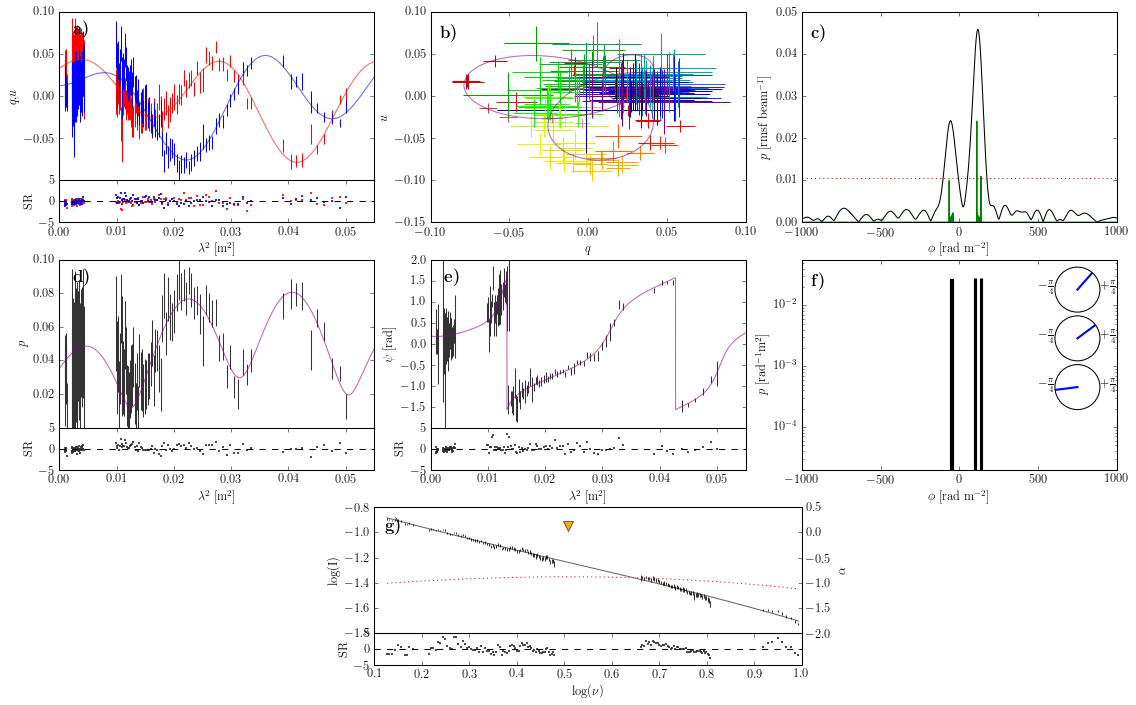}
\caption{As for Fig. \ref{fig:lmc_s11}. Source: lmc\_c15.}
\label{fig:lmc_c15}
\end{figure*}

\begin{figure*}[htpb]
\centering
\includegraphics[width=0.95\textwidth]{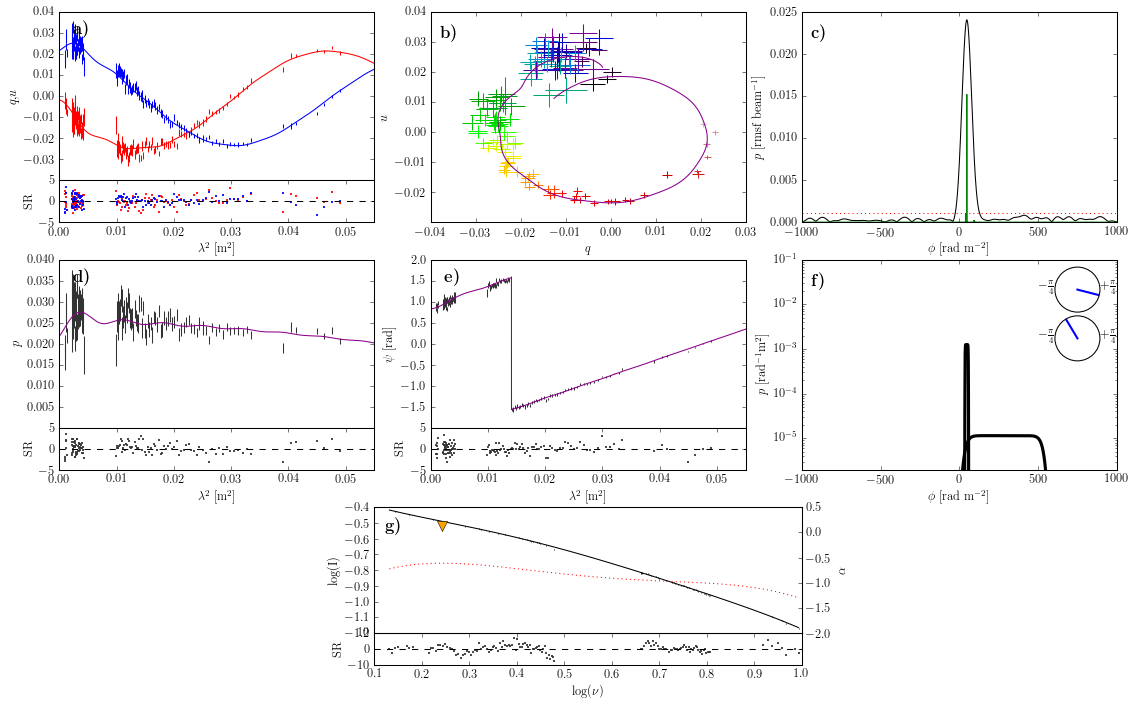}
\caption{As for Fig. \ref{fig:lmc_s11}. Source: lmc\_c16.}
\label{fig:lmc_c16}
\end{figure*}

\begin{figure*}[htpb]
\centering
\includegraphics[width=0.95\textwidth]{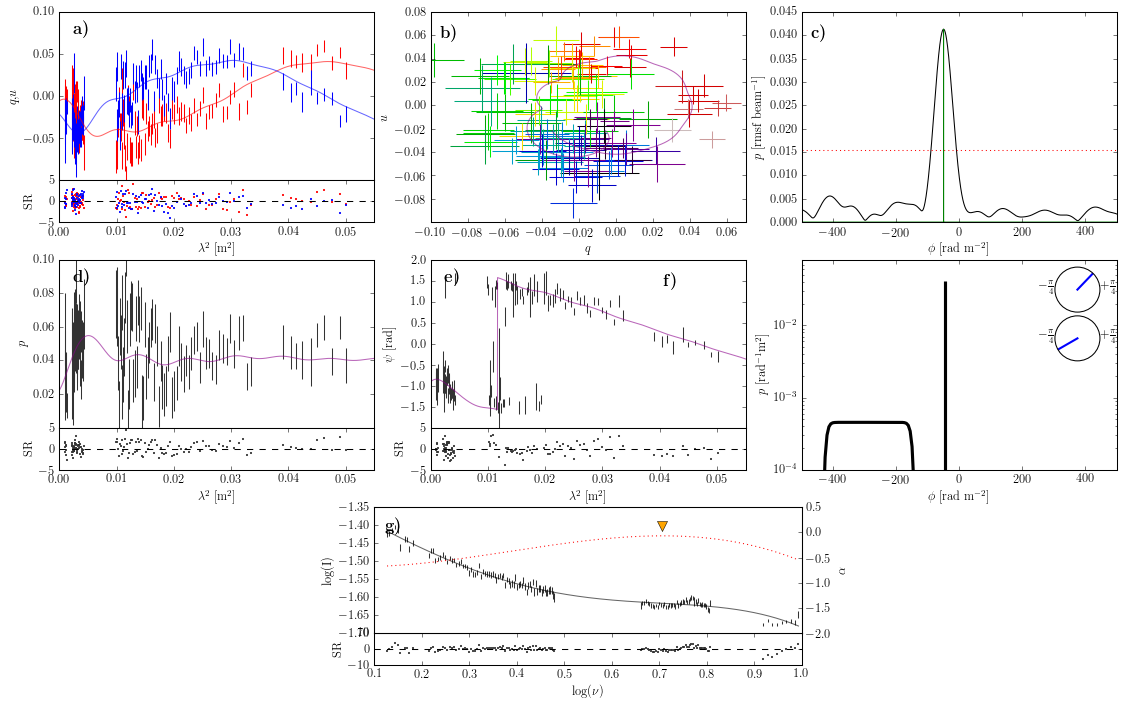}
\caption{As for Fig. \ref{fig:lmc_s11}. Source: cen\_s1290.}
\label{fig:cena_s1290}
\end{figure*}

\begin{figure*}[htpb]
\centering
\includegraphics[width=0.95\textwidth]{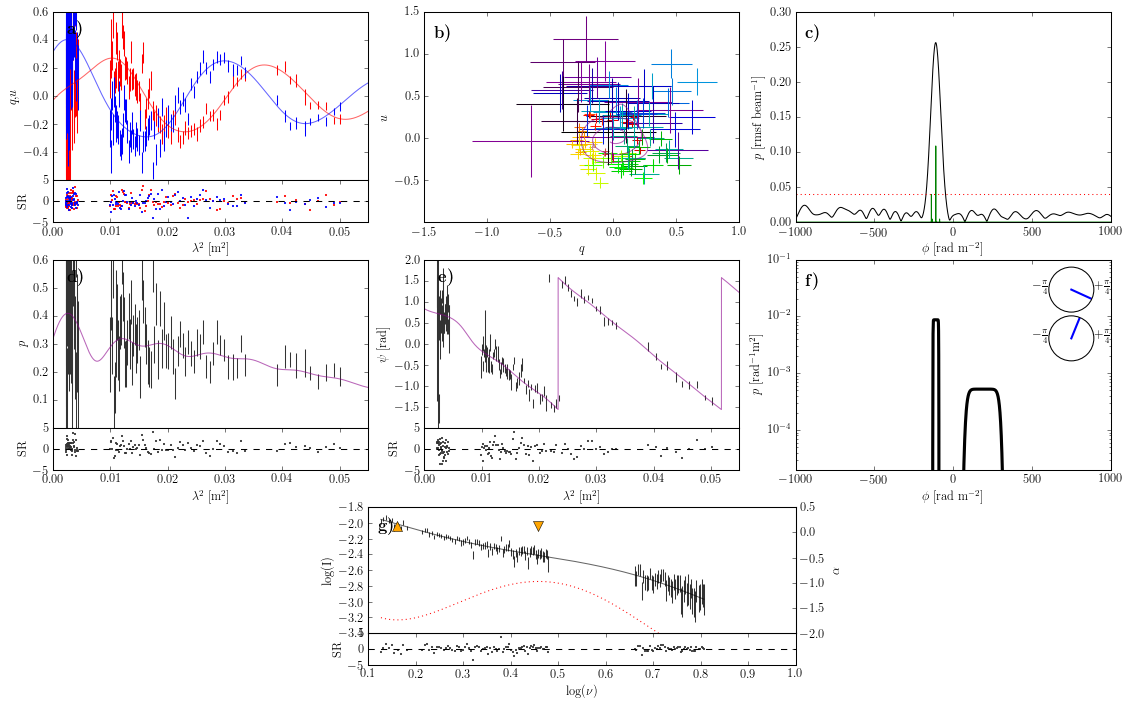}
\caption{As for Fig. \ref{fig:lmc_s11}. Source: cen\_c1636.}
\label{fig:cena_c1636}
\end{figure*}

\begin{figure*}[htpb]
\centering
\includegraphics[width=0.95\textwidth]{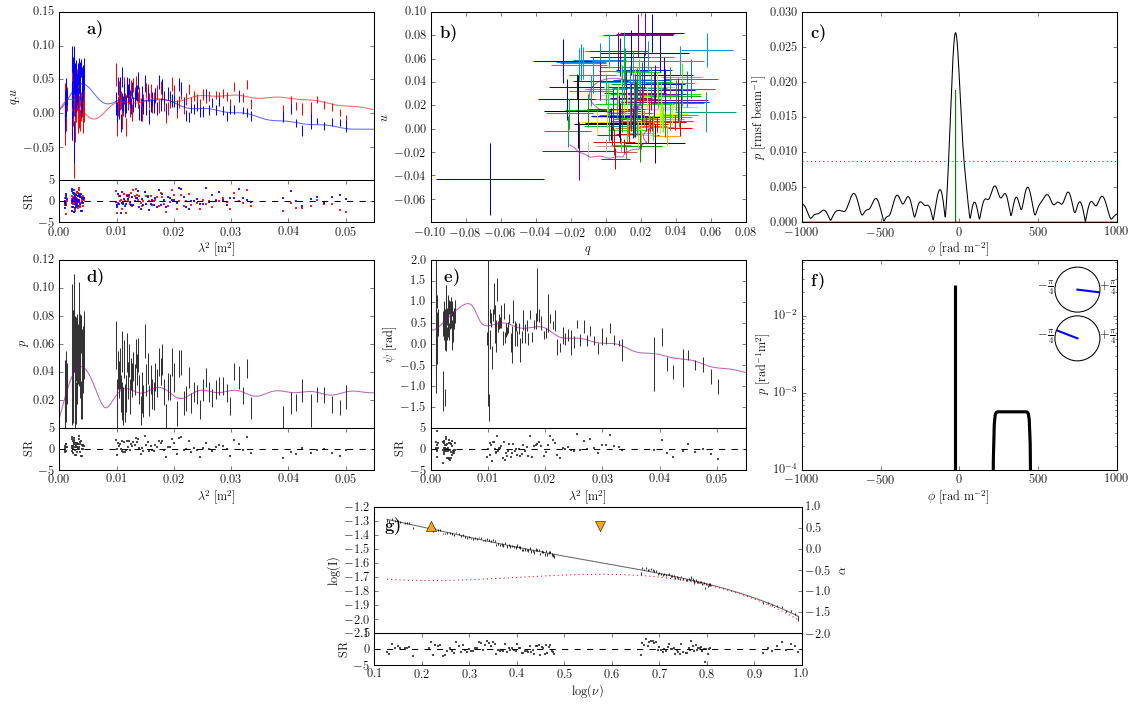}
\caption{As for Fig. \ref{fig:lmc_s11}. Source: cen\_s1437.}
\label{fig:cena_c1437}
\end{figure*}

\begin{figure*}[htpb]
\centering
\includegraphics[width=0.95\textwidth]{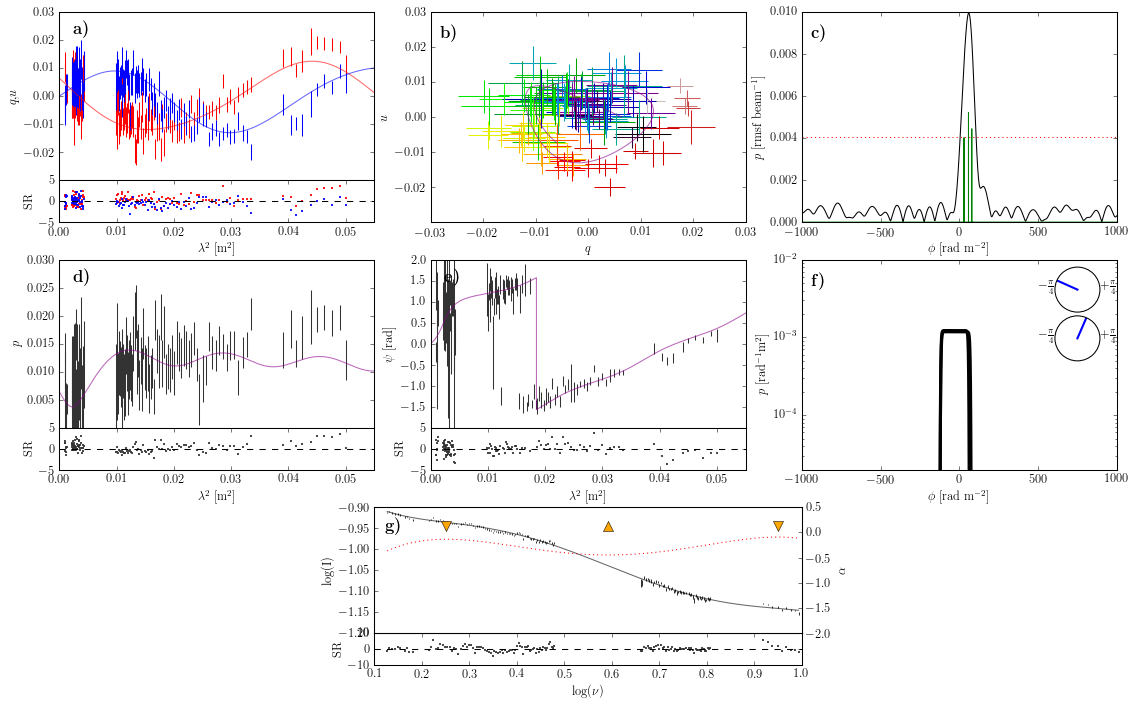}
\caption{As for Fig. \ref{fig:lmc_s11}. Source: lmc\_c07.}
\label{fig:lmc_c07}
\end{figure*}

\begin{figure*}[htpb]
\centering
\includegraphics[width=0.95\textwidth]{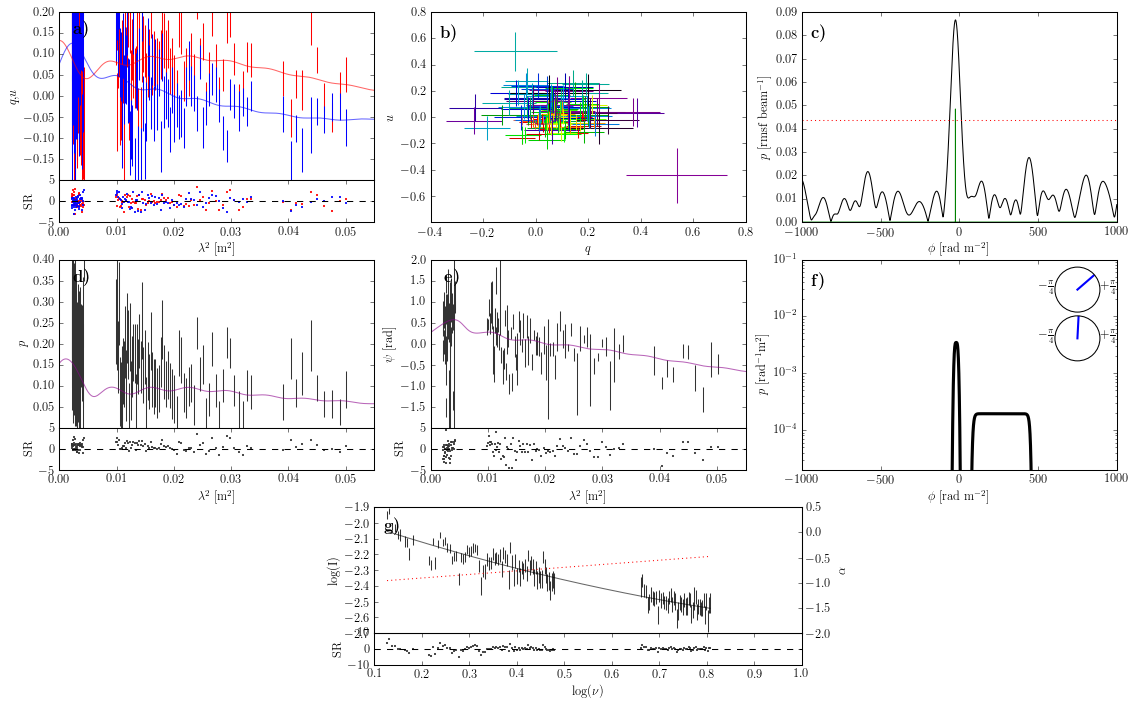}
\caption{As for Fig. \ref{fig:lmc_s11}. Source: cen\_s1443.}
\label{fig:cena_s1443}
\end{figure*}

\begin{figure*}[htpb]
\centering
\includegraphics[width=0.95\textwidth]{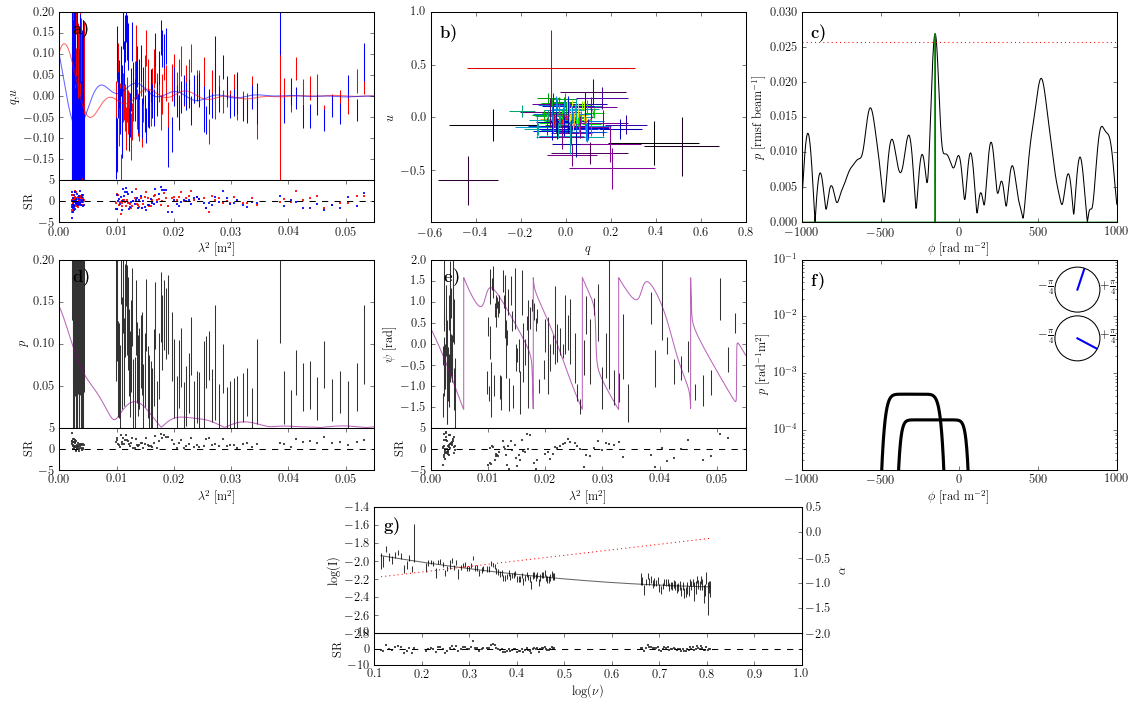}
\caption{As for Fig. \ref{fig:lmc_s11}. Source: cen\_c1640.}
\label{fig:cena_c1640}
\end{figure*}

\begin{figure*}[htpb]
\centering
\includegraphics[width=0.95\textwidth]{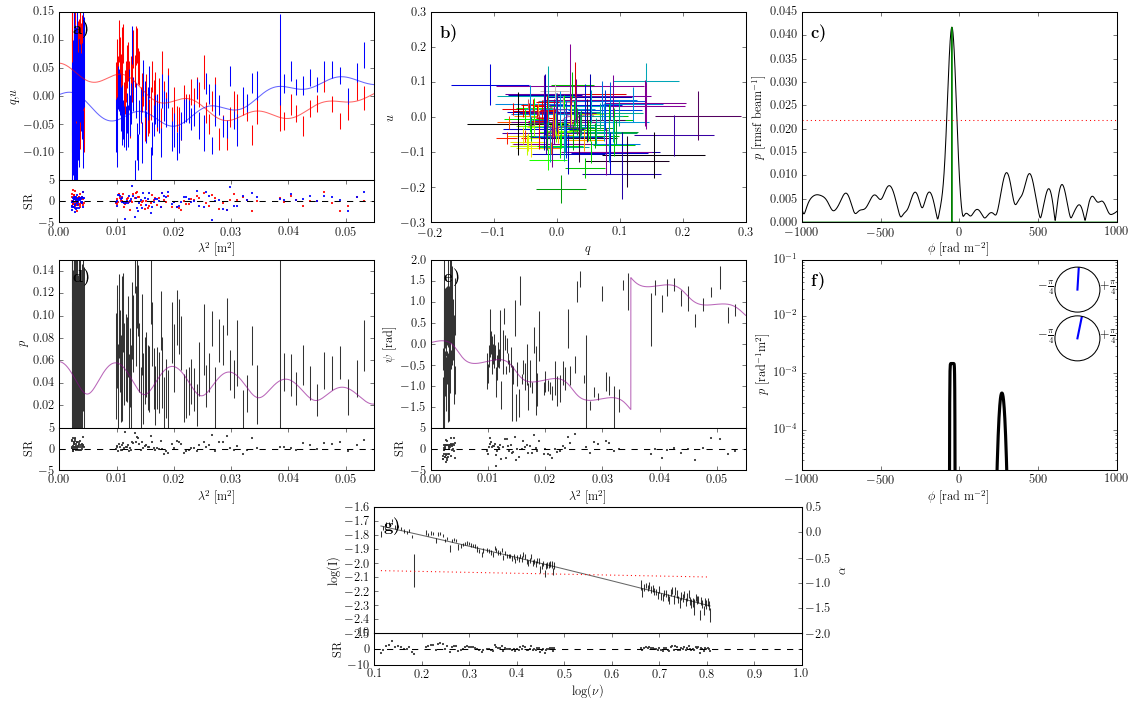}
\caption{As for Fig. \ref{fig:lmc_s11}. Source: cen\_s1031.}
\label{fig:cena_s1031}
\end{figure*}

\begin{figure*}[htpb]
\centering
\includegraphics[width=0.95\textwidth]{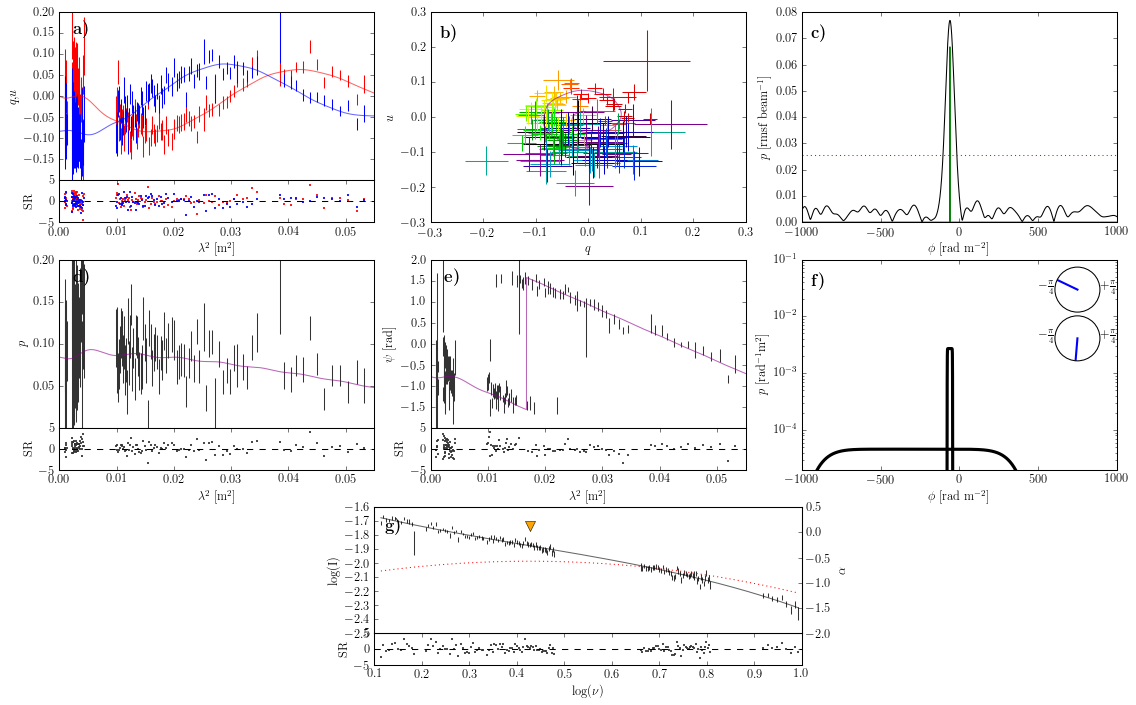}
\caption{As for Fig. \ref{fig:lmc_s11}. Source: cen\_c1093.}
\label{fig:cena_c1093}
\end{figure*}

\begin{figure*}[htpb]
\centering
\includegraphics[width=0.95\textwidth]{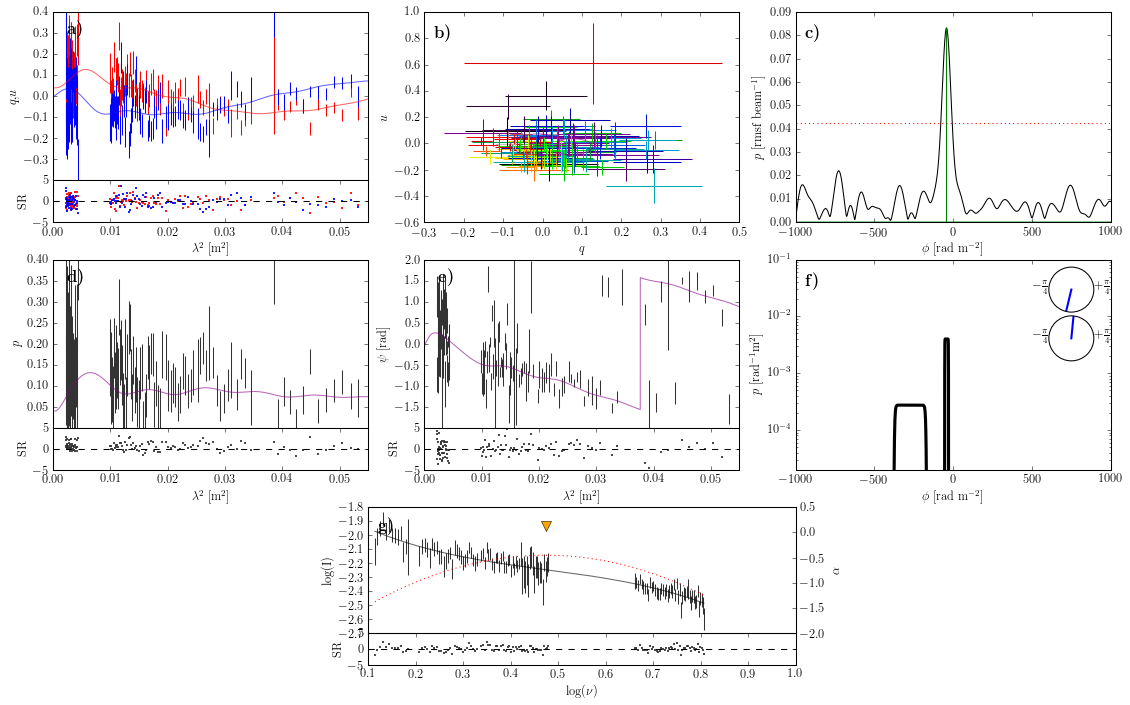}
\caption{As for Fig. \ref{fig:lmc_s11}. Source: cen\_s1605.}
\label{fig:cena_s1605}
\end{figure*}

\begin{figure*}[htpb]
\centering
\includegraphics[width=0.95\textwidth]{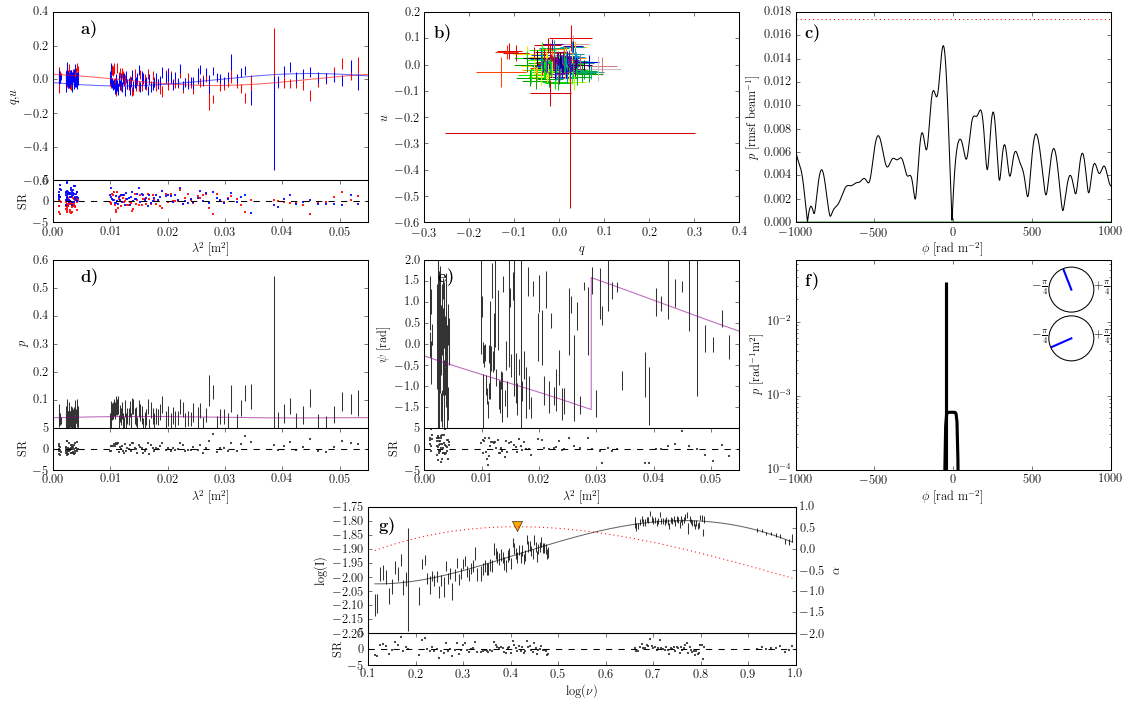}
\caption{As for Fig. \ref{fig:lmc_s11}. Source: cen\_s1568.}
\label{fig:cena_s1568}
\end{figure*}

\begin{figure*}[htpb]
\centering
\includegraphics[width=0.95\textwidth]{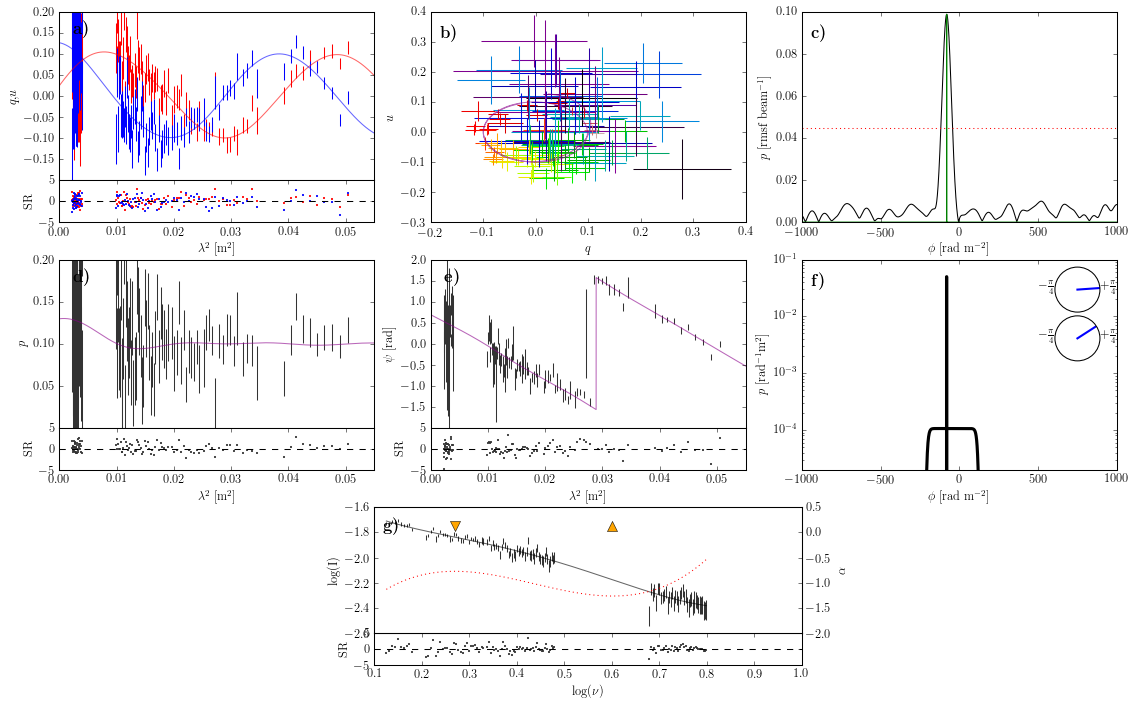}
\caption{As for Fig. \ref{fig:lmc_s11}. Source: cen\_s1681.}
\label{fig:cena_s1681}
\end{figure*}

\begin{figure*}[htpb]
\centering
\includegraphics[width=0.95\textwidth]{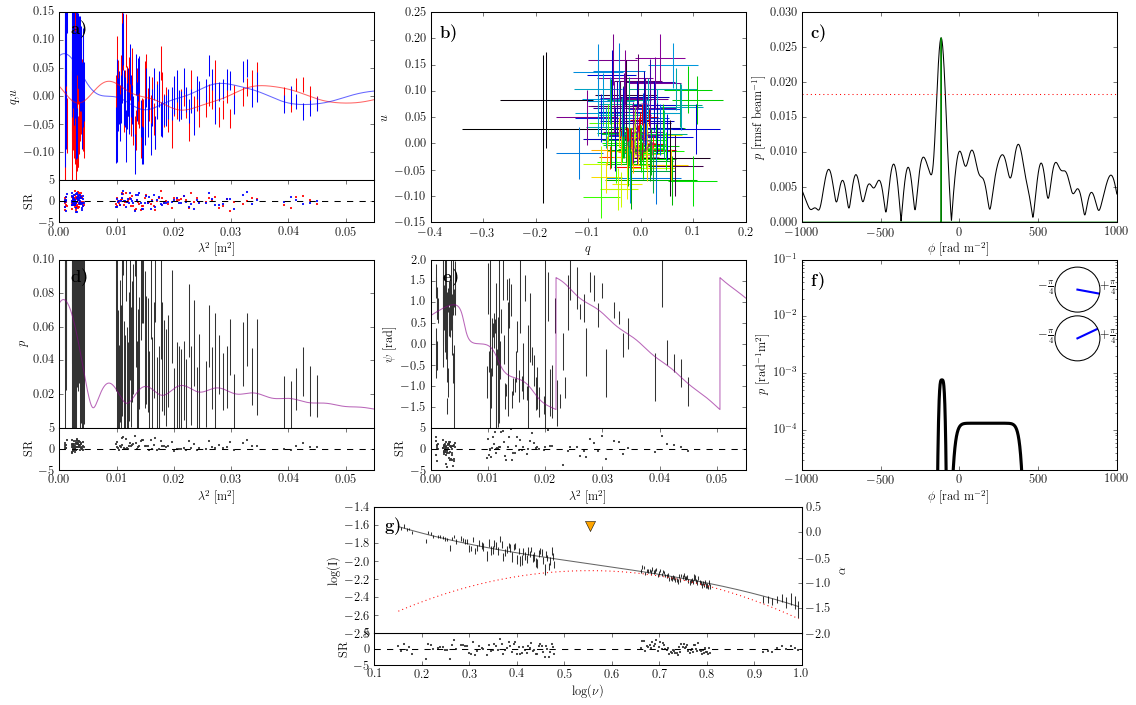}
\caption{As for Fig. \ref{fig:lmc_s11}. Source: cen\_c1748.}
\label{fig:cena_c1748}
\end{figure*}

\begin{figure*}[htpb]
\centering
\includegraphics[width=0.95\textwidth]{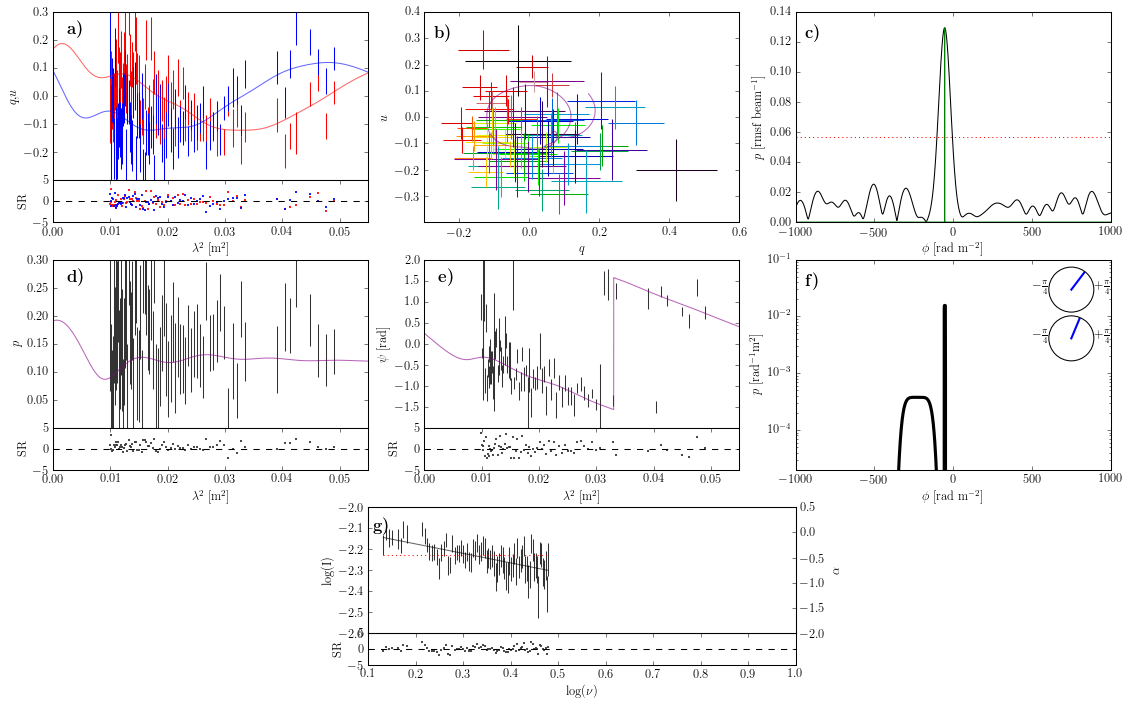}
\caption{As for Fig. \ref{fig:lmc_s11}. Source: cen\_s1382.}
\label{fig:cena_s1382}
\end{figure*}

\begin{figure*}[htpb]
\centering
\includegraphics[width=0.95\textwidth]{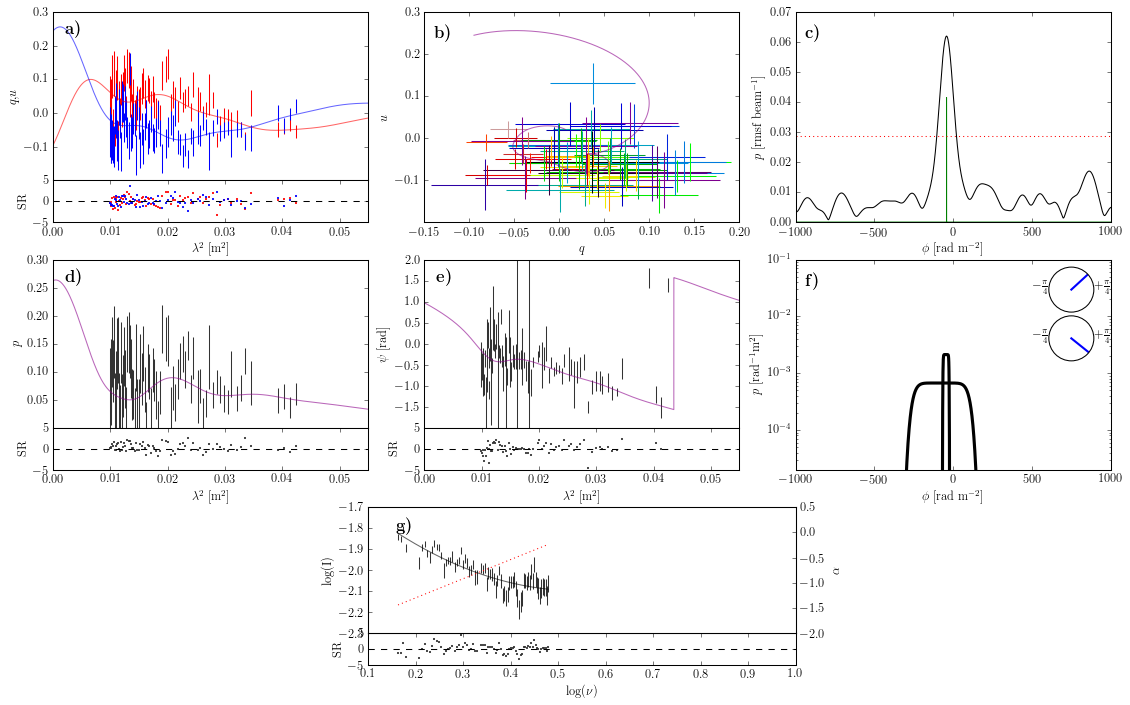}
\caption{As for Fig. \ref{fig:lmc_s11}. Source: cen\_c1152.}
\label{fig:cena_c1152}
\end{figure*}

\begin{figure*}[htpb]
\centering
\includegraphics[width=0.95\textwidth]{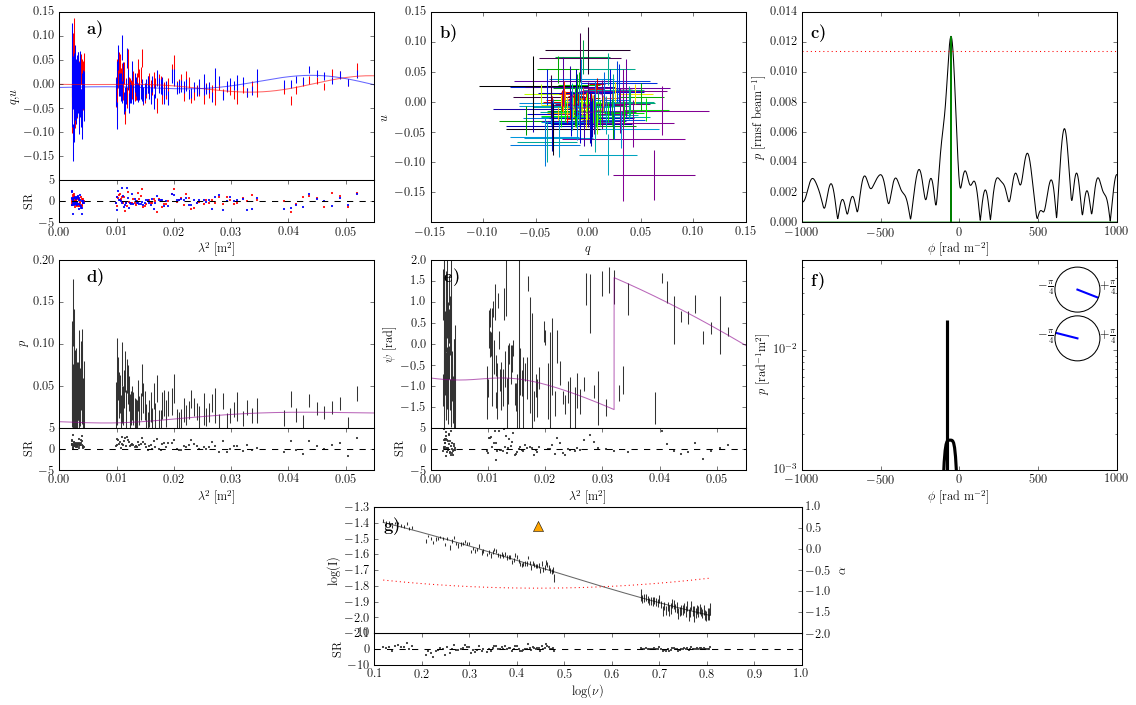}
\caption{As for Fig. \ref{fig:lmc_s11}. Source: cen\_c1832.}
\label{fig:cena_c1832}
\end{figure*}

\begin{figure*}[htpb]
\centering
\includegraphics[width=0.95\textwidth]{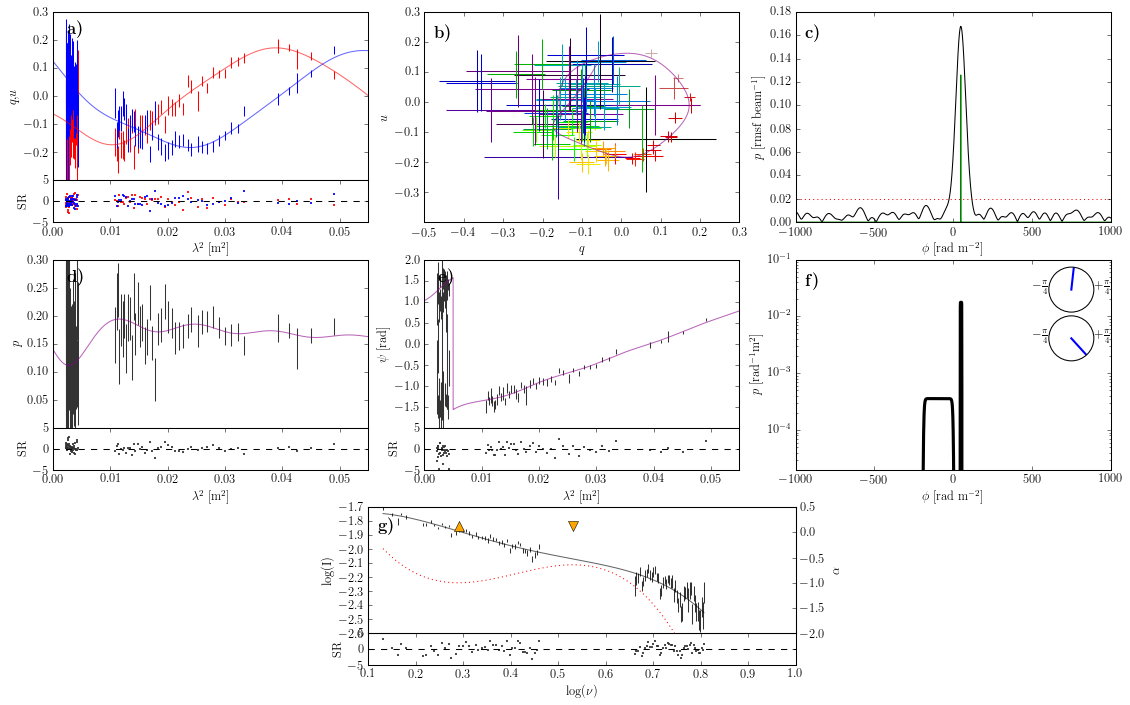}
\caption{As for Fig. \ref{fig:lmc_s11}. Source: lmc\_s14.}
\label{fig:lmc_s14}
\end{figure*}

\begin{figure*}[htpb]
\centering
\includegraphics[width=0.95\textwidth]{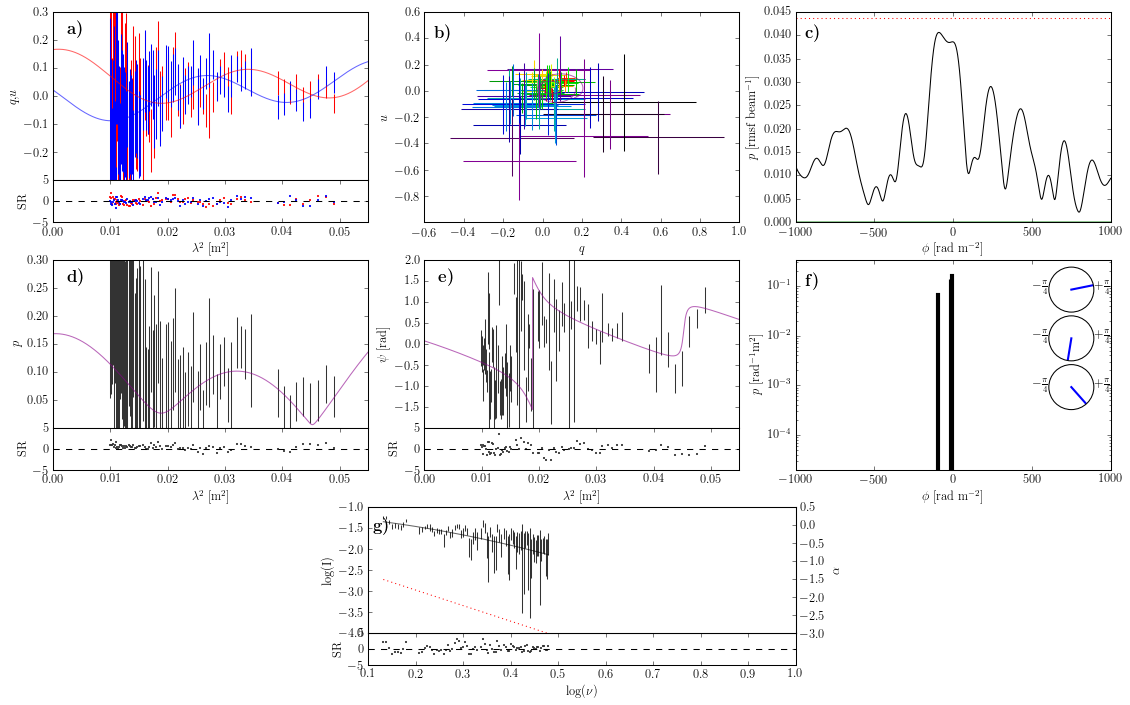}
\caption{As for Fig. \ref{fig:lmc_s11}. Source: cen\_s1014.}
\label{fig:cena_s1014}
\end{figure*}

\begin{figure*}[htpb]
\centering
\includegraphics[width=0.95\textwidth]{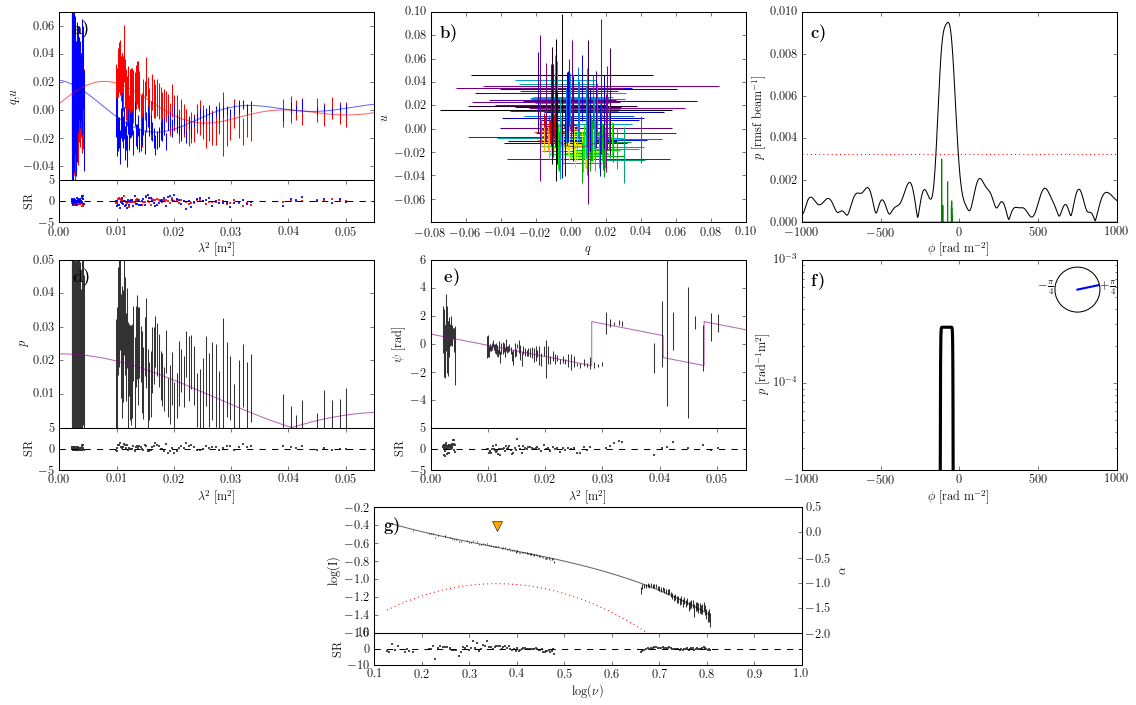}
\caption{As for Fig. \ref{fig:lmc_s11}. Source: cen\_c1764.}
\label{fig:cena_c1764}
\end{figure*}

\begin{figure*}[htpb]
\centering
\includegraphics[width=0.95\textwidth]{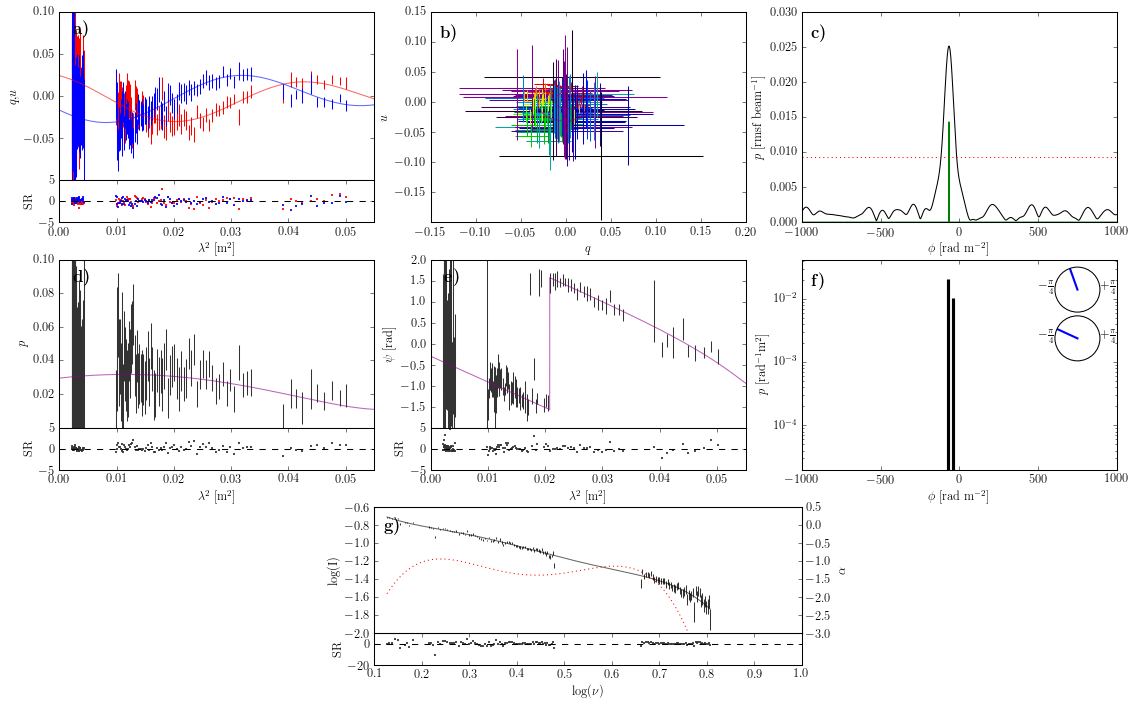}
\caption{As for Fig. \ref{fig:lmc_s11}. Source: cen\_c1435.}
\label{fig:cena_c1435}
\end{figure*}

\begin{figure*}[htpb]
\centering
\includegraphics[width=0.95\textwidth]{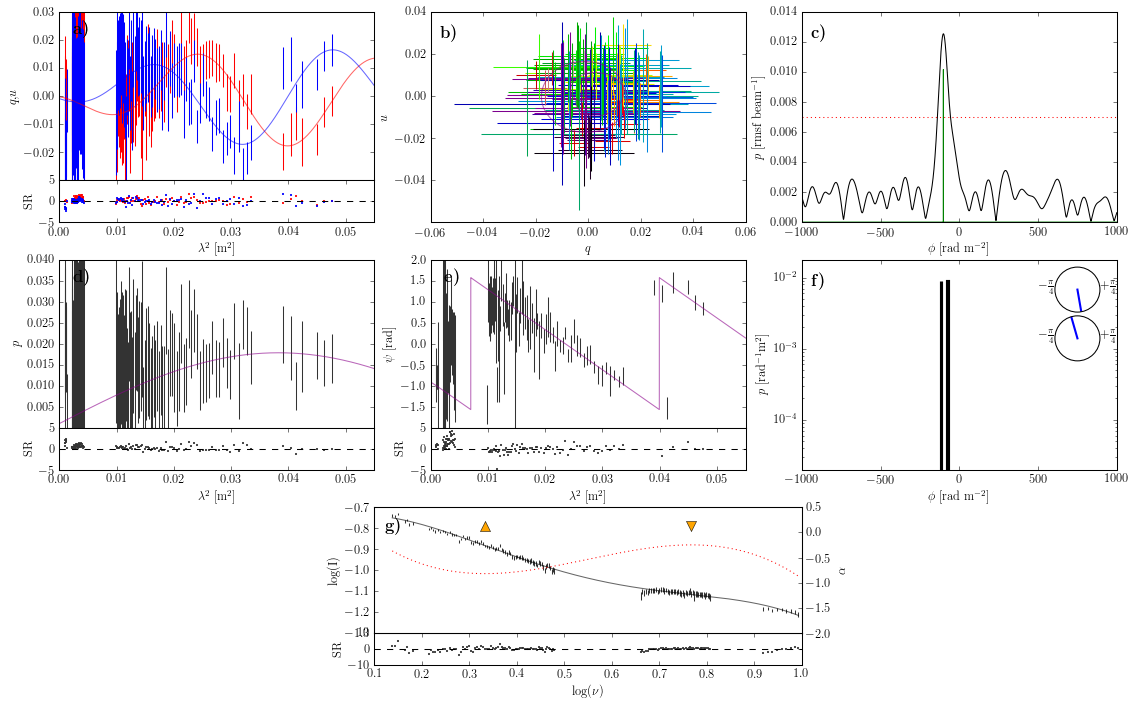}
\caption{As for Fig. \ref{fig:lmc_s11}. Source: cen\_s1349.}
\label{fig:cena_s1349}
\end{figure*}

\begin{figure*}[htpb]
\centering
\includegraphics[width=0.95\textwidth]{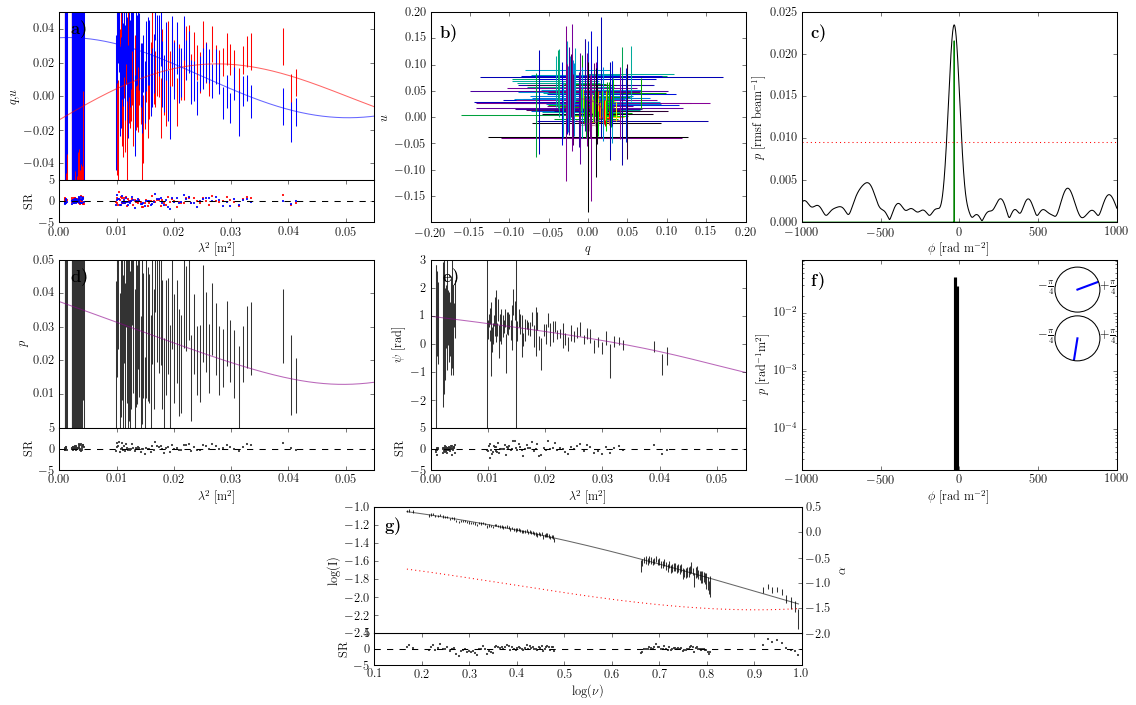}
\caption{As for Fig. \ref{fig:lmc_s11}. Source: cen\_s1803.}
\label{fig:cena_s1803}
\end{figure*}

\begin{figure*}[htpb]
\centering
\includegraphics[width=0.95\textwidth]{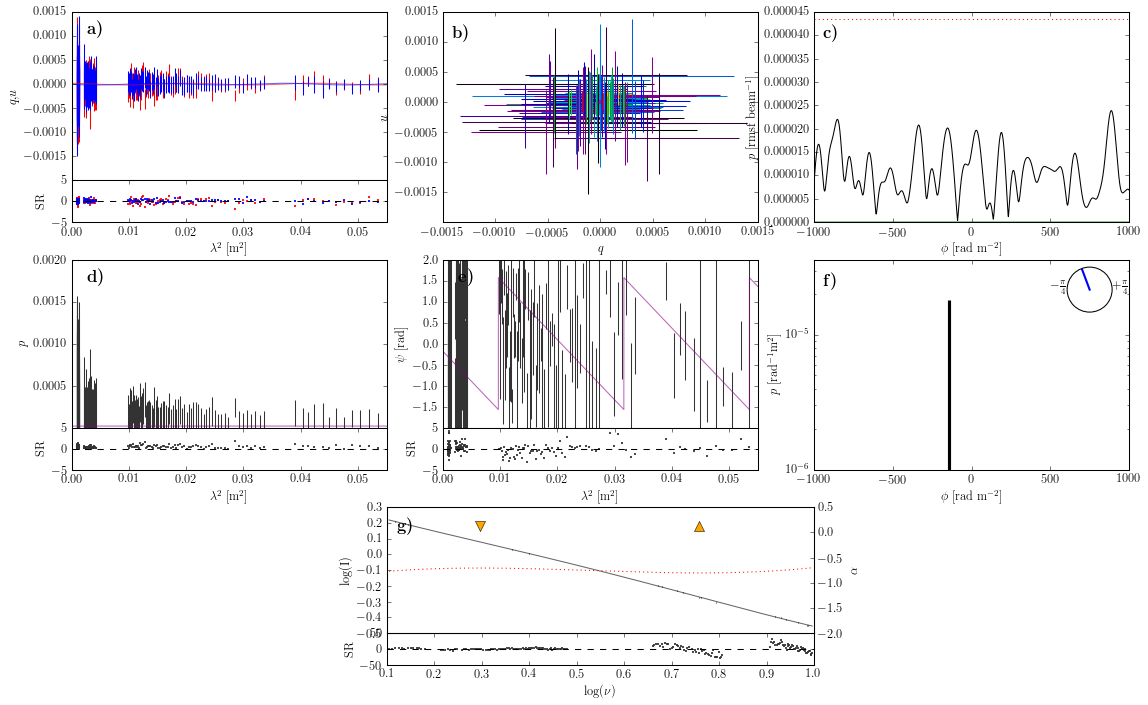}
\caption{As for Fig. \ref{fig:lmc_s11}. Source: 0515-674 (LMC dataset phase calibrator).}
\label{fig:0515-674}
\end{figure*}

We begin by presenting our spectropolarimetric data, and by pointing out several obvious and noteworthy features found therein. Our data (and our best-fit polarization models --- discussed in the next section) are plotted on separate panels in Figs. \ref{fig:lmc_s11}--\ref{fig:cena_s1803}. The sources are presented in order such that the $\tilde{\chi}^2$ value of the best fit S model decreases --- i.e. roughly in order of decreasing Faraday complexity. In panels (a), (d) \& (e) of these figures, we plot ($q$,$u$), $p$ \& $\psi$ (respectively) against $\lambda^2$. The $\lambda^2$ coverage is dominated the 16 cm band data which spans $0.01<\lambda^2<0.05$ m$^2$, while the 6 cm \& 3 cm data correspond to $\lambda^2<0.005$ m$^2$ and $\lambda^2<0.0015$ m$^2$ respectively. The uncertainty in the plotted quantities is larger at frequencies where our sources are resolved (see column 5 of Table \ref{tab:SourceDat}), reflecting the the loss of information brought about by the image convolution procedure described in Section \ref{sec-extract}. Panel (b) plots Stokes $u$ vs. $q$, while panel (c) plots the magnitude of the FDS \& {\sc rmclean} components, along with the {\sc rmclean} cutoff (see Section \ref{sec-extras-RMS}; also Heald et al. 2009). Panel (f) plots the overall best fit model FDF; the magnitude of each model component is plotted on the main axes, while the intrinsic polarization angle of each component is plotted on the inset polar axes. Panel (g) plots the total intensity spectrum and a polynomial model fit to these data (see Section \ref{sec-extract}, also col. 7 of Table \ref{tab:SourceDat}). We also plot the instantaneous spectral index associated with the polynomial model, calculated as $\partial \text{log}(I_{\text{model}}(\nu))/\partial \text{log}(\nu)$. Panels (a), (d), (e) \& (g) additionally contain inset axes below the main axes, on which we plot the standardized residual (SR) between the data and the model fit, where SR $=(d_i-M_i)/\sigma_i$, and $d_i$, $M_i$ \& $\sigma_i$ are the values of the data, model and standard uncertainty (respectively) in channel $i$. For panels (a) \& (g), these data should be Gaussian distributed about SR $=0$ with a standard deviation of 1. 

The plots described above immediately reveal the existence of substantial variety in the spectropolarimetric behaviour of the sample. Particular features of note include:

\begin{itemize}
\item The presence of Faraday thick structures: These are both obvious and resolved in the FDS for several sources in our sample (see panel (c) of relevant figures, noting that our observations have $\phi_{\text{Max-scale}}=3400$ rad m$^{-2}$ but $\delta\phi\approx70$ rad m$^{-2}$ --- see Table \ref{table:observations}). lmc\_c04 (Fig. \ref{fig:lmc_c04}) is particularly striking in this regard. lmc\_s11 (Fig. \ref{fig:lmc_s11}), lmc\_c03 (Fig. \ref{fig:lmc_c03}), and 1315-46 (Fig. \ref{fig:1315-46}) provide further examples.

\item Varied $p(\lambda^2$) behaviour: $p(\lambda^2$) (panel (d) decreases monotonically to zero for some sources (e.g. cen\_c1827; Fig. \ref{fig:cena_c1827}), apparently to finite limiting values for others (e.g. cen\_c1466; Fig. \ref{fig:cena_c1466}), and oscillates for still others (e.g. lmc\_c15; Fig. \ref{fig:lmc_c15}). lmc\_s13 (Fig. \ref{fig:lmc_s13}) repolarizes (i.e. $\partial p / \partial \lambda^2$ is positive) over the full $\lambda^2$ band.

\item Behavioural transitions in ($q$,$u$)/$p$/$\psi$ (panels (a), (d) \& (e)): Several sources show abrupt changes in spectropolarimetric behaviour which occur over comparatively small $\lambda^2$ ranges, particularly at short wavelengths --- e.g. lmc\_s11 (Fig. \ref{fig:lmc_s11}) at $\lambda^2\approx 0.005$ m$^2$, lmc\_c02 (Fig. \ref{fig:lmc_c02}) at both $\lambda^2\approx 0.005$ m$^2$ \& $\lambda^2\approx 0.01$ m$^2$, lmc\_c03 (Fig. \ref{fig:lmc_c03}) at $\lambda^2< 0.003$ m$^2$. 

\item Linearity (or lack thereof) of $\psi(\lambda^2)$: For some sources, near-linearity of $\psi(\lambda^2)$ (panel (e) is maintained over broad $\lambda^2$ ranges despite strong concurrent depolarization --- e.g. lmc\_s11 (Fig. \ref{fig:lmc_s11}), lmc\_c03 (Fig. \ref{fig:lmc_c03}), lmc\_c02 (Fig. \ref{fig:lmc_c02}), cen\_c1827 (Fig. \ref{fig:cena_c1827}), cen\_c1972 (Fig. \ref{fig:cena_c1972}) \& lmc\_c01 (Fig. \ref{fig:lmc_c01}). For other sources, $\psi(\lambda^2)$ is strongly frequency dependent (and so therefore is the RM) --- e.g. lmc\_c15 (Fig. \ref{fig:lmc_c15}) and lmc\_c04 (Fig. \ref{fig:lmc_c04}).

\item Structure in the total intensity spectra (panel (g)): $\partial\text{log}(I)$ vs. $\partial\text{log}(\nu)$ is rarely well described by a pure power law, but rather, often shows substantial structure. Concave (e.g cen\_c1972; Fig. \ref{fig:cena_c1972}) and convex (e.g lmc\_c04; Fig. \ref{fig:lmc_c04}) features are apparent --- often in the same spectrum (e.g lmc\_s11; Fig. \ref{fig:lmc_s11}). Several sources such as lmc\_c02 (Fig. \ref{fig:lmc_c02}) and lmc\_s13 (Fig. \ref{fig:lmc_s13}) possess particularly complicated Stokes I spectra.
\end{itemize}

We interpret these behaviours in light of our polarimetric models in Section \ref{sec-interp}.

\subsection{1.3--10 GHz spectropolarimetric modelling}\label{sec-results-broadbandmodelling}

We now report on the results of fitting the 1.3--10 GHz polarization data presented in Section \ref{sec-results-observed}, using the modelling techniques described in Sections \ref{sec-modelling} \& \ref{sec-fitting}. Table \ref{tab:fitgoodness} contains a summary of the fitting results. The sources are ordered in the table by the $\tilde{\chi}^2$ value of the best fitting S model, from highest to lowest (i.e. most to least Faraday complex). We include results for the best fit S model for each source by default, but only include results for Faraday complex models if they have a better (i.e. lower) AIC value than the best fit S model, and are either a) the best fitting model overall, or b) indistinguishable from the overall best fit model on the basis of our AIC criteria (i.e. $-10<$AIC$_{\text{best fit}}-$ AIC$_2<0$). 

The results for each source occupy several rows in the table and are arranged as follows. The source names in column 1 indicate where the fitting data for each source begins and ends. In the first row entry for each source, the figure number displaying the source data, the polarized signal-to-noise ratio of the source, the overall best fit model type, and the corresponding $\tilde{\chi}^2$ \& AIC values are listed in columns 2, 3, 4, 5 \& 6 (respectively). For the first component in the best fit model, we list its type in column 7, contribution to fractional polarization as $\lambda\to0$ in column 8, and best fit parameter values $P_1$--$P_5$ in columns 9--13. This is repeated for each of the $n$ remaining best fit model components in columns 9--f $n$ subsequent rows of the table. Further complex models warranting inclusion are presented in subsequent rows in the same way, followed by the best fit S model for each source. In all cases, the first row entry for a given model is indicated in column 4 by listing its model type. For all quantities, the uncertainties are indicated using standard parentheses notation. Note that for one source only in our sample (0515-674), a model with zero polarized emission provides an excellent fit to the data. We have entered `Unp' as the model type in this isolated case.

\begin{turnpage} 
\begin{table*} 
\caption{Results for model fitting described in Sections \ref{sec-modelling} \& \ref{sec-fitting}} 
\label{tab:fitgoodness} 
\scriptsize 
\tabcolsep=0.3cm 
\begin{tabular}{l c c c c l c l l l l l l} 
\hline 
\hline 
\multicolumn{1}{c}{(1)} & \multicolumn{1}{c}{(2)} & \multicolumn{1}{c}{(3)} & \multicolumn{1}{c}{(4)} & \multicolumn{1}{c}{(5)} & \multicolumn{1}{c}{(6)} & \multicolumn{1}{c}{(7)} & \multicolumn{1}{c}{(8)} & \multicolumn{1}{c}{(9)} & \multicolumn{1}{c}{(10)} & \multicolumn{1}{c}{(11)} & \multicolumn{1}{c}{(12)} & \multicolumn{1}{c}{(13)}
\\ 
\multicolumn{1}{c}{Source} & \multicolumn{1}{c}{Figure} & \multicolumn{1}{c}{Polarized} & \multicolumn{1}{c}{Model} & \multicolumn{1}{c}{$\tilde{\chi}^2$} & \multicolumn{1}{c}{AIC} & \multicolumn{1}{c}{Component} & \multicolumn{1}{c}{$p_{\lambda=0}$} & \multicolumn{5}{c}{Best fit parameters for model component}
\\ 
\multicolumn{1}{c}{} & \multicolumn{1}{c}{no.} & \multicolumn{1}{c}{S/N} & \multicolumn{1}{c}{Type} & \multicolumn{1}{c}{} & \multicolumn{1}{c}{} & \multicolumn{1}{c}{Type} & \multicolumn{1}{c}{} & \multicolumn{1}{c}{P$_1$} & \multicolumn{1}{c}{P$_2$} & \multicolumn{1}{c}{P$_3$} & \multicolumn{1}{c}{P$_4$} & \multicolumn{1}{c}{P$_5$} 
\\ 
\multicolumn{1}{c}{} & \multicolumn{1}{c}{} & \multicolumn{1}{c}{} & \multicolumn{1}{c}{} & \multicolumn{1}{c}{} & \multicolumn{1}{c}{} & \multicolumn{1}{c}{} & \multicolumn{1}{c}{} & \multicolumn{1}{c}{[ ]\tablenotemark{a}} & \multicolumn{1}{c}{[ ]\tablenotemark{a}} & \multicolumn{1}{c}{[ ]\tablenotemark{a}} & \multicolumn{1}{c}{[ ]\tablenotemark{a}} & \multicolumn{1}{c}{[ ]\tablenotemark{a}} 
\\
\hline 

lmc\_s11 & \ref{fig:lmc_s11} & 157.1 & SSTT & 4.4 & -2807.2 & S & 0.031 & 0.0313(7)& 53.7(6) & - & - & -0.46(2)  \\[1pt] 
  & &  &  &  &  & S & 0.004 & 0.0040(8) & 0(3) &- &- & -0.3(1) \\[1pt] 
  &  &  &  &  &  & T & 0.012 & 0.0123(8) & 109(7) & 98(5) & 2.1(1) & -0.94(4) \\[1pt] 
  &  &  &  &  &  & T & 0.025 & 0.0246(3) & -133(5) & 132(3) & 2.13(2) & 0.65(1)\\[1pt] 
  &  & & S &315.8 &$>50000$ &S &0.036 &0.03626(4) &51.01(3) &- &- &-0.3492(7) \\[1pt] 
lmc\_c04 & \ref{fig:lmc_c04} & 24.8 & STT & 3.7 & -3699.6 & S & 0.007 & 0.0075(4) & 5(1) &- &- & 0.92(4) \\[1pt] 
  &  &  &  &  &  & T & 0.042 & 0.050(1) & 124(2) & 67(1) & 6.4(4) & -1.42(2) \\[1pt] 
  &  &  &  &  &  & T & 0.003 & 0.0031(5) & 310(3) & 16(2) & 6(3) & -1.5(1) \\[1pt] 
  &  &  & S &259.5 &$>50000$ &S &0.017 &0.01673(5) &193.7(2) &- &- &1.443(2) \\[1pt] 
lmc\_c03 &  \ref{fig:lmc_c03} & 334.1 & SSTT & 10.2 & -630.6 & S & 0.163 & 0.163(6) & 39.29(8) & - & - & 0.692(7)  \\[1pt] 
  &  &  &  &  &  & S & 0.162 & 0.162(6) & 41.5(1) &- &- & -1.15(1) \\[1pt] 
  &  &  &  &  &  & T & 0.869 & 0.87(2) & 287(20) & 1439(4) & 4.59(1) & -0.42(6) \\[1pt] 
  &  &  &  &  &  & T & 0.166 & 0.166(6) & 269(10) & 503(3) & 3.59(1) & -0.47(4)\\[1pt] 
  &  &  &S &223.3 &$>50000$ &S &0.070 &0.06970(5) &39.56(2) &- &- &1.3646(6) \\[1pt] 
lmc\_s13 & \ref{fig:lmc_s13} & 54.5 & ST &2.1 &-3204.9 &S &0.048 &0.048(2) &61(2) &- &- &-0.28(8) \\[1pt] 
  &  &  &  &  &  &T &0.034 &0.042(1) &77.3(9) &31(3) &16(7) &2.1(2) \\[1pt] 
  &  &  &S &74.1 &21208.3 &S &0.025 &0.02480(6) &51.18(9) &- &- &0.031(2) \\[1pt] 
cena\_c1466  & \ref{fig:cena_c1466} & 160.9 & SS & 4.2 & -1756.7 &S &0.004 &0.0102(2) & -3.7(7) &- &- & 0.09(4) \\[1pt] 
  &  &  &  &  &  &S &0.004 & 0.058(4) & -38(1) &- &- & 0.01(8) \\[1pt] 
  &  &  &S &61.4 &11615.4 &S &0.061 &0.06115(8) &-37.91(5) &- &- &0.059(1) \\[1pt] 
lmc\_c02  & \ref{fig:lmc_c02} & 40.4 &TT &2.1 &-2520.1 &T &0.010 &0.0105(5) &147(8) &71(6) &2.0(1) &-1.52(6) \\[1pt] 
  &  &  &  &  &  &T &0.028 &0.0291(5) &41.9(3) &6.6(5) & 2(8) &0.21(2) \\[1pt] 
  &  &  &S &26.8 &3659.7 &S &0.022 &0.02152(6) &42.8(1) &- &- &0.184(2) \\[1pt] 
1315-46  & \ref{fig:1315-46} & 3.1 &TT & 0.8 & -4144.3 & T & 0.023 & 0.024(1) & -70(30) & 470(15) & 2.5(2) & 1.05(6) \\[1pt] 
  &  &  &  &  &  &T & 0.004 & 0.005(1) & 420(40) & 300(15) & 6(4) & -1.2(2)\\[1pt] 
  &  &  &S &18.0 &1256.1 &S &0.0003 & 0.00028(2) & -650(5) &- &- &1.73(3) \\[1pt] 
cena\_c1827  & \ref{fig:cena_c1827} & 20.41 &TT &0.5 &-1787.5 &T &0.039 &0.048(6) &-18(7) &19(2) &10(9) &0.1(0.5) \\[1pt] 
  &  &  &  &  &  &T &0.085 &0.10(1) &-32(3) &37(1) &4(2) &1.26(9) \\[1pt] 
  &  &  &S &12.0 &799.1 &S &0.027 &0.0274(3) &-52.3(5) &- &- &1.19(1) \\[1pt] 
cena\_c1972 & \ref{fig:cena_c1972} & 104.0 &TT &1.2 &-1488.5 &T &0.023 &0.028(6) &-20.0(50.0) &195(60) &6(4) &0.5(4) \\[1pt] 
  &  &  &  &  &  &T &0.098 &0.120(5) &-75.2(2) &8.2(4) &10(1) & 1.23(1) \\[1pt] 
  &  & &S &11.0 &306.7 &S &0.089 &0.0889(2) &-75.04(7) &- &- &1.218(2) \\[1pt] 
cena\_c1573 & \ref{fig:cena_c1573} & 31.1 &ST &1.6 &-1624.9 &S &0.022 & 0.013(2) &-54(5) &- &- &1.0(2) \\[1pt] 
  &  &  &  &  &  &T &0.046 &0.071(6) &-88(4) &22(2) &2.1(2) &-0.23(5) \\[1pt] 
  &  & &S &8.9 &29.4 &S &0.032 &0.0322(3) &-94.7(4) &- &- &-0.20(1) \\[1pt] 
lmc\_c01 & \ref{fig:lmc_c01} & 114.2 &TT &2.1 &-1694.0 &T &0.010 &0.013(2) &299(8) &113(8) &16(7) &-1.5(1) \\[1pt] 
  &  &  &  &  &  &T &0.094 &0.104(3) &18.3(2) &8.0(4) &3(2) &0.63(1) \\[1pt] 
  &  &  &S &7.5 &-541.3 &S &0.085 &0.0854(3) &18.2(1) &- &- &0.626(4) \\[1pt] 
lmc\_c06& \ref{fig:lmc_c06} & 149.0 &TT &1.9 &-1033.3 &T &0.660 &0.82(6) &29.9(5) &10.3(6) &28(8) &0.01(1) \\[1pt] 
  &  &  &  &  &  &T &0.418 &0.51(5) &26.6(7) &17(2) &7.0(9.0) &-1.25(8) \\[1pt] 
  &  &  &ST &2.0 &-1025.2 &S &0.325 &0.325(3) &39.2(4) &- &- &-0.49(1) \\[1pt] 
  &  &  &  &  &  &T &0.103 &0.11(1) &40(20) &30(6) &3.0(9.0) &0.08(0.2) \\[1pt] 
  &  &  &S &7.4 &71.1 &S &0.344 &0.344(1) &36.3(1) &- &- &-0.359(3) \\[1pt] 
  \hline 
\end{tabular} 
\end{table*} 
\end{turnpage}    

\addtocounter{table}{-1}
\begin{turnpage} 
\begin{table*} 
\caption{\emph{continued}} 
\label{tab:fitgoodness} 
\scriptsize 
\tabcolsep=0.3cm 
\begin{tabular}{l c c c c l c l l l l l l} 
\hline 
\hline 
\multicolumn{1}{c}{(1)} & \multicolumn{1}{c}{(2)} & \multicolumn{1}{c}{(3)} & \multicolumn{1}{c}{(4)} & \multicolumn{1}{c}{(5)} & \multicolumn{1}{c}{(6)} & \multicolumn{1}{c}{(7)} & \multicolumn{1}{c}{(8)} & \multicolumn{1}{c}{(9)} & \multicolumn{1}{c}{(10)} & \multicolumn{1}{c}{(11)} & \multicolumn{1}{c}{(12)} & \multicolumn{1}{c}{(13)}
\\ 
\multicolumn{1}{c}{Source} & \multicolumn{1}{c}{Figure} & \multicolumn{1}{c}{Polarized} & \multicolumn{1}{c}{Model} & \multicolumn{1}{c}{$\tilde{\chi}^2$} & \multicolumn{1}{c}{AIC} & \multicolumn{1}{c}{Component} & \multicolumn{1}{c}{$p_{\lambda=0}$} & \multicolumn{5}{c}{Best fit parameters for model component}
\\ 
\multicolumn{1}{c}{} & \multicolumn{1}{c}{no.} & \multicolumn{1}{c}{S/N} & \multicolumn{1}{c}{Type} & \multicolumn{1}{c}{} & \multicolumn{1}{c}{} & \multicolumn{1}{c}{Type} & \multicolumn{1}{c}{} & \multicolumn{1}{c}{P$_1$} & \multicolumn{1}{c}{P$_2$} & \multicolumn{1}{c}{P$_3$} & \multicolumn{1}{c}{P$_4$} & \multicolumn{1}{c}{P$_5$} 
\\ 
\multicolumn{1}{c}{} & \multicolumn{1}{c}{} & \multicolumn{1}{c}{} & \multicolumn{1}{c}{} & \multicolumn{1}{c}{} & \multicolumn{1}{c}{} & \multicolumn{1}{c}{} & \multicolumn{1}{c}{} & \multicolumn{1}{c}{[ ]\tablenotemark{a}} & \multicolumn{1}{c}{[ ]\tablenotemark{a}} & \multicolumn{1}{c}{[ ]\tablenotemark{a}} & \multicolumn{1}{c}{[ ]\tablenotemark{a}} & \multicolumn{1}{c}{[ ]\tablenotemark{a}} 
\\
\hline 
lmc\_c15 & \ref{fig:lmc_c15} & 22.6 &SSS &1.5 &-1584.4 &S &0.028 &0.0283(6) & 98(1) &- &- &0.44(3) \\[1pt] 
  &  &  &  &  &  &S &0.028 &0.0279(6) &136(3) &- &- &-0.85(4) \\[1pt] 
  &  &  &  &  &  &S &0.026 &0.0257(8) &-50(4) &- &- &0.38(3) \\[1pt] 
  &  &  &S &6.7 &-205.7 &S &0.046 &0.0463(6) &111.5(8) &- &- &-0.09(2) \\[1pt] 
lmc\_c16 & \ref{fig:lmc_c16} & 118.0 & TT &1.6 &-2370.9 &T &0.023 &0.031(1) &46.8(5) &7(1) &11(2) & 0.91(3) \\[1pt] 
  &  &  &  &  &  &T &0.005 &0.0069(9) &284(9) & 163(9) &13(4) &-0.3(2) \\[1pt] 
  &  &  &ST &1.6 &-2365.1 &S &0.022 &0.0223(4) &47.3(6) &- &- &0.89(3) \\[1pt] 
  &  &  &  &  &  &T &0.005 &0.0057(7) &74(4) &31(6) &4.0(7.0) &1.0(3) \\[1pt] 
  &  &  &S &6.4 &-1411.5 &S &0.024 &0.02416(9) &48.0(1) &- &- &0.867(3) \\[1pt] 
cena\_s1290 & \ref{fig:cena_s1290} & 13.8 &ST &2.1 &-1354.2 &S &0.041 &0.041(1) &-45(1) &- &- &-1.05(4) \\[1pt] 
  &  &  &  &  &  &T &0.020 &0.025(5) &-290(10) &89(40) &23(7) &0.4(2) \\[1pt] 
  &  &  &S &6.1 &-564.1 &S &0.041 &0.0413(6) &-49.9(5) &- &- &-0.90(1) \\[1pt] 
cena\_c1636 & \ref{fig:cena_c1636} & 32.8 &TT &1.8 &-273.2 &T &0.307 &0.38(2) &-110(1) &11.6(8) &20(10) &1.00(6) \\[1pt] 
 &  &  &  &  &  &T &0.107 &0.13(6) &190(90) &70(40) &9(7) &0.2(7) \\[1pt] 
  &  &  &S &6.0 &488.6 &S &0.260 &0.260(3) &-110.8(5) &- &- &1.00(2) \\[1pt] 
cena\_s1437& \ref{fig:cena_c1437} & 15.9 & ST & 1.9 & -1411.4 & S &0.025 &0.025(2) &-27.9(2) &- &- &0.84(8) \\[1pt] 
  &  &  &  &  &  &T &0.021 &0.027(1) & 334.0(9) &75(3) &26(7) &-0.6(2) \\[1pt] 
  &  &  &S &6.0 &-612.0 &S &0.027 &0.0271(5) &-24.0(7) &- &- &0.74(2) \\[1pt] 
lmc\_c07 & \ref{fig:lmc_c07} & 12.6 &TT &1.1 &-2174.8 &T &0.21 &0.263(1) &-30(20) &56(10) &29(8) &-1.15(4) \\[1pt] 
  &  &  &  &  &  &T &0.26 &0.268(1) &-24(20) &60(10) &28(8) &0.41(4) \\[1pt] 
  &  &  &S &5.4 &-1331.0 &S &0.010 &0.0101(2) &58.4(8) &- &- &0.56(2) \\[1pt] 
cena\_s1443 & \ref{fig:cena_s1443} & 10.3 &TT &1.6 &-477.6 &T &0.069 &0.09(4) &3(1)e2 &120(60) &28(8) &0.03(0.7) \\[1pt] 
  &  &  &  &  &  &T &0.097 &0.11(1) &-20(3) &11(2) &3.0(9.0) &0.4(2) \\[1pt] 
  &  &  &ST &1.7 &-468.5 &S &0.079 &0.079(6) &-18(3) &- &- &0.35(9) \\[1pt] 
  &  &  &  &  &  &T &0.094 &0.12(4) &260(60) &120(40) &23(9) &0.05(0.2) \\[1pt] 
  &  &  &S &4.6 &6.3 &S &0.086 &0.086(3) &-24(1) &- &- &0.53(4) \\[1pt] 
cena\_c1640 & \ref{fig:cena_c1640} & 5.5 &TT &1.6 &-587.7 &T &0.173 &0.173(3) &-295(60) &114(40) &9(10) &0.2(2) \\[1pt] 
  &  &  &  &  &  &T &0.128 &0.08(5) &-160(60) &135(40) &14(10) &1.0(2) \\[1pt] 
  &  &  &S &4.6 &-94.4 &S &0.027 &0.027(2) &-153(3) &- &- &-0.10(7) \\[1pt] 
cena\_s1031 & \ref{fig:cena_s1031} & 10.0 &TT &1.6 &-845.1 &T &0.014 &0.015(7) &3(1)e2 &10.0(50.0) &2.0(9.0) &0.1(1.0) \\[1pt] 
  &  &  &  &  &  &T &0.045 &0.055(6) &-45(2) &10(4) &13(10) &0.03(0.1) \\[1pt] 
  &  &  &ST &1.7 &-838.8 &S &0.047 &0.047(3) &-47(2) &- &- &0.08(7) \\[1pt] 
  &  &  &  &  &  &T &0.015 &0.018(7) &3(2)e2 &20.0(70.0) &14(9) &0.04(1.0) \\[1pt] 
  &  &  &S &4.5 &-379.6 &S &0.041 &0.041(2) &-47(1) &- &- &0.08(4) \\[1pt] 
cena\_c1093 & \ref{fig:cena_c1093} & 15.5 &TT &1.5 & -1063.4 &T &0.090&0.011(6) &-59.9(9) & 11(2) &26(4) &-0.56(1)\\[1pt] 
  &  &  &  &  &  &T & 0.056 &0.068(9) & -273(8) & 419(5) &9(6) &-1.5(1) \\[1pt] 
  &  &  &S &4.5 &-547.7 &S &0.077 &0.077(1) &-59.3(5) &- &- &-0.58(1) \\[1pt] 
cena\_s1605& \ref{fig:cena_s1605} & 10.3 &TT &1.5 &-576.6 &T &0.053 &0.07(3) &-3(2)e2 &60.0(70.0) &28(7) &-1(1) \\[1pt] 
  &  &  &  &  &  &T &0.092 &0.12(1) &-42(2) &8(3) &23(8) &0.1(0.2) \\[1pt] 
  &  &  &S &4.1 &-167.3 &S &0.083 &0.083(3) &-43(1) &- &- &0.13(4) \\[1pt] 
cena\_s1568 & \ref{fig:cena_s1568} & 4.5 &ST &1.4 &-1100.8 &S &0.035 &0.035(5) &-48(8) &- &- &-0.2(0.2) \\[1pt] 
  &  &  &  &  &  &T & 0.008 &0.01(1) &-10.0(70.0) & 26.0(50.0) &10(9) &-1(5) \\[1pt] 
  &  & &S &4.1 &-654.5 &S &0.015 &0.015(1) &-63(3) &- &- &0.20(6) \\[1pt] 
 \hline
\end{tabular} 
\end{table*} 
\end{turnpage}    

\addtocounter{table}{-1}
\begin{turnpage} 
\begin{table*} 
\caption{\emph{continued}} 
\label{tab:fitgoodness} 
\scriptsize 
\tabcolsep=0.3cm 
\begin{tabular}{l c c c c l c l l l l l l} 
\hline 
\hline 
\multicolumn{1}{c}{(1)} & \multicolumn{1}{c}{(2)} & \multicolumn{1}{c}{(3)} & \multicolumn{1}{c}{(4)} & \multicolumn{1}{c}{(5)} & \multicolumn{1}{c}{(6)} & \multicolumn{1}{c}{(7)} & \multicolumn{1}{c}{(8)} & \multicolumn{1}{c}{(9)} & \multicolumn{1}{c}{(10)} & \multicolumn{1}{c}{(11)} & \multicolumn{1}{c}{(12)} & \multicolumn{1}{c}{(13)}
\\ 
\multicolumn{1}{c}{Source} & \multicolumn{1}{c}{Figure} & \multicolumn{1}{c}{Polarized} & \multicolumn{1}{c}{Model} & \multicolumn{1}{c}{$\tilde{\chi}^2$} & \multicolumn{1}{c}{AIC} & \multicolumn{1}{c}{Component} & \multicolumn{1}{c}{$p_{\lambda=0}$} & \multicolumn{5}{c}{Best fit parameters for model component}
\\ 
\multicolumn{1}{c}{} & \multicolumn{1}{c}{no.} & \multicolumn{1}{c}{S/N} & \multicolumn{1}{c}{Type} & \multicolumn{1}{c}{} & \multicolumn{1}{c}{} & \multicolumn{1}{c}{Type} & \multicolumn{1}{c}{} & \multicolumn{1}{c}{P$_1$} & \multicolumn{1}{c}{P$_2$} & \multicolumn{1}{c}{P$_3$} & \multicolumn{1}{c}{P$_4$} & \multicolumn{1}{c}{P$_5$} 
\\ 
\multicolumn{1}{c}{} & \multicolumn{1}{c}{} & \multicolumn{1}{c}{} & \multicolumn{1}{c}{} & \multicolumn{1}{c}{} & \multicolumn{1}{c}{} & \multicolumn{1}{c}{} & \multicolumn{1}{c}{} & \multicolumn{1}{c}{[ ]\tablenotemark{a}} & \multicolumn{1}{c}{[ ]\tablenotemark{a}} & \multicolumn{1}{c}{[ ]\tablenotemark{a}} & \multicolumn{1}{c}{[ ]\tablenotemark{a}} & \multicolumn{1}{c}{[ ]\tablenotemark{a}} 
\\
\hline 
cena\_s1681 & \ref{fig:cena_s1681} & 11.4 &TT &1.3 &-758.0 &T &0.033 &0.04(2) &-40.0(100.0) &100(60) &21(9) &0.5(5) \\[1pt] 
  &  &  &  &  &  &T &0.100 &0.123(7) &-80(1) &0.7(2.0) &14(8) &0.75(9) \\[1pt] 
  &  &  &ST &1.4 &-752.6 &S &0.100 &0.100(3) &-81(1) &- &- &0.76(4) \\[1pt] 
  &  &  &  &  &  &T &0.048 &0.06(2) &-0.2(100.0) &110(60) &19(10) &0.6(0.8) \\[1pt] 
  &  &  &S &3.7 &-447.0 &S &0.099 &0.099(2) &-79.9(7) &- &- &0.73(2) \\[1pt] 
cena\_c1748 & \ref{fig:cena_c1748} & 7.3 &TT &1.2 &-1015.7 &T &0.052 &0.06(2) &180(90) &130(40) &13(8) &0.6(4) \\[1pt] 
  &  &  &  &  &  &T &0.025 &0.029(5) &-110(8) &13(6) &3.0(9.0) &0.9(4) \\[1pt] 
  &  &  &ST &1.2 &-1010.3 &S &0.021 &0.021(3) &-102(7) &- &- &0.8(2) \\[1pt] 
  &  &  &  &  &  &T &0.056 &0.07(2) &190(80) &110(30) &20(9) &1.0(4) \\[1pt] 
  &  &  &S &3.7 &-612.1 &S &0.027 &0.027(2) &-115(2) &- &- &1.11(5) \\[1pt] 
cena\_s1382 & \ref{fig:cena_s1382} & 11.1 &TT &1.3 &-320.9 &T &0.069 &0.1(1) &-2(2)e2 &65(40) &6.0(6.0) &0.3(1.0) \\[1pt] 
  &  &  &  &  &  &T &0.124 &0.16(1) &-53(3) &3.0(3.0) &30(9) &0.2(2) \\[1pt] 
  &  &  &ST &1.4 &-317.0 &S &0.130 &0.130(8) &-54(3) &- &- &0.25(9) \\[1pt] 
  &  &  &  &  &  &T &0.142 &0.2(1) &-2(2)e2 &80(40) &4.0(10.0) &2(1) \\[1pt] 
  &  &  &S &3.7 &-136.3 &S &0.129 &0.129(5) &-54(2) &- &- &0.26(5) \\[1pt] 
cena\_c1152 & \ref{fig:cena_c1152} & 10.4 &TT &1.2 &-503.8 &T &0.080 &0.10(1) &-46(6) &13(4) &12(9) &0.4(3) \\[1pt] 
  &  &  &  &  &  &T &0.238 &0.3(1) &-80(70) &120(30) &8.0(9.0) &1.1(8) \\[1pt] 
  &  &  &ST &1.3 &-498.8 &S &0.059 &0.059(6) &-47(6) &- &- &0.5(2) \\[1pt] 
  &  &  &  &  &  &T &0.105 &0.1(0.1) &2(1)e2 &70(30) &12(9) &1.4(9) \\[1pt] 
  &  &  &S &3.6 &-325.0 &S &0.062 &0.062(2) &-42(2) &- &- &0.31(7) \\[1pt] 
cena\_c1832 & \ref{fig:cena_c1832} & 5.7 &ST &1.2 &-1202.6 & S &0.018 &0.018(2) &-77(5) &- &- &1.0(2) \\[1pt] 
  &  &  &  &  &  &T &0.024 &0.03(1) &-60.0(100.0) &30(40) &4(8) &-0.7(6) \\[1pt] 
  &  &  &TT &1.2 &-1199.8 &T &0.032 &0.04(2) &-90(80) &10.0(50.0) &5.0(6.0) &3.5(7) \\[1pt] 
  &  &  &  &  &  &T &0.036 &0.04(2) &-80.0(100.0) &20.0(30.0) &16(4) &0.1(0.3) \\[1pt] 
  &  &  &S &3.5 &-902.8 &S &0.012 &0.0124(8) &-53(2) &- &- &0.02(0.05) \\[1pt] 
lmc\_s14 & \ref{fig:lmc_s14} & 44.9 &TT &1.0 &-580.4 &T &0.065 &0.08(3) &-90(20) &60(30) &25(5) &0.1(1) \\[1pt] 
  &  &  &  &  &  &T &0.173 &0.22(1) &50(1) &3(2) &19(8) &1.20(7) \\[1pt] 
  &  &  &ST &1.0 &-572.3 &S &0.170 &0.170(5) &50(1) &- &- &1.19(4) \\[1pt] 
  &  &  &  &  &  &T &0.070 &0.09(3) &-90(80) &60(50) &8(7) &0.01(0.2) \\[1pt] 
  &  &  &S &3.1 &-391.1 &S &0.168 &0.168(2) &48.3(6) &- &- &1.24(2) \\[1pt] 
cena\_s1014 & \ref{fig:cena_s1014} & 4.1 &SSS &1.6 &-332.1 &S &0.171 &0.17(8) &-11(10) &- &- &0.4(1.0) \\[1pt] 
  &  &  &  &  &  &S &0.142 &0.14(7) &-24(10) &- &- &-1.2(10.0) \\[1pt] 
  &  &  &  &  &  &S &0.068 &0.07(2) &-99(30) &- &- &-0.2(20.0) \\[1pt] 
  &  &  &SS &1.6 &-330.4 &S &0.053 &0.053(5) &-102(6) &- &- &0.3(2) \\[1pt] 
  &  &  &  &  &  &S &0.050 &0.050(5) &117(6) &- &- &-0.6(3) \\[1pt] 
  &  &  &S &2.2 &-240.5 &S &0.042 &0.042(5) &-90(10) &- &- &0.05(0.3) \\[1pt] 
cena\_c1764 & \ref{fig:cena_c1764} & 15.3 &T &1.1 &-1660.7 &T &0.027 &0.027(6) &-80(3) &25(1) &28(8.0) &0.68(5) \\[1pt] 
  &  & &S &1.9 &-1468.8 &S &0.010 &0.0104(5) &-78(2) &- &- &0.65(5) \\[1pt] 
cena\_c1435 & \ref{fig:cena_c1435} & 14.2 &SS &1.4 &-1398.1 &S &0.021 &0.021(3) &-72(9) &- &- &-0.2(1) \\[1pt] 
  &  &  &  &  &  &S &0.011 &0.011(3) &-37(20) &- &- &-0.6(2.0) \\[1pt] 
  &  &  &T &1.4 &-1397.3 &T &0.137 &0.169(7) &-65(1) &14.5(7) &12(9) &1.46(3) \\[1pt] 
  &  &  &SSS &1.4 &-1390.8 &S &0.255 &0.3(1) &-75(8) &- &- &-1.3(3) \\[1pt] 
  &  &  &  &  &  &S &0.153 &0.15(8) &-68(3) &- &- &0.09(7) \\[1pt] 
  &  &  &  &  &  &S &0.129 &0.13(3) &-67(5) &- &- &0.3(1) \\[1pt] 
  &  &  &S &1.6 &-1342.7 &S &0.025 &0.0254(7) &-66(1) &- &- &-0.20(4) \\[1pt] 
 \hline
\end{tabular} 
\end{table*} 
\end{turnpage}    

\addtocounter{table}{-1}
\begin{turnpage} 
\begin{table*} 
\caption{\emph{continued}} 
\label{tab:fitgoodness} 
\scriptsize 
\tabcolsep=0.3cm 
\begin{tabular}{l c c c c l c l l l l l l} 
\hline 
\hline 
\multicolumn{1}{c}{(1)} & \multicolumn{1}{c}{(2)} & \multicolumn{1}{c}{(3)} & \multicolumn{1}{c}{(4)} & \multicolumn{1}{c}{(5)} & \multicolumn{1}{c}{(6)} & \multicolumn{1}{c}{(7)} & \multicolumn{1}{c}{(8)} & \multicolumn{1}{c}{(9)} & \multicolumn{1}{c}{(10)} & \multicolumn{1}{c}{(11)} & \multicolumn{1}{c}{(12)} & \multicolumn{1}{c}{(13)}
\\ 
\multicolumn{1}{c}{Source} & \multicolumn{1}{c}{Figure} & \multicolumn{1}{c}{Polarized} & \multicolumn{1}{c}{Model} & \multicolumn{1}{c}{$\tilde{\chi}^2$} & \multicolumn{1}{c}{AIC} & \multicolumn{1}{c}{Component} & \multicolumn{1}{c}{$p_{\lambda=0}$} & \multicolumn{5}{c}{Best fit parameters for model component}
\\ 
\multicolumn{1}{c}{} & \multicolumn{1}{c}{no.} & \multicolumn{1}{c}{S/N} & \multicolumn{1}{c}{Type} & \multicolumn{1}{c}{} & \multicolumn{1}{c}{} & \multicolumn{1}{c}{Type} & \multicolumn{1}{c}{} & \multicolumn{1}{c}{P$_1$} & \multicolumn{1}{c}{P$_2$} & \multicolumn{1}{c}{P$_3$} & \multicolumn{1}{c}{P$_4$} & \multicolumn{1}{c}{P$_5$} 
\\ 
\multicolumn{1}{c}{} & \multicolumn{1}{c}{} & \multicolumn{1}{c}{} & \multicolumn{1}{c}{} & \multicolumn{1}{c}{} & \multicolumn{1}{c}{} & \multicolumn{1}{c}{} & \multicolumn{1}{c}{} & \multicolumn{1}{c}{[ ]\tablenotemark{a}} & \multicolumn{1}{c}{[ ]\tablenotemark{a}} & \multicolumn{1}{c}{[ ]\tablenotemark{a}} & \multicolumn{1}{c}{[ ]\tablenotemark{a}} & \multicolumn{1}{c}{[ ]\tablenotemark{a}} 
\\
\hline 
cena\_s1349 & \ref{fig:cena_s1349} & 9.2 &SS &1.4 &-1711.1 &S &0.009 &0.009(0.02) &-76(4) &- &- &1.49(10) \\[1pt] 
  &  &  &  &  &  &S &0.009 &0.009(0.02) &-115(4) &- &- &-0.14(4) \\[1pt] 
  &  &  &S &1.6 &-1660.2 &S &0.013 &0.0127(6) &-100(2) &- &- &-0.74(6) \\[1pt] 
cena\_s1803 & \ref{fig:cena_s1803} & 12.5 &SS &1.1 &-1284.5 &S &0.042 &0.042(10) &-27(5) &- &- &0.6(2) \\[1pt] 
  &  &  &  &  &  &S &0.029 &0.029(9) &-17(7) &- &- &1.6(4) \\[1pt] 
  &  &  &S &1.1 &-1273.8 &S &0.023 &0.023(1) &-32(2) &- &- &1.06(6) \\[1pt] 
0515-674 & \ref{fig:0515-674} & 3.5 & S & 0.2 & -4957.5 & S & 0.00002 & 0.00002(2) & -100(700) &- &- & 0(1) \\[1pt] 
  &  &  & Unp & 0.2 & -4956.4 & - & - & - & - &- & - & - \\[1pt] 
  &  &  & S & 0.2 & -4957.5 & S & 2$\times10^{-5}$ & 2(2)$\times10^{-5}$ & -100(700) &- &- & 0(1) \\[1pt] 
 \hline 

\end{tabular} 
\tablecomments{$^a$ The fitted model parameters P$_1$--P$_5$ refer to different quantities possessing different units for S (Faraday simple/thin) and T (Faraday thick) model components. For T model components, P$_1$, P$_2$, P$_3$, P$_4$ \& P$_5$ correspond to $A$ [non-dim], $\phi$ [rad m$^{-2}$], $\sigma_0$ [rad m$^{-2}$], $N$ [non-dim] \& $\psi_0$ [rad] in Eqn. \ref{eq:supergauss} respectively. For S model components, P$_1$, P$_2$ \& P$_5$ correspond to $p$ [non-dim], RM [rad m$^{-2}$] \& $\psi_0$ [rad] respectively. The column positions for S components are chosen such that similar quantities align between S \& T components. P$_3$ \& P$_4$ are therefore left blank. Note that a model with zero polarization provides an excellent fit to 0515-674. In this table for this isolated case, we designate this model type as `Unp' (unpolarized).} 
\end{table*} 
\end{turnpage}

\subsubsection{Overall best fit models: Goodness of fit and model type}\label{sec-results-bestfit} 

The overall best fit model for 30/36 sources has $\tilde{\chi}^2<2.1$ (column 5, Table \ref{tab:fitgoodness}), with an average $\tilde{\chi}^2$ value of 1.49. We used a Komogorov-Smirnov (K-S) test to compare the distribution of the standardized ($q$,$u$) residuals (defined in Section \ref{sec-results-observed}, see also inset axes in panel (a) of Figs. \ref{fig:lmc_s11}--\ref{fig:cena_s1803}) to the standard normal distribution, and found no statistically significant difference for any of these sources. The remaining four sources --- lmc\_s11 (Fig. \ref{fig:lmc_s11}), lmc\_c04 (Fig. \ref{fig:lmc_c04}), lmc\_c03 (Fig. \ref{fig:lmc_c03}) \& cen\_c1466 (Fig. \ref{fig:cena_c1466}) --- had higher $\tilde{\chi}^2$ values of 3.7, 4.4, 7.8 \& 4.2 (respectively), but each have high polarized S/N and the fractional data-model residual values are low.

 With the exception of a single source (0515-674), the overall best fit model improves on the AIC value of the best fitting S model by between 10 \& $\sim$23000. Thus, we positively detect Faraday complex polarization structure in almost our entire sample, including those sources selected to be part of our Faraday simple control group (Section \ref{sec-sample}). The bright calibrator source 0515-674 represents the sole, remarkable exception (Fig. \ref{fig:0515-674}). The best-fit S model has a fractional polarization of less than 2$\times10^{-5}$, though an unpolarized model is also consistent with the data across the entire band.
 
 The best fit model type is unambiguous for 23 sources, while the remaining thirteen sources --- which are generally faint or heavily resolved --- have multiple best fit candidates. The overall best fit model for each source contains between 1 \& 4 emission components. Only in three cases was the polarization data well-described by a model with a single component (0515-674; Fig. \ref{fig:0515-674}, cen\_c1764; Fig. \ref{fig:cena_c1764} \& cen\_c1435; Fig. \ref{fig:cena_c1435}). 29 sources required two model components, while five sources required either 3 or 4 model components. Three of the sources in this latter group (lmc\_s11, Fig. \ref{fig:lmc_s11}; lmc\_c04, Fig. \ref{fig:lmc_c04}; lmc\_c03, Fig. \ref{fig:lmc_c03}) also had larger $\tilde{\chi}^2$ values for the overall best fit models, suggesting that an even greater number of model components might be required to adequately describe their polarization behaviour. However, we were unsuccessful in identifying any such model. Alternately, it may indicate that a more sophisticated functional form for T model components is required to fit the subtle $\boldsymbol{P}(\lambda^2)$ behaviour in data with high S/N.

Of the T, SS, ST, TT \& SSS model types fit to the entire sample, each provided the best fit for at least one source. TT models were most commonly selected by the AIC (19 sources), followed by ST \& SS models (6 \& 4 sources respectively), SSS models (2 sources) and a T model (1 source). The custom models required for lmc\_c04, lmc\_s11 \& lmc\_c03 were of type STT, SSTT \& SSTT respectively. In general then, both S \& T components were required to model our data, with 14 (28) of the 36 sources requiring at least one S (T) component.

\subsubsection{Spectropolarimetric structure of resolved sources}\label{sec-results-resolved} 

In Appendix \ref{sec-appendA}, we provide images of the resolved sources in our sample in both total and linearly polarized intensity. A range of morphological types are present. In both quantities, there are sources which resolve cleanly into multiple subcomponents which are themselves unresolved, sources which resolve into one or more extended components, and various other combinations of these possibilities.

It is interesting to consider whether the number and type of spectropolarimetric model components fit to these sources (remembering that our spectropolarimetric analysis is performed on the integrated flux for each source) shows any clear relationship with the morphological components present in images of the sources. In general, we do not find this to be the case. Sources with two FDF model components (either S or T-type) often show two spatial components in the appendix images (e.g. lmc\_c16 --- Figs. \ref{fig:lmc_c16} \& \ref{fig:lmc_c16_showsrc}; lmc\_s14 --- Figs. \ref{fig:lmc_s14} \& \ref{fig:lmc_s14_showsrc}; lmc\_c06 --- Figs. \ref{fig:lmc_c06} \& \ref{fig:lmc_c06_showsrc}; cen\_c1832 --- Figs. \ref{fig:cena_c1832} \& \ref{fig:cena_c1832_showsrc}; cen\_c1573 --- Figs. \ref{fig:cena_c1573} \& \ref{fig:cena_c1573_showsrc}). However, there are also sources which require a greater number of components in the FDF model than there are visible morphological components (e.g. lmc\_c15 --- Figs. \ref{fig:lmc_c15} \& \ref{fig:lmc_c15_showsrc}; lmc\_s11 --- Figs. \ref{fig:lmc_s11} \& \ref{fig:lmc_s11_showsrc}; lmc\_c03 --- Figs. \ref{fig:lmc_c03} \& \ref{fig:lmc_c03_showsrc}; cen\_s1681 --- Figs. \ref{fig:cena_s1681} \& \ref{fig:cena_s1681_showsrc}; cen\_c1972 --- Figs. \ref{fig:cena_c1972} \& \ref{fig:cena_c1972_showsrc}), and vice versa (e.g. cen\_c1764 --- Figs. \ref{fig:cena_c1764} \& \ref{fig:cena_c1764_showsrc}; cen\_c1827 --- Figs. \ref{fig:cena_c1827} \& \ref{fig:cena_c1827_showsrc}). Furthermore, we are unable to identify any clear relationship between the presence or absence of unresolved / extended subcomponents in the images, and the presence or absence of S / T components in the best fit FDF models. For example, spatially unresolved components in the source images give rise to both S components (e.g. lmc\_c15 --- Figs. \ref{fig:lmc_c15} \& \ref{fig:lmc_c15_showsrc}) and T components (e.g. lmc\_c16 --- Figs. \ref{fig:lmc_c16} \& \ref{fig:lmc_c16_showsrc}), while extended components in the source images can do likewise (e.g. cen\_s1803 --- Figs. \ref{fig:cena_s1803} \& \ref{fig:cena_s1803_showsrc} \& lmc\_c06 --- Figs. \ref{fig:lmc_c06} \& \ref{fig:lmc_c06_showsrc}, for S \& T components respectively).

However, it is likely that this analysis is significantly hampered by a lack of spatial resolution and the range of source brightnesses present in our sample.

\subsubsection{Requirement for broad components in the best-fit FDF models}\label{sec-results-fthick}

A primary motivation for our study was to search for emission components with large peak Faraday depths and dispersions, since this can place strong constraints on the physical origin of the emission. Such components are required to fit the polarization data in the FDF models of 1315-46, lmc\_s11, lmc\_c02, lmc\_c03 \& lmc\_c04. We now describe the characteristics of these components.

The phase calibrator source 1315-46 shows very strong depolarization in the 3 \& 6 cm bands. Its FDF model (panel (f) of Fig. \ref{fig:1315-46}) possesses two broad emission components, centered on $\phi_{\text{peak}}=-70\pm30$ \& $+420\pm40$ rad m$^{-2}$ with $\sigma_\phi=470\pm15$ \& $300\pm15$ rad m$^{-2}$ respectively. The Super-Gaussian shape parameter is strongly constrained for the dominant T component with a value of $N=2.5\pm0.2$. Thus, this T component depolarizes in a manner close to that expected for the Burn (1966) foreground screen until $\lambda^2\approx0.01$ m$^2$, by which point their contribution to the integrated polarization is negligible. 

The FDF model for lmc\_s11 (panel (f) of Fig. \ref{fig:lmc_s11}) possesses two components with large Faraday depth/dispersion, centered on $\phi_{\text{peak}}=+109\pm7$ \& $-133\pm5$ rad m$^{-2}$ with $\sigma_\phi=98\pm5$ \& $132\pm3$ rad m$^{-2}$ respectively. This model achieves an excellent fit to the data with the exception of Stokes $u$ in the 3 cm band, which deviates from our model significantly (see inset axis in panel (a) of Fig. \ref{fig:lmc_s11}). While we could broadly reproduce this behaviour using models containing components with similar Faraday depths \& dispersions, the goodness of fit at 6 cm was compromised and the result was strongly disfavoured by the AIC. The Super-Gaussian shape parameter is strongly constrained for both T components in the best fit model with $N=2.1\pm0.1$ \& $2.13\pm0.02$. Thus, each T component essentially depolarizes $\propto\text{exp}(-k\lambda^4)$ (where $k$ is some constant) until $\lambda^2\approx0.01$ m$^2$, beyond which the polarized signal is effectively undetectable.

For lmc\_c03 (Fig. \ref{fig:lmc_c03}), two T components successfully reproduce most aspects of the ($q$,$u$) behaviour at 6/3 cm, though minor discrepancies are also evident (see inset axis in panel (a) of Fig. \ref{fig:lmc_c03}). The components have $\sigma_\phi=1439\pm4$ \& $503\pm3$ rad m$^{-2}$ --- by far the the most highly Faraday dispersed emission components detected in our data --- centered on $\phi_{\text{peak}}=+287\pm20$ \& $+269\pm10$ rad m$^{-2}$ respectively. Both components are mostly depolarized by $\lambda^2\approx0.003$ m$^2$, but since $N>2$ for each, small oscillations persist in $q$ \& $u$ out to $\lambda^2\approx0.02$ m$^2$ (see panel (b)). For $\lambda^2<0.0009$ m$^2$ and $0.0014<\lambda^2<0.0021$ m$^2$ where we have no data, the model predicts pronounced structure which is likely an artefact of the unconstrained fit in these regions. 

The model for lmc\_c02  (Fig. \ref{fig:lmc_c02}) contains two T components. The component with the smaller amplitude has $\sigma_\phi=71\pm6$ rad m$^{-2}$, $\phi_{\text{peak}}=+147\pm8$ rad m$^{-2}$, and $N=2.0\pm0.1$. As with lmc\_s11, this component depolarizes smoothly and monotonically $\propto\text{exp}(-k\lambda^4)$ for $0<\lambda^2<0.012$ m$^2$.
 
 lmc\_c04 (Fig. \ref{fig:lmc_c04}) shows arguably the most spectacular $\boldsymbol{P}(\lambda^2)$ behaviour, which is primarily caused by a dominant T component possessing $\sigma_\phi\approx67\pm1$, centred on $\phi_{\text{peak}}=+124\pm2$ rad m$^{-2}$ with $N\approx6.4\pm0.4$ (panel (f)). The $\sigma_\phi$ and $N$ values combine to produce strong oscillatory depolarization in $p(\lambda^2)$, with a deep null at $\lambda^2\approx0.022$ m$^2$ and several local maxima and minima evident (panel (d)). While this behaviour is reminiscent of that associated with the Burn (1966) slab, our model differs in its detailed behaviour (e.g. see Fig. \ref{fig:sgdemo}, which shows the subtle effect which the $N$ parameter has on depolarization behaviour).

\subsubsection{Detection \& characteristics of bright emission components with low Faraday dispersion}\label{sec-results-fthin}

The overall best fit models for the sources lmc\_s13 (Fig. \ref{fig:lmc_s13}), cen\_c1466 (Fig. \ref{fig:cena_c1466}), lmc\_s11 (Fig. \ref{fig:lmc_s11}) \&  lmc\_c15 (Fig. \ref{fig:lmc_c15}) contain at least one bright S component (with band-averaged S/N of between 20 \& 160). Though components which are truly Faraday-thin will never be generated in Nature, the emission modelled by these S components must nevertheless approach this ideal, otherwise T components would have been selected over S components by the AIC criteria. At the same time, $|\phi_{\text{peak}}|$ values of 10--120 rad m$^{-2}$ indicate that substantial Faraday rotation has occurred along the LOS. Together, this suggests that the Faraday rotating plasmas lying along the LOS possess comparatively \emph{uniform} $n_e$ and $\boldsymbol{B}$ structure. In this section, we derive upper limits on the Faraday dispersion of these components to more fully explore this possibility.

To derive these limits, we refit our data over $\lambda^2$ ranges where the low-Faraday-dispersion components dominate $\boldsymbol{P}(\lambda^2)$ (i.e. $\lambda^2>0.01$ m$^{2}$ for lmc\_s13, cen\_c1466 \& lmc\_s11, and the entire band for lmc\_c15) using models consisting solely of T components. We present the results in Table \ref{tab:fitgoodnessthincomp}. They show that emission components which contribute substantially to the fractional polarization of these sources (column 5) typically possess very low Faraday dispersions --- the upper limits are generally a few rad m$^{-2}$, while the lower limits are in some cases consistent with 0 rad m$^{-2}$ (column 7). We discuss these results further in Section \ref{sec-physnaturethin}.


\begin{table} 
\caption{Results for T component model fitting described in Section \ref{sec-results-fthin}} 
\label{tab:fitgoodnessthincomp} 
\scriptsize 
\tabcolsep=0.15cm 
\begin{tabular}{l c l c c c c} 
\hline 
\hline 
\multicolumn{1}{c}{(1)} & \multicolumn{1}{c}{(2)} & \multicolumn{1}{c}{(3)} & \multicolumn{1}{c}{(4)} & \multicolumn{1}{c}{(5)} & \multicolumn{1}{c}{(6)} & \multicolumn{1}{c}{(7)} 
\\ 
\multicolumn{1}{c}{Source} & \multicolumn{1}{c}{Model} & \multicolumn{1}{c}{$\tilde{\chi}^2$} & \multicolumn{1}{c}{Comp.} & \multicolumn{1}{c}{$A$} & \multicolumn{1}{c}{$\phi_{\text{peak}}$} & \multicolumn{1}{c}{$\sigma_\phi$} 
\\ 
\multicolumn{1}{c}{} & \multicolumn{1}{c}{} & \multicolumn{1}{c}{} & \multicolumn{1}{c}{Type} & \multicolumn{1}{c}{} & \multicolumn{1}{c}{} & \multicolumn{1}{c}{} 
\\ 
\multicolumn{1}{c}{} & \multicolumn{1}{c}{} & \multicolumn{1}{c}{} & \multicolumn{1}{c}{} & \multicolumn{1}{c}{} & \multicolumn{1}{c}{[rad m$^{-2}$]} & \multicolumn{1}{c}{[rad m$^{-2}$]}
\\
\hline 

lmc\_s13 & TT & 1.3 & T & 0.038 & 75(5) & $35(1)$\\
 & & & T & 0.018 & 69(5) & $0.8\substack{+1.8 \\ -0.8}$\\
cen\_c1466 & TT & 2.8 & T & 0.058 & $-39(2)$ & $0.8\substack{+1.7 \\ -0.6}$\\
 & & & T & 0.011 & $-1(3)$ & $1.7\substack{+2.3 \\ -1.3}$\\
lmc\_s11 & TT & 2.5 & T & 0.032 & 54.5(6) & $0.1\substack{+1.0 \\ -0.1}$\\
 & & & T & 0.005 & 88(5) & $1.3\substack{+3.0 \\ -1.3}$\\
lmc\_c15 & TTT & 1.9 & T & 0.045 & 119.3(8) & $0.7\substack{+0.8 \\ -0.5}$\\
 & & & T & 0.025 & -58(2) & $1.4\substack{+2.2 \\ -1.0}$\\
 & & & T & 0.007 & 79(7) & $19\substack{+1 \\ -17}$\\
\hline 
\end{tabular} 
\tablecomments{This table is formatted similarly to Table \ref{tab:fitgoodness} --- see Section \ref{sec-results-broadbandmodelling} for a detailed description of the layout. The superscript and subscript values quoted in column 7 are the $1\sigma$ standard errors on the fitted model parameters.} 
\end{table} 


\subsection{Spectropolarimetric variability over 1.3--1.5 GHz}\label{sec-results-temporal} 

Law et al. (2011) \& O'Sullivan et al. (2012) detected multiple components in the broadband polarization spectra of a number of radio sources, and argued that these emission components likely arise in the inner parsec-scale regions of the associated AGN. Given that parsec-scale magnetoionic structure in AGN is observed to evolve over timescales of months to years (e.g. Hovatta et al. 2012, Zavala \& Taylor 2001), these authors predict that the integrated polarization spectra of core-dominated radio sources may also vary with time. If this behaviour is commonplace, time domain broadband spectropolarimetry could become a powerful tool for studying magnetoionic structure in the immediate vicinity of AGN using non-VLBI radio telescopes.

We searched for the predicted variability by fitting polarization models to both the archival and 2012 data (for the non-calibrator sample sources only) over the 1.3--1.5 GHz band common to both epochs. We used models consisting of a single T component to account for the small amount of depolarization apparent in some sources over this narrow band. However, the $\sigma_\phi$ \& $N$ parameters (Eqn. \ref{eq:supergauss}) were poorly constrained by the narrowband data, so we ignore them in the subsequent analysis. We identified variability in the remaining parameters --- $A$, $\phi_{\text{peak}}$ \& $\psi_0$ (see Eqn. \ref{eq:supergauss}) --- by calculating the standardized difference $\xi$ between the best fit model parameter values for each epoch, defined by:

\begin{eqnarray}
\xi=\frac{|\mu_1 - \mu_2|}{\sqrt{\sigma_1^2+\sigma_2^2}}
\end{eqnarray}

where $\mu_1$ and $\mu_2$ represent the best fit parameters values for $A$, $\phi_{\text{peak}}$ \& $\psi_0$ in the archival and 2012 epochs respectively, and $\sigma_1$ and $\sigma_2$ are the standard uncertainties on those parameter values. We flagged sources as potentially variable when $\xi>3$ (i.e. the best fit model parameter values differ between epochs with $>99$\% confidence) for one or more of the best fit model parameters.


\begin{table*} 
\caption{Best fit model parameters, uncertainties and standardized differences in parameter values between epochs ($\xi$; defined in main text) for S model fits to the archival and 2012 data} 
\label{table:variability}
\scriptsize 
\tabcolsep=0.2cm 
\begin{tabular}{l l l l l l l l l l} 
\hline 
\hline 
\multicolumn{1}{c}{(1)} & \multicolumn{1}{c}{(2)} & \multicolumn{1}{c}{(3)} & \multicolumn{1}{c}{(4)} & \multicolumn{1}{c}{(5)} & \multicolumn{1}{c}{(6)} & \multicolumn{1}{c}{(7)} & \multicolumn{1}{c}{(8)} & \multicolumn{1}{c}{(9)} & \multicolumn{1}{c}{(10)} 
\\ 
\multicolumn{1}{c}{Source} & \multicolumn{1}{c}{$A$ (2012)} & \multicolumn{1}{c}{$A$ (Archival)} & \multicolumn{1}{c}{$\phi_{\text{peak}}$ (2012)} & \multicolumn{1}{c}{$\phi_{\text{peak}}$ (Archival)} & \multicolumn{1}{c}{$\psi_0$ (2012)} & \multicolumn{1}{c}{$\psi_0$ (Archival)} & \multicolumn{1}{c}{$\xi_A$} & \multicolumn{1}{c}{$\xi_{\phi_{\text{peak}}}$} & \multicolumn{1}{c}{$\xi_{\psi_{0}}$} 
\\ 
\multicolumn{1}{c}{} & \multicolumn{1}{c}{} & \multicolumn{1}{c}{} & \multicolumn{1}{c}{[rad m $^{-2}$]} & \multicolumn{1}{c}{[rad m $^{-2}$]} & \multicolumn{1}{c}{[rad]} & \multicolumn{1}{c}{[rad]} & \multicolumn{1}{c}{} & \multicolumn{1}{c}{} & \multicolumn{1}{c}{}
\\
\hline 

cen\_c1972 & $7.94(7)\times10^{-2}$ & $6.89(8)\times10^{-2}$  & -76(1) & -74(2) & 1.30(5) & 1.36(8) & 9.8 & 1.0  & 1.4  \\
\tableline
lmc\_c03 & $5.71(2)\times10^{-2}$  & $5.87(5)\times10^{-2} $& 40.6(6) & 47(2) & 1.34(4) & 0.8(1) & 2.8  & 3.1 & 4.7  \\ 
lmc\_s13 & $5.00(7)\times10^{-2}$ & $2.12(9)\times10^{-2}$  & 66(2) & 39(9) & -0.4(1) & 0.9(8) & 27.7 & 2.9 & 2.6\\
lmc\_s11 & $2.81(2)\times10^{-2}$ & $5.8(1)\times10^{-3} $& 45.5(1)&17(3)& -0.11(4) & 1.0(1) & 108 & 9.5 & 7.9\\
\hline 
\end{tabular} 
\tablecomments{The line ruled under the first row of the table distinguishes sources in which we detect temporal variability (sources below the line) from cen\_c1972 --- an example of a source which we argue shows temporal change but not true polarization varability (see main text).} 
\end{table*}

\begin{figure*}[htpb]
\centering
\includegraphics[width=0.95\textwidth]{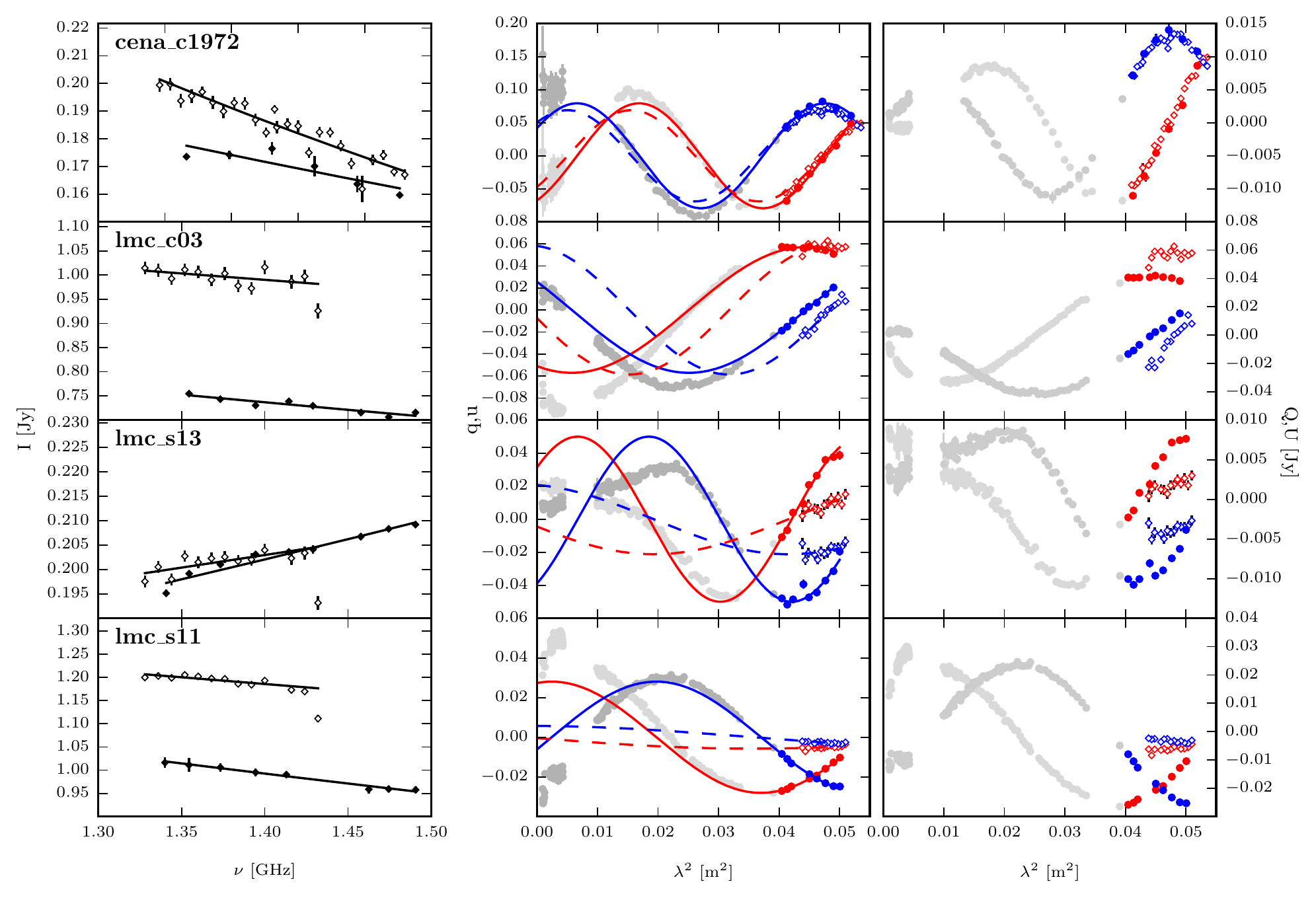}
\caption{Two-epoch comparison of polarization data for selected sources. Each row contains plots for a particular source (labelled on plots axes). Column 1 contains the Stokes I fit plus degree 1 polynomial fit to the Stokes I data for the archival (unfilled diamonds) and 2012 (filled diamonds) data. Column 2 contains plots of the restricted band Stokes $q$ (red) and $u$ (blue) data (see main text) for the archival (unfilled diamonds) and 2012 (filled diamonds) epochs. The greyscale points are the full 1.3--10 GHz 2012 data. The best fit S model to the archival (dashed) and 2012 (solid line) data is also plotted. The column 3 plots are identical to the column 2 plots with the exception that ($Q$,$U$) rather than ($q$,$u$) data is plotted.}
\label{fig:archcomparo}
\end{figure*}

31 of the 34 sources either have $\xi\leq3$ for all model parameters and thus show no evidence of spectropolarimetric variability, or show significant changes which are nevertheless more readily attributed to other factors. An example of the latter case is cen\_c1972. Its Stokes I, ($q$,$u$) \& ($Q$,$U$) data are plotted in row 1 of Fig. \ref{fig:archcomparo}, while the fitted parameter values and uncertainties are listed in row 1 of Table \ref{table:variability}. The polarization data and models are evidently similar between epochs. $\xi\approx 1$ for both $\phi_{\text{peak}}$ and $\psi_0$, indicating no substantial change in either polarization property. At first, the 13\% increase in $A$ appears significant with $\xi_A=9.8$. However, the total intensity decreases by a similar amount over the same period, meaning the changes are probably unrelated to the polarization structure of the source. The Stokes ($Q$,$U$) data show no statistically significant differences between epochs.

The remaining three sources --- lmc\_c03, lmc\_s11 \& lmc\_s13 --- vary in one or more of the fitted model parameters. lmc\_s11 shows the biggest changes (4th row of Table \ref{table:variability} \& Fig. \ref{fig:archcomparo}), with $\xi_A\approx108$, $\xi_{\phi_{\text{peak}}}\approx10$ and $\xi_{\psi_{0}}\approx8$. Between the archival and 2012 epochs, the fractional polarization (i.e. $A$) increases by 2.23(3)\%, $\phi_{\text{peak}}$ increases by $\sim29(3)$ rad m$^{-2}$, and $\psi_0$ by $\sim1.1(1)$ rad. The total intensity changes during the same period, but the polarization changes are evident in both ($q$,$u$) and ($Q$,$U$), and cannot be explained by total intensity variation. lmc\_s13 (3rd row of Table \ref{table:variability} \& Fig. \ref{fig:archcomparo}) shows highly significant change in $A$ with $\xi_A\approx28$, with moderately significant accompanying changes in $\phi_{\text{peak}}$ ($\xi_{\phi_{\text{peak}}}=2.9$) \& $\psi_0$ ($\xi_{\psi_{0}}=2.6$). From the archival to 2012 epochs, the fractional polarization has increased by 2.9(1)\%, $\phi_{\text{peak}}$ by 27(11) rad m$^{-2}$, an $\psi_0$ by 1.3(9) rad. Finally, for lmc\_c03 (2nd row of Table \ref{table:variability} \& Fig. \ref{fig:archcomparo}), Stokes I changes by $\sim25$\% between epochs and the Stokes ($Q$,$U$) data differ more between epochs than the ($q$,$u$) data. Nonetheless, moderately significant differences remain in the ($q$,$u$) data. The source shows small but significant changes of 6.4(2.5) rad m$^{-2}$ in $\phi_{\text{peak}}$ ($\xi_{\phi_{\text{peak}}}=3.1$) and 0.5(1) rad in $\psi_0$ ($\xi_{\psi_{0}}=4.7$), with a tiny, marginal change of 0.16(7)\% in $A$ ($\xi_A=2.8$).

\section{Discussion}\label{sec-discussion} 
 
 \subsection{Interpreting complex $\boldsymbol{P}(\lambda^2)$ behaviour using model FDFs}\label{sec-interp}
 
The FDF models derived using our method can often assist in interpreting $\boldsymbol{P}(\lambda^2)$ behaviour which might otherwise remain cryptic. In this section we briefly illustrate this point by considering examples of the notable polarization behaviours  described in Section \ref{sec-results-observed}.

In Section \ref{sec-results-observed}, we noted the existence of transitions in $(q,u)$, $p$ or $\psi$ behaviour for the sources lmc\_s11 (Fig. \ref{fig:lmc_s11}) at $\lambda^2\approx 0.005$ m$^2$, lmc\_c02 (Fig. \ref{fig:lmc_c02}) at both $\lambda^2\approx 0.005$ m$^2$ \& $\lambda^2\approx 0.01$ m$^2$ and lmc\_c03 (Fig. \ref{fig:lmc_c03}) at $\lambda^2< 0.003$ m$^2$. This behaviour is invariably caused by a transition in the emission component(s) dominating the net polarization as a function of $\lambda^2$. Typically, one or more strongly depolarizing T components dominates at small $\lambda^2$ values, transitioning to dominance by fainter component(s) with lower Faraday dispersion(s) at larger $\lambda^2$ values. The resulting $\boldsymbol{P}(\lambda^2)$ behaviour can be counter-intuitive: In the case of lmc\_c02, over the $\lambda^2$ range 0--0.01 m$^2$, two depolarizing T components (panel (f) of Fig. \ref{fig:lmc_c02}) interfere to produce a pronounced \emph{increase} in the integrated fractional polarization (panel (d)), before the thicker of the two components depolarizes completely at $\lambda^2\approx0.01$ m$^2$ and net depolarization associated with the remaining component ensues. 

The linearity of $\psi(\lambda^2)$ depends on the relative symmetry of the FDF (Burn 1966), after depolarization effects are accounted for. Sources with asymmetric FDFs exhibit non-linearity in $\psi(\lambda^2)$ (e.g. lmc\_c15; Fig. \ref{fig:lmc_c15}) and vice versa (e.g. cen\_s1349; Fig. \ref{fig:cena_s1349}). lmc\_s11 (Fig. \ref{fig:lmc_s11}) provides an example of more complicated behaviour: Its FDF model is asymmetric at high frequencies due to the presence of two broad T components, but these depolarize by $\lambda^2\approx0.01$ m$^2$. Two S components remain, providing comparative symmetry in the FDF and approximately linear $\psi(\lambda^2)$ behaviour.
 
The observed $p(\lambda^2)$ behaviour depends on the number, shape, and type of components in the model FDF. If the source contains T components, their shape in $\phi$ space strongly affects whether they depolarize monotonically to $p=0$ (e.g. cen\_c1827; Fig. \ref{fig:cena_c1827}) or repolarize (e.g. lmc\_c04; Fig. \ref{fig:lmc_c04}). The presence of an additional S component is associated with depolarization to a finite limiting value (e.g. cen\_s1681; Fig. \ref{fig:cena_s1681}). 
    
 \subsection{The origin of broad components in the FDF models}\label{sec-highdisperse}
 
 In Section \ref{sec-results-fthick}, we described how broad FDF model components were required to model the polarization data for several of our sources. There are two possible sources for these polarization behaviours: the first is associated with Faraday rotation, while the second, which may apply to flat-spectrum core-dominated source, is associated with optical depth effects, whereby different parts of the radio core are observed at different frequencies. In the following, we assume that Faraday effects are responsible for the observed polarization properties, while acknowledging that latter effect may indeed be acting to some extent (most plausibly in the flat spectrum source lmc\_c02). Disentangling the relative contribution of these effects in typical sources will require broadband polarimetric studies of sources at high spatial resolution, and is beyond the scope of this paper.  
 
\subsubsection{The location of the dispersive medium along the LOS}\label{sec-physnaturethick}

 The Faraday thick emission components detected in 1315-46 (Fig. \ref{fig:1315-46}), lmc\_s11 (Fig. \ref{fig:lmc_s11}), lmc\_c02 (Fig. \ref{fig:lmc_c02}), lmc\_c03 (Fig. \ref{fig:lmc_c03}), \& lmc\_c04 (Fig. \ref{fig:lmc_c04}) have $\sigma_\phi$ in the range $\sim$65--1450 rad m$^{-2}$ and $|\phi_{\text{peak}}|$ in the range $\sim$40--420 rad m$^{-2}$. These values are lower limits --- they must multiplied by a factor of $(1+z)^2$ to get the values in the rest frame of a Faraday rotating medium at redshift $z$. In this section we discuss the origin of these components, the information they convey about magnetoionic structure along the LOS, and the nature of the associated radio sources.

Other than the Faraday rotating medium associated with the host galaxy, the most plausible origin of the large Faraday dispersions observed in 1315-46, lmc\_s11, lmc\_c02, lmc\_c03, \& lmc\_c04 is variation in the RM across each source, caused by the the hot thermal plasma associated with the intracluster medium (ICM) of galaxy clusters. Typically, RM variations across radio sources embedded in ICMs are observed to be $\sim$ a few hundred rad m$^{-2}$ in the emitting frame (e.g. Murgia et al. 2004, Govoni et al. 2010, Bonafede et al. 2010), but can exceed thousands of rad m$^{-2}$ for the inner-most regions of cooling core clusters (Perley \& Taylor 1991, Vogt \& En{\ss}lin 2006). However, lmc\_s11, lmc\_c02 \& lmc\_c04 are completely spatially unresolved over 1.3--10 GHz, while lmc\_c03 is dominated by an unresolved component in the 6/3 cm bands (see Fig. \ref{fig:lmc_c03_showsrc}). The corresponding upper limit on physical extent is $\sim$ a few kpc regardless of redshift. Moreover, the visibility ratios from the AT20G high angular resolution catalogue (col. 9, Table \ref{tab:SourceDat}) indicate that lmc\_s11 and lmc\_c04 are compact on angular scales of 0.15" ($\sim1$ kpc maximum). The auto-correlation length for the magnetic field in cooling core clusters is typically estimated to be $\sim$5 kpc, with minimum fluctuation scales of $\sim$0.7--2 kpc (e.g. Bonafede 2010, Vacca et al. 2012). Thus, even if these sources were deeply embedded in such a cluster, their small physical extent means they are unlikely to intercept a sufficient number of depolarizing cells to account for their Faraday dispersion. Independent evidence against the the ICM-based origin is provided by the spectral index of the sources: Sources embedded in dense cluster environments typically have $\alpha\lesssim-1.0$ (e.g. Slee, Siegman \& Wilson 1983, Klamer et al. 2006), whereas lmc\_s11, lmc\_c04 \& lmc\_c02 have $\langle\alpha_{1.3}^{10}\rangle>-0.55$, and lmc\_c03 has $\langle\alpha_{1.3}^{10}\rangle=-0.83$ (Table \ref{tab:SourceDat} , col. 8). The observed values of $\sigma_\phi$ \& $\phi_{\text{peak}}$ are generally lower than seen at parsec-scales in AGN jets (with the exception of lmc\_c03), but are broadly consistent with structure seen at kpc scales (e.g. Algaba et al. 2012). Thus, for the sources lmc\_s11, lmc\_c02, lmc\_c03, \& lmc\_c04, we claim that the Faraday dispersion of the broad T components detected in our modelling is generated either inside, or in the immediate vicinity of, AGN jets on $\sim$kpc scales or below. 

The source 1315-46 is somewhat different to the other four discussed above. Though it is also unresolved over 1.3--10 GHz, it possesses a steep spectral index of $\alpha=-1.0$, and two emission components with large Faraday dispersion ($\sim$several hundred rad m$^{-2}$). The brightest of these has an $N$ value consistent with depolarization by a turbulent foreground screen (see Section \ref{sec-results-fthick}). Thus, we suggest that 1315-46 is most plausibly a young, compact source embedded in a strong depolarising medium, possibly a cluster environment.

Finally, we note that Anderson et al. (2015) implicated the Galactic ISM in causing depolarization amongst sources which were partially resolved at $\sim$arcminute resolution between 1.3 \& 2 GHz. We reject a Galactic origin for the depolarization associated with the emission components discussed in this section. Given their angular size, the sources intercept structures no larger than a tiny fraction of a parsec within the Galaxy --- a distance over which the required RM variations are unlikely to be generated (but see e.g., Bannister et al. 2016).

\subsubsection{Use in characterizing magnetoionized structure in AGN}\label{sec-specpolconstraints}

If the Faraday thick components do originate due to a Faraday rotating medium in the vicinity of AGN, our FDF model fits place direct constraints on the magnetoionized structure of these regions. In this section we explain which structures are ruled out by our data, and speculate on some remaining possibilities. 

For 1315-46, lmc\_s11 \& lmc\_c02, the dominant T components in the FDF model have $N\approx2$ (Section \ref{sec-results-fthick}; panel (f) of Figs. \ref{fig:1315-46}, \ref{fig:lmc_s11} \& \ref{fig:lmc_c02}), depolarizing proportional to $\text{exp}(-k\lambda^4)$. This strongly suggests depolarization by numerous independent  `cells' in a random, turbulent Faraday rotating plasma (Burn 1966), which must be separated from the synchrotron emitting source, but could be located in close proximity to it. In contrast, lmc\_c04 shows oscillatory $p(\lambda^2)$ depolarization (Fig. \ref{fig:lmc_c04}, panel (d) produced by a dominant T component with $N=6.4\pm0.3$ (described in Section \ref{sec-physnaturethick}). This immediately rules out magnetoionic structures that cause monotonic depolarization, such as those described by Burn (1966), Tribble (1992), Rossetti et al. (2008), Bernet et al. (2012) \& Farnes et al. (2014). While the behaviour is reminiscent of that expected for a cubic volume of mixed emitting and rotating plasma (i.e. the Burn slab 1966), the precise location, spacing and amplitude of maxima and minima in $p(\lambda^2)$ for the two T components are not consistent with the Burn slab model. The data are also inconsistent with a spherical volume of mixed emitting and rotating plasma (Burn 1966), linear RM gradients across a uniform background source or a single layer of depolarizing turbulent cells in front of a background source (Schnitzeler et al. 2015). Similar arguments apply to the T components with $N=$ 3.4 \& 4.6 for lmc\_c03 (Fig. \ref{fig:lmc_c03}) described in the Section \ref{sec-results-fthick}. 

\begin{figure}[htpb]
\centering
\includegraphics[width=0.38\textwidth]{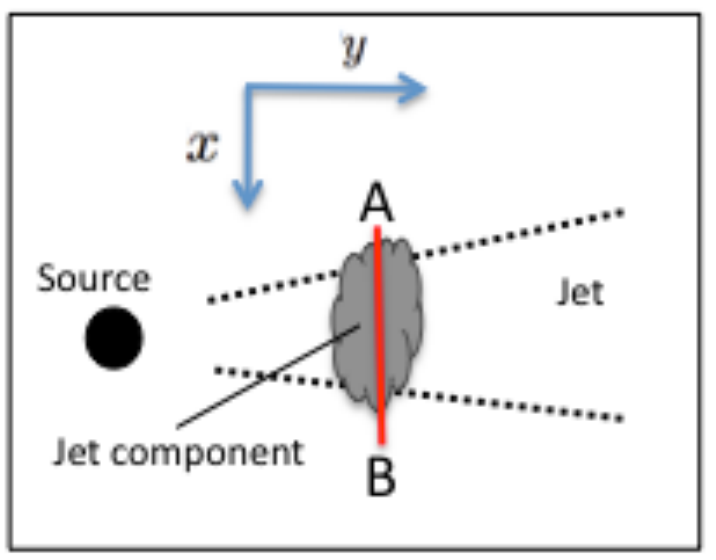}
\caption{A diagram of the hypothetical jet scenario described in Section \ref{sec-specpolconstraints}. The dotted lines indicate the boundaries of a jet emanating from the labelled source. A bright polarized component within the jet is also labelled. The axes orientation is defined above the diagram. The red line from points A to B shows the path along which an RM gradient might be observed, as described in the main text.} 
\label{fig:Jet}
\end{figure}

\begin{figure}[htpb]
\centering
\includegraphics[width=0.465\textwidth]{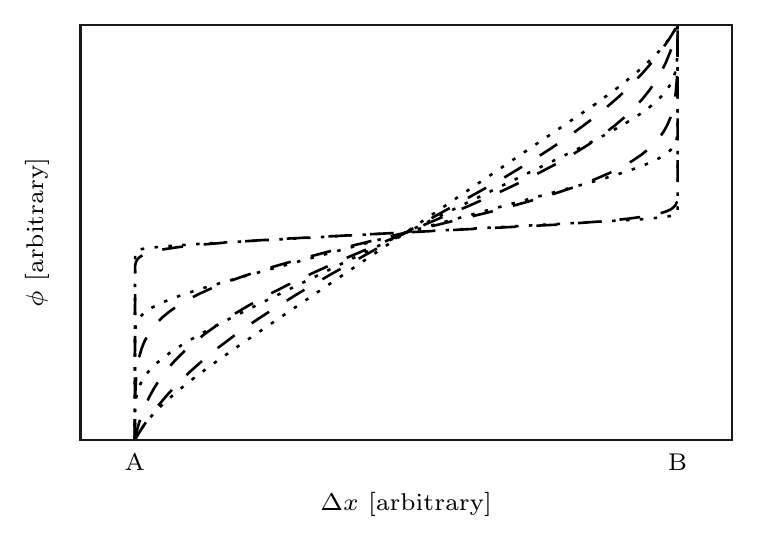}
\caption{Functional forms for $\phi(x)$, where $x$ increases from point A to point B in Figure \ref{fig:Jet}, which yield Super-Gaussian FDF components under the assumptions described in the main text. The curves correspond to Super-Gaussians possessing four arbitrary values of $\sigma_\phi$ and $N=2$ (dashed lines) \& 6 (dotted lines).} 
\label{fig:sgcomptoBrodMcKin}
\end{figure}

In spite of this, numerous plausible magnetoionic configurations remain consistent with the lmc\_c03 \& lmc\_c04 data. It is not our intention to explore these possibilities exhaustively. Rather, we choose to outline one intriguing possibility, which is that the Faraday dispersions are associated with non-linear, monotonic RM gradients across AGN jets. Let us assume for simplicity that an emission component in a jet provides uniform illumination of a magnetized plasma in the immediate foreground. We depict this scenario in Figure \ref{fig:Jet}. We designate $x$ \& $y$ to be the transverse and longitudinal coordinates (relative to the jet axis) over this illuminating component. We also assume that $\partial\phi/\partial x$ is positive definite and that $\partial\phi/\partial y=0$. We then pose the question: How must $\phi(x)$ vary across the jet (i.e. from point A to point B in Fig. \ref{fig:Jet}) to produce a T component with the functional form of a Super-Gaussian in the source FDF? The answer is plotted in Fig. \ref{fig:sgcomptoBrodMcKin} for Super-Gaussian FDF components possessing $N=2$ \& 6 and two different (arbitrary) values of $\sigma_\phi$. These curves are comparable in qualitative form to the RM gradients observed across jets in VLBI observations (e.g. Gabuzda et al. 2015, cf. Fig. 1; Mahmud et al. 2013, cf. Figs. 3 \& 4) and which are predicted by theoretical models (e.g. Broderick \& McKinney 2010, cf. Figs. 4--10). RM gradients, particularly those where the sign of the RM changes from negative to positive or vice versa, are invoked as evidence of helical $\boldsymbol{B}$ fields in jets --- an important part of the theory of jet launching and propagation. We suggest that by using broadband polarimetric analysis, these structures could be identified in unresolved sources by searching for broad T components which emit over both positive and negative Faraday depths (e.g. for lmc\_c03; see panel (f) of Fig. \ref{fig:lmc_c03}). Sources suspected of having RM gradients could then be re-observed with targeted high resolution observations.

Of course, since jets generally propagate from regions of higher to lower average plasma density, RM gradients may also occur along jets (i.e. $\partial\phi/\partial y\neq0$ in the scenario above). For individual unresolved sources, this may be impossible to unambiguously distinguish from gradients across jets. However, we suggest that that on average, the character of the resulting Faraday thick components will differ. In the former scenario for example, the symmetry of an RM gradient about the jet axis is more likely to produce a T component spanning both positive and negative Faraday depths (e.g. lmc\_c03; see panel (f) of Fig. \ref{fig:lmc_c03}), while in the latter scenario, the T component will perhaps be more likely to span only positive or negative Faraday depths (e.g. lmc\_c02; see panel (f) of Fig. \ref{fig:lmc_c02}). In addition, the specific shape of the T component in $\phi$-space may also differ on average between the scenarios, though we make no attempt to calculate the nature of that difference here.

Since magnetic fields in AGN jets are highly structured, we also searched for relationships between the initial polarization angles of model components within a single source. No relationship was found when S components were compared to one another or between S \& T components. Intriguingly however, the difference in $\psi_0$ between the two T components detected in each of lmc\_c03, lmc\_c04, lmc\_s11 \& lmc\_c02 --- the sources where we attributed the polarized emission to jets directly above --- is $0.05\pm0.1$, 0.09(15), 1.59(5) \& 1.73(8) radians respectively. This is consistent with a difference in angle of either $\sim0$ or $\pi/2$ radians (the $\psi_0$ values for these sources are plotted on the inset polar axes in panel (f) of Figs. \ref{fig:lmc_c03}, \ref{fig:lmc_c04}, \ref{fig:lmc_s11} \& \ref{fig:lmc_c02} respectively). Components with parallel and orthogonal de-Faraday-rotated polarization angles are commonly observed in AGN jets (e.g. Attridge, Roberts \& Wardle 1999, Lyutikov et al. 2005, Pushkarev et al. 2005). Thus, we suggest that we have been able to discern the initial polarization angles of multiple components in unresolved AGN jets spectropolarimetrically. 

If magnetoionic structure in AGN, including RM gradients and the polarization angles of components in jets, can be detected and constrained with broadband spectropolarimetric data, this will have major applications in upcoming broadband spectropolarimetric surveys such as POSSUM \& VLASS. However, the correspondence between magnetized structure inferred from low resolution spectropolarimetric data and data with VLBI-scale resolution is yet to be robustly explored. We intend to undertake such an analysis in future work.

\subsubsection{Stokes I SEDs and the nature of the AGN}\label{sec-hostnature}

Of the five sources possessing highly Faraday dispersed emission components, lmc\_c02 (Fig. \ref{fig:lmc_c02}, panel (g)), lmc\_s11 (Fig. \ref{fig:lmc_s11}) \& lmc\_c03 (Fig. \ref{fig:lmc_c03}) each show multiple oscillations in spectral index through the band, while lmc\_c04 (Fig. \ref{fig:lmc_c04}) possesses a convex spectrum which flattens from $\alpha=-0.7$ to $\alpha=-0.3$ at 10 GHz. These results are consistent with the recent findings of Pasetto et al. (2015), who found that compact sources which possess high RMs often show substantial structure in their Stokes I spectrum. The observed Stokes I structures imply that contributions are made by multiple populations of synchrotron emitting particles. This may suggest that the sources have undergone episodic periods of AGN activity, or recent re-triggering. In future, broadband spectropolarimetry may therefore provide a unique and powerful avenue for studying magnetized plasmas intimately associated with AGN, possibly extending to, we speculate, material which will eventually fuel the AGN, or which is closely associated with AGN feedback processes. However, the multiple emitting regions implied by the Stokes I data also mean that our ($q$,$u$) spectra must be interpreted cautiously. This is because the polarized emission properties of each emitting region will likely differ, and dividing Stokes Q \& U by Stokes I to form Stokes q \& u introduces inaccuracies. This remains a fundamental ambiguity with the interpretation of low angular resolution spectropolarimetric analysis.

\subsection{The nature of bright emission components with low Faraday dispersion}\label{sec-physnaturethin}

In Section \ref{sec-results-fthin} we placed upper limits on the Faraday dispersion of emission components from lmc\_s13, cen\_c1466, lmc\_s11 \& lmc\_c15 of order $\sim$1 rad m$^{-2}$. 

Each of these sources is either bright and unresolved at all frequencies (Table \ref{tab:SourceDat}), or is otherwise dominated by such components at 10 GHz (lmc\_c15). Furthermore, lmc\_s13 and cen\_c1466 both have inverted spectra (see Table \ref{tab:SourceDat} and panel (g) of Figs. \ref{fig:lmc_s13} \& \ref{fig:cena_c1466}) while lmc\_s13 \& lmc\_s11 show spectropolarimetric variability (Section \ref{sec-results-temporal}). Thus, each source shows evidence of being dominated by sub-kpc or pc-scale emission regions --- regions which typically have comparatively high Faraday dispersions (e.g. Algaba et al. 2012, Zavala \& Taylor 2004). However, Hovatta et al. (2012) found that the Faraday rotation and depolarization of pc-scale jet components was best explained by interception of fewer than 10 depolarizing cells in a random foreground Faraday rotating plasma structure in the host galaxy. We suggest that our low Faraday dispersion components are either shining through comparatively uniform `gaps' in such structure in these AGN, or are otherwise sufficiently compact that they do not intercept multiple depolarizing cells in these screens. 

\subsection{Prospects for studying magnetized structure in AGN with time domain broadband spectropolarimetry}\label{sec-variabilitysource}

For the time-variable sources discussed in Section \ref{sec-results-temporal}, we can consider whether the narrowband archival data can be fit by making minor changes to the model parameters for our 2012 1.3--10 GHz broadband data (Section \ref{sec-results-bestfit}). We find that this is indeed the case for each source. For example, consider the 2012 best fit FDF model for lmc\_s13 (panel (f) of Fig. \ref{fig:lmc_s13}). If the difference in $\phi_{\text{peak}}$ between the two dominant emission components is either 5 rad m$^{-2}$ or 70 rad m$^{-2}$, rather than our measured value of 16(3) rad m$^{-2}$, the 1996 ($q$,$u$) data is well fit. Alternatively, the 1996 data is well-modelled if one emission component is switched off completely, leaving the parameter values of the other component unchanged. For lmc\_s11 the archival data can be fit by changing $p$ and $\psi_0$ by 1\% and 0.2 rad (respectively) for one of the S components in the best-fit SSTT model (panel (f) of Fig. \ref{fig:lmc_s11}). The archival lmc\_c03 data can, for example, be fit by a 1\% change to $p$ and a 0.3 rad change to $\psi_0$ for one of the Faraday thin components in the full band 2012 model (panel (f) of Fig. \ref{fig:lmc_c03}). 

Thus, only minor changes in the polarization structure of each source are required to explain the changes in the spectropolarimetric data. Since all three of the variable sources are unresolved at 0.15" (col. 9, Table \ref{tab:SourceDat}) with corresponding $\sim$kpc upper size limits on the emitting regions, and have $\langle\alpha_{1.3}^{10}\rangle>-0.5$ (col. 8, Table \ref{tab:SourceDat}), it is tempting to attribute the observed spectropolarimetric variability to changes in pc-scale polarization structure, as suggested by Law et al. (2011) and O'Sullivan et al (2012). While the narrow bandwidth of our current archival dataset prevents us from discerning the precise nature of the changes, broadband polarization data would provide much more stringent constraints. If our claim in Section \ref{sec-physnaturethin} is correct --- namely, that components possessing low Faraday dispersion represent highly compact emitting regions in AGN jets --- tracking the temporal evolution of the polarization angle, Faraday depth \& dispersion of these components spectropolarimetrically  may provide an extremely fine probe of the magnetoionic properties of material in AGN at sub-pc scales.

Based on our data alone, it is difficult to say how common such `spectropolarimetric variables' might be. While we have obtained only 3 detections from 34 sources, we note that all of the detections are associated with bright and unresolved objects, and that for the mostpart, the signal level was only just sufficient to allow for statistically significant detections of changes in their polarization properties. If we consider only the unresolved, flat spectrum (i.e. core-dominated) sources in our sample, which are not much fainter than those for which we achieve detections ($\sim$ 100 mJy), then variability is detected in about half of the viable candidates (see Table \ref{tab:SourceDat}). Sources of this type might therefore be common, and high cadence, broadband polarization monitoring of numerous such sources using survey telescopes such as the Australian Square Kilometre Array Pathfinder (ASKAP, Johnston et al. 2007) might be possible. 

\subsection{The impact of attenuated bandwidth and implications for upcoming spectropolarimetric surveys}\label{sec-impact}

While upcoming spectropolarimetric surveys will have broad $\lambda^2$ coverage by historical standards --- e.g. POSSUM [1.1--1.4 GHz], POSSUM early science [approximately 0.7--1.8 GHz], \& VLASS [2--4 GHz] --- the 1.3--10 GHz coverage explored in this work will not be achieved in survey-style observations for the foreseeable future. In this section, we comment on the impact that limited $\lambda^2$ coverage has on the detection and characterization of Faraday complex polarization structure in our sources.

\begin{figure}[htpb]
\centering
\hspace{-0.5em}
\includegraphics[width=0.52\textwidth]{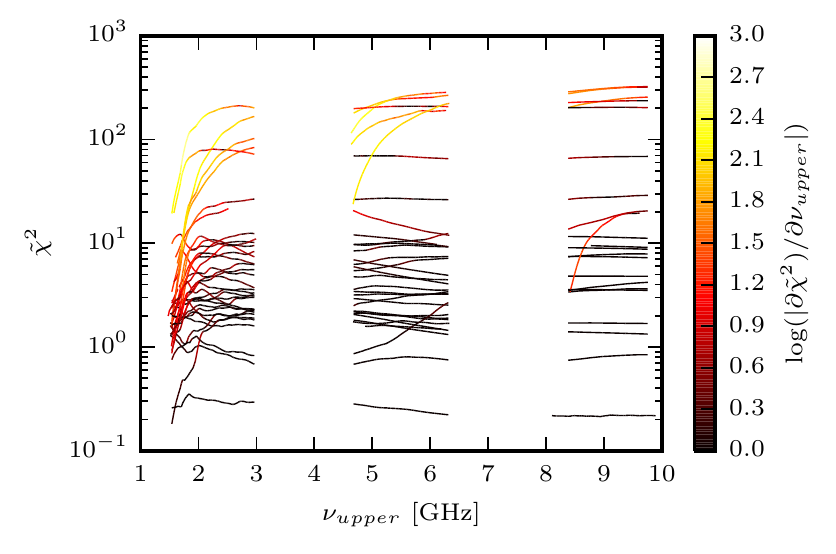}
\caption{Plots of $\tilde{\chi}^2$ vs. $\nu_{\text{upper}}$ for each source in our sample, where $\tilde{\chi}^2$ and $\nu_{\text{upper}}$ are explained in the main text. The curves are colored according to the instantaneous value of the logarithm of $\partial\tilde{\chi}^2/\partial\nu_{\text{upper}}$.}
\label{fig:ReducedChiSquaredGradientVsBandwidth}
\end{figure}

First, we consider the degree of deviation from Faraday simple behaviour shown by our sources as a function of frequency over 1.3--10 GHz. Figure \ref{fig:ReducedChiSquaredGradientVsBandwidth} plots a measure of this deviation. To generate this plot, we first fit the following Faraday simple model to each source over the narrow 1.3--1.5 GHz band: 

\begin{equation}
\boldsymbol{p}(\lambda^2) = p_0\text{e}^{2i(\psi_0+\text{RM}\lambda^2)}
\label{eq:simplemod}
\end{equation}

where $p_0$ is a constant and all other parameters have been previously defined. For each source, we then calculated $\tilde{\chi}^2$ for the best-fit simple model and the polarization data between 1.3 GHz and an upper frequency limit $\nu_{\text{upper}}$ GHz, then plotted $\tilde{\chi}^2$ vs. $\nu_{\text{upper}}$. The resulting curves show the increase in $\tilde{\chi}^2$ from a simple model fit as a function of upper bound on the considered frequency band for each source. The curves are colored according to the logarithm of $\partial\tilde{\chi}^2/\partial\nu_{\text{upper}}$. The plot shows the following:

\begin{itemize}
\item Several sources show a strong increase in $\tilde{\chi}^2$ with $\nu_{\text{upper}}$ in each of the 16, 6 \& 3 cm bands. Thus, data from each band is necessary to fully characterize the Faraday complex polarization behaviour of these sources. Furthermore, it demonstrates that polarization data at $\nu<1$ GHz and $\nu>10$ GHz will be necessary for characterizing the polarization structure of at least some types of sources. Given the way in which our sample was selected, it is difficult to say how common such sources might be. 
\item Strong increases in $\tilde{\chi}^2$ vs. $\nu_{\text{upper}}$ ($\partial\tilde{\chi}^2/\partial\nu_{\text{upper}}\gtrsim10$) are most common in the 16 cm band, and least common in the 3 cm band. This is unsurprising given the way in which our sample was selected. However, a Spearman rank comparison of the $\tilde{\chi}^2$ values obtained for the sample at $\nu_{\text{upper}}=2$ GHz \& 10 GHz yields a coefficient of 0.75, indicating that the $\tilde{\chi}^2$ value calculated for $\nu_{\text{upper}}=2$ GHz is a reasonably good predictor of relative Faraday complexity over the full 1.3--10 GHz band. 
\item The four sources with the highest $\tilde{\chi}^2$ values at $\nu_{\text{upper}}=10$ GHz all reach $\partial\tilde{\chi}^2/\partial\nu_{\text{upper}}>100$ GHz$^{-1}$ at $\nu_{\text{upper}}<2$ GHz. Despite the small sample, this suggests that a strong increase in $\tilde{\chi}^2$ with $\nu_{\text{upper}}$ at low frequencies might be a useful predictor of the presence of Faraday thick components at higher frequencies, which would otherwise be unobservable in frequency-limited observations. 
\end{itemize}

In summary, both 6 \& 3 cm band observations are crucial for \emph{characterizing} the Faraday structure of some sources. Conversely, indications of complex behaviour in the 6/3 cm bands can, at least sometimes, be \emph{identified} at frequencies as low as 2 GHz.

\begin{figure}[htpb]
\centering
\includegraphics[width=0.51\textwidth]{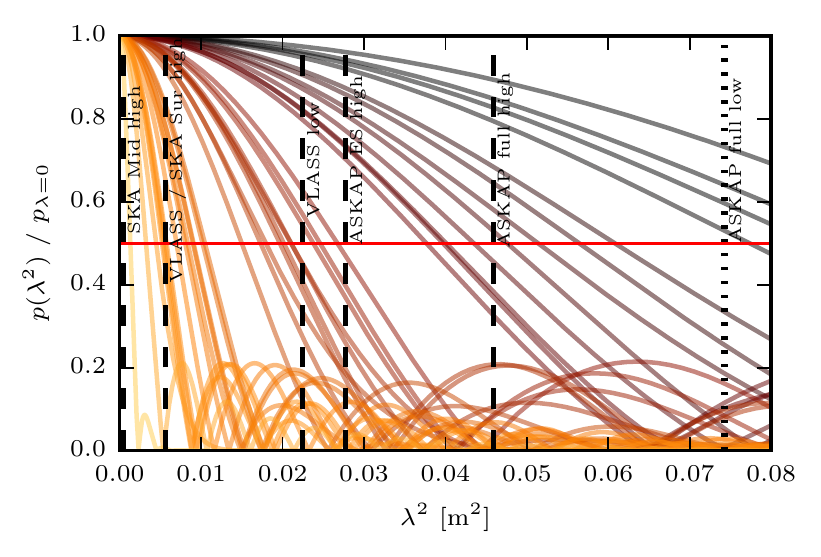}
\caption{Plots of the fractional depolarization of every T component in the overall best fit FDF model of every source in our sample. Shading of the curves changes linearly from yellow to red with increasing $\lambda_{1/2}$ (described in the main text), and serves only to aid in distinguishing the behaviour of components with different $\sigma_\phi$ \& $N$ values. Several reference frequencies are indicated on the plot with vertical dotted and dashed lines. These correspond to the lower (indicated with a `low' label) and upper (indicated with a `hi' label) frequencies of several major upcoming spectropolarimetric surveys for reference.}
\label{fig:TCompDepol}
\end{figure}

Deviations from the Faraday simple model in the 6/3 cm bands can generally be attributed to the presence of Faraday thick emission components. To gauge the extent to which these components would manifest in data possessing the $\lambda^2$ coverage of upcoming surveys, we plot the fractional depolarization of each of the T components detected in our FDF models in Figure \ref{fig:TCompDepol}. We include upper and lower $\lambda^2$ limits for the surveys mentioned above (see caption). It shows:

\begin{itemize}

\item Oscillatory depolarization continues to $\lambda^2$ values that are substantially higher than $\lambda_{1/2}$ --- i.e. the wavelength at which the fractional depolarization first drops below 0.5. This is caused by T components with $N\neq2$. For low S/N data at low frequencies, these oscillations could be misconstrued as bandpass, calibration or deconvolution errors. In high S/N data, such structure can used to infer the presence of high $\sigma_\phi$ T components which would otherwise be completely depolarized for $N=2$.

\item Over 1.1--1.4 GHz (i.e. the POSSUM survey band), oscillatory ringing from T components is generally not of sufficiently high frequency to result in multiple maxima and minima in the band. For components with initial minima at frequencies just above 1.4 GHz ($\lambda^2\approx0.045$ m$^2$), the resulting $p(\lambda^2)$ essentially mimics the behaviour expected from interfering Faraday thin components --- completely analogous to the equivalent situation in aperture synthesis imaging where, for example, an extended circular source will appear as a ring when short baseline information is not present. Care must be taken in interpreting such data. 

\item In combination, POSSUM (both the full survey and early science) and VLASS will cover almost the full range of depolarization behaviour caused by Faraday thick components in our sample (lmc\_c03 is an exception). Thus we anticipate that the types of sources and behaviours we have detected will be prevalent in these surveys, and that the type of analysis we have conducted will represent powerful methods for studying magnetized structure in AGN.
\end{itemize}

\begin{figure}[htpb]
\centering
\includegraphics[width=0.52\textwidth]{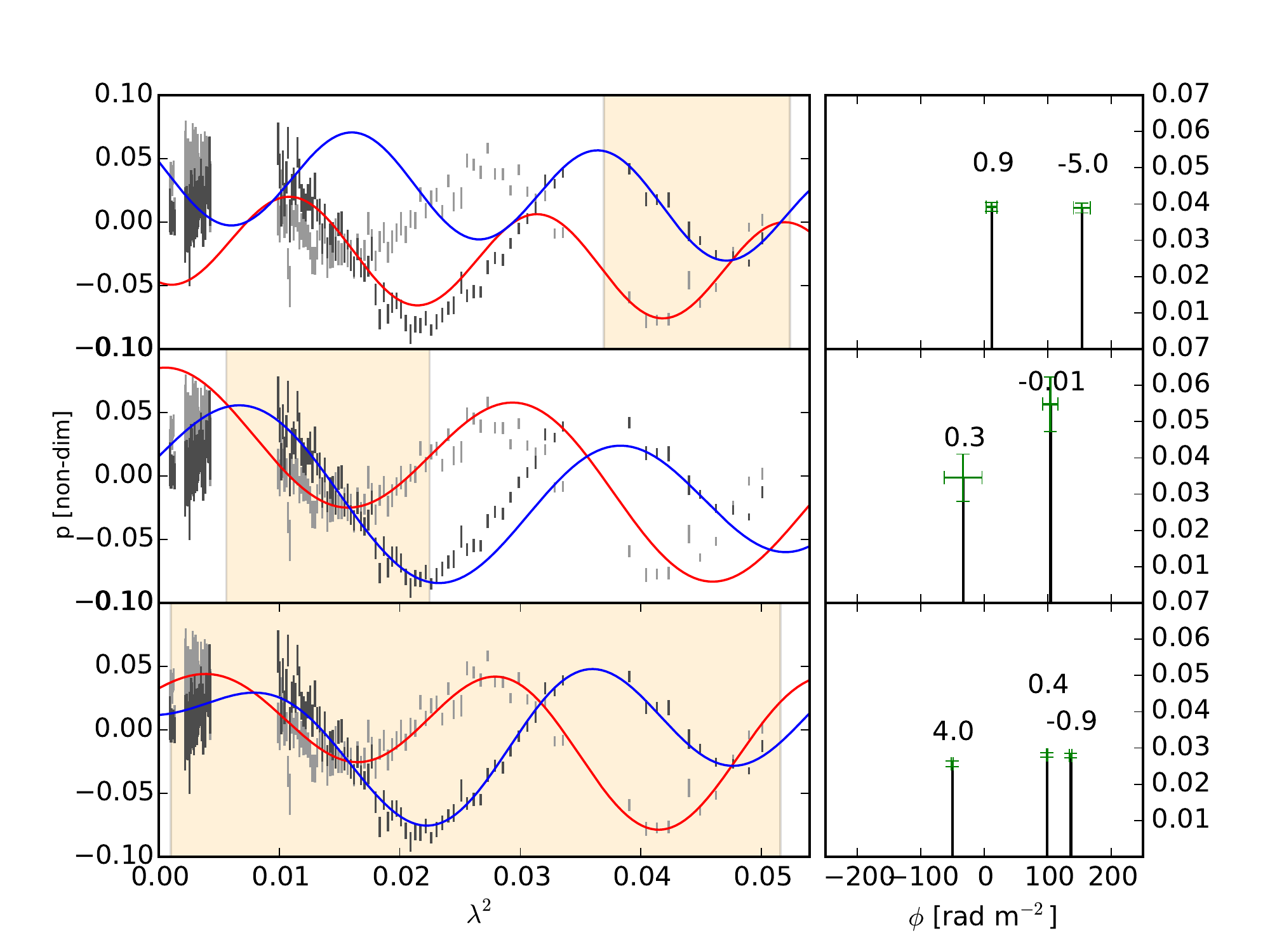}
\caption{The polarization models required to fit lmc\_c15 over the bands $1.3<\nu<1.5$ GHz, $2<\nu<4$ GHz, and $1.3<\nu<10$ GHz. The left most column of plots show the bands over which the fits were performed (beige shading), the ($q$,$u$) data (grey error bars) and the best fit ($q$,$u$) model (red \& blue curves respectively). The right most column of plots shows the best fit model FDFs including green error bars on $\phi_{\text{peak}}$ and $p$. The numbers above the FDF model components are the best-fit initial polarization angles of each.}
\label{fig:lmc_c15_bandeffect}
\end{figure}

For sources possessing multiple S components, the primary issue for interpretation of the polarization data is degeneracy of the fitted polarization models over narrow bands (O'Sullivan et al. 2012, Sun et al. 2015). The frequency-dependent instantaneous RM values (i.e. $\partial\psi/\partial\lambda^2$) of many of our sources can deviate from the RM value determined from a Faraday simple model fit to low frequency, narrowband data by 10s--100s of rad m$^{-2}$. Such large deviations typically occur over small $\lambda^2$ ranges, but inaccuracies and biases can persist when fitting to larger bands, especially at higher frequencies. As an example, in Figure \ref{fig:lmc_c15_bandeffect} we plot the 1.3--10 GHz data for lmc\_c15, including the best fit models resulting from fits to the data over the restricted bands $\nu<1.5$ GHz and $2<\nu<4$ GHz. Good fits to the data are achieved in each individual band. However, the number of interfering components in the source, their RMs and the fractional polarizations are mischaracterized for both sub-bands. Note that this is true even when observed using 2 GHz of continuous bandwidth in the case of the 2--4 GHz band. For this particular source, the addition of 4--6 GHz data is required to obtain the `correct' RMs, though there is no guarantee that additional low frequency data would not modify the RMs or even the number of polarization components required even further. 

\section{Summary \& conclusions}\label{sec-conclusion}

We have obtained polarization data for 36 discrete, mostly unresolved radio sources spanning a total frequency range of 1.3 to 10 GHz with 6 GHz of the spectral range densely sampled. The polarization data show remarkable structure and variety, with complicated frequency-dependent changes in Stokes $q$ \& $u$ observed in a number of sources. After finding that we were unable to reproduce these behaviours with existing depolarization models, we developed an alternative spectropolarimetric modelling method. This involves constructing model Faraday Dispersion Functions (FDFs) in Faraday depth space from simple parametric basis functions, which can then be Fourier-transformed and assessed for goodness of fit to Stokes $q(\lambda^2)$ \& $u(\lambda^2$) polarization data. Our main findings are as follows:

\begin{enumerate}[i)]
\item Using our modelling technique, we were able to achieve good fits to complex frequency-dependent polarization behaviour over three densely sampled 2 GHz bands which, in total, spanned nearly 9 GHz (a fractional bandwidth of $\sim200\%$). Subtle discrepancies between the data and best fit models remained for the very brightest sources in our sample. However even in these cases, we were able to broadly reproduce the observed polarization behaviour. The resulting FDF models can greatly aid in interpreting ($q$,$u$), $p$ \& $\psi$ vs. $\lambda^2$ behaviours.
\item Only one source in our sample is best fit by a Faraday-rotation-only polarization model over the 1.3--10 GHz band (though a zero-polarization model does almost as well in this case). All other sources show some level of deviation from idealized, Faraday-rotation-only behaviour (i.e. show Faraday complexity). 
\item The presence of multiple individual emission components in spatially unresolved radio sources can be discerned, and their individual Faraday depth distributions and initial polarization angles can be measured, using broadband polarization data. These measurements can place strong constraints on the structure of magnetized material in the radio sources.
\item For five sources in our sample, we detect Faraday thick emission components at high signal-to-noise, which emit over a Faraday depth range of between $\sim$130 \& 2890 rad m$^{-2}$. In contrast to our previous work, where we attributed the Faraday complex behaviour observed in extended sources at low frequencies to the Galactic ISM (Anderson et al. 2015), constraints supplied by the linear size and spectral index of four of these sources, coupled with their large measured Faraday dispersions, strongly implicate a Faraday dispersive medium which is intrinsic to the source. We believe the most satisfactory explanation of these observations is that the Faraday thick emission components are generated by magnetoionic material in the vicinity of kpc-scale AGN jets. 
\item Three sources show variability in their broadband polarization data on $\sim$15 year timescales. This represents a detection rate of about 50\% amongst viable candidate sources (i.e. sufficiently bright, core-dominated sources). As such, broadband polarization variability might be commonly observed in future. Only small changes in the physical polarization structure of each source were required to account for these changes, though the limited bandwidth of the archival observations prevented us from uniquely determining their nature. Multi-epoch comparison of $\sim$ GHz bandwidth data would allow the nature of the changes to be determined far more precisely.
\item Observations at wavelengths as short as 3 cm are crucial for \emph{characterizing} the full polarization behaviour of some sources. Conversely, complex behaviour in the 6/3 cm bands can sometimes be \emph{inferred} from polarization behaviour at frequencies as low as 2 GHz. In particular, oscillatory depolarization from Faraday thick components `rings' down to substantially lower frequencies than might be expected under the assumption of monotonic depolarization, even for components with $\sigma_\phi\gg100$ rad m$^{-2}$. This has important implications for upcoming polarization surveys such as POSSUM and VLASS.
\end{enumerate}

In future work, we intend to obtain broadband radio polarization data and observations at other wavelengths for a sample of sources, in order to facilitate a deeper investigation of the physics of magnetized structures in AGN. We intend to compare magnetized structure observed at VLBI-scale resolution with that inferred through broadband spectropolarimetry, to ascertain the reliability of these inferences. Finally, we plan to undertake multi-epoch broadband polarimetric observations of core-dominated AGN, in order to explore the utility of broadband polarization variability for studying magnetized structure in the vicinity of AGN jets.

\section{Acknowledgments}\label{sec-acknowledgements}

The authors would like to thank the anonymous referee for their thoughtful and constructive critique of our work. This paper has been much improved as a result. C.~S.~A. would like to thank Joe Callingham for insightful conversations regarding aspects of this work. B.~M.~G. and C.~S.~A.  acknowledge the support of the Australian Research Council (ARC) through grant FL100100114. The Dunlap Institute is funded through an endowment established by the David Dunlap family and the University of Toronto. The Australia Telescope Compact Array is part of the Australia Telescope National Facility which is funded by the Commonwealth of Australia for operation as a National Facility managed by CSIRO.


\section{REFERENCES}
Akaike, H., IEEE Transactions on Automatic Control 19 (6) (1974) 716Ð723\\
Algaba, J. C., Asada, K., Nakamura, M., 2012, Arxiv:1308.5429\\
Anderson, C. S., Gaensler, B. M., Feain, I. J., Franzen, T. M. O., 2015, ApJ, 815, 49\\
Attridge, J. M., Roberts, D. H., Wardle, J. F. C., 1999, ApJ, 518, 87\\
Bannister, K, W., Stevens, J., Tuntsov, A. V., 2016, Science, 351, 354\\
Bernet, M. L., Miniati, F., \& Lilly, S. J. 2012, ApJ, 761, 144\\
Bonafede, A. et al. 2010, A\&A, 513, 30\\ 
Brentjens, M. A., de Bruyn A. G., 2005, A\&A, 441, 1217\\
Briggs, D. S. 1995, BAAS, 27,1444\\
Broderick, A. E., McKinney, J. C., 2010, ApJ, 725, 750\\
Burn, B. J., 1966, MNRAS, 133, 67\\
Chhetri, R., Ekers, R. D., Jones, P. A., Ricci, R., 2013, MNRAS, 434, 956\\
Decker, F. J., 1994, \\
Farnes, J. S. , O'Sullivan, S. P., Corrigan, M. E., Gaensler, B. M, 2014, ApJ, 795, 63\\ 
Feain, I. J., Ekers, R. D., Murphy, T., et al. 2009, ApJ, 707, 114\\
Feain, I. J., Cornwell, T. J., Ekers, R. D., et al., 2011, ApJ, 740, 17\\
Fomalont, E. B., Ekers, R. D., 1989, ApJ, 346, L17\\
Foreman-Mackey D., Hogg D. W., Lang D., Goodman J., 2013, PASP,125, 306\\
Gabuzda, D. C., Kneuttel, S., Reardon, B., 2015, MNRAS, 450, 244\\
Gaensler, B. M. et al., 2005, Science, 307, 1610\\
Gaensler, B. M. 2009, in IAU Symp. Cosmic Magnetic Fields: From Planets, to Stars and Galaxies, ed. K. G. Strassmeier, 259, 645\\ 
G\'{o}mez, J. L., Roca-Sogorb M., Agudo I., Marscher A. P., Jorstad S. G., 2011, ApJ, 733, 11\\
Govoni, F. et al., 2010, A\&A, 522, 105\\
Heald, G., Braun R., Edmonds R., 2009, A\&A, 503, 409\\
Hovatta, T., Lister, M. L., Aller, M. F., et al. 2012, AJ, 144, 105\\
Johnston S. et al., 2007, Publ. Astron. Soc. Aust., 24, 174\\
Kim, S. et al., 1998, ApJ 503, 674\\
Klamer, I. J., Ekers, R. D., Bryant, J. J., et al. 2006, MNRAS, 371, 852\\
Kronberg, P. P., Perry, J. J., Zukowski, E. L. H., 1992, ApJ, 387, 528\\
Law, C. J. et al., 2011, ApJ, 728, 57\\
Lyutikov, M., Pariev, V. I., Gabuzda, D. C., 2005, MNRAS, 360, 869L\\
Mahmud, M., Coughlan, C. P., Murphy, E., Gabuzda, D. C., Hallahan, D. R., 2013, MNRAS, 431, 695\\
Murgia, M. et al., 2004, A\&A, 424, 429\\
O'Sullivan, S. P. et al., 2012, MNRAS, 421, 3300\\
Offringa, A. R. et al., 2010, MNRAS, 405, 155\\
Pasetto, A., Kraus, A., Mack, K-H., Bruni, G., Carrasco-Gonzalez, C., 2015, A\&A in press\\
Perley, R. A., Taylor, G. B., 1991, AJ, 101, 1623\\
Pushkarev, A. B., Gabuzda, D. C., Vetukhnovskaya, Y. N., Yakimov, V. E., 2005, MNRAS, 356, 859\\
Rossetti, A., Dallacasa, D., Fanti, C., Fanti, R., Mack, K-H., 2008, A\&A 487, 865\\
Schnitzeler, D. H. F. M., et al., 2011, Australia Telescope Technical memo \# AT/39.9/129\\
Schnitzeler, D. H. F. M., Banfield, J . K., Lee, K. J., 2015, MNRAS, 450, 3579\\
Simmons, J. F. L., \& Stewart, B. G. 1985, A\&A, 142, 100\\
Slee, O. B., Siegman, C. B., \& Wilson, I. R. G. 1983, Australian Journal of Physics, 36, 101\\
Slysh V. I., 1965, AZh, 42, 689\\
Sun, X. H. et al. 2015, AJ, 149, 60\\
Tribble, P. C., 1991, MNRAS, 250, 726\\
VSSG 2015, VLA Survey Science Group 2015, VLASS ``final" proposal, https://safe.nrao.edu/wiki/pub/JVLA/VLASS/VLASS
final.pdf\\
Vacca, V., et al., 2012, A\&A, 540, A38\\
Vogt, C., En{\ss}lin, T. A., 2006, AN, 327, 595\\
Wilson, W. E. et al., 2011, MNRAS, 416, 832\\
Zavala, R. T., \& Taylor, G. B. 2001, ApJ, 550, L147\\
Zavala R. T., Taylor G. B., 2004, ApJ, 612, 749\\

\appendix 

\section{Images of resolved sources}\label{sec-appendA}

Here we provide images of our resolved sample sources at both the low and high frequency ends of each CABB band. The sources are presented in the order that they are listed in Table \ref{tab:SourceDat} --- i.e. in order of increasing RA. We only include images that span the frequency range used in the spectropolarimetric analysis of each source (see Section \ref{sec-obscal}). Each image represents a {\sc clean}ed and restored multi-frequency synthesis image of a 100 MHz band centred on (for image positions running from left to right) 1.4, 2.8, 4.8, 6.1, 8.3 or 9.6 GHz. For each source, the top row contains total intensity images, for which the saturation points for the grayscale are set to -3 \& 15 mJy. The bottom row contains images of linearly polarized intensity, for which the saturation points for the grayscale are set to -1 \& 5 mJy. The source name is provided in the top left-hand corner of the lowest frequency Stokes I image of each source, while the maximum flux density, noise level, beam, and angular scale are provided in all images. Note that the coordinate labels on the 1.4 GHz images apply to these images only --- all images at frequencies greater than 1.4 GHz have different spatial scales, which can be judged using the scale bar at the bottom right of each image. For the Stokes I images only, we have drawn green contours at 50\% of the peak measured flux density, in order to facilitate comparison of source morphology with the primary beam. Brief notes on individual sources are provided in the figure captions.

\begin{figure*}[htpb]
\includegraphics[width=0.95\textwidth]{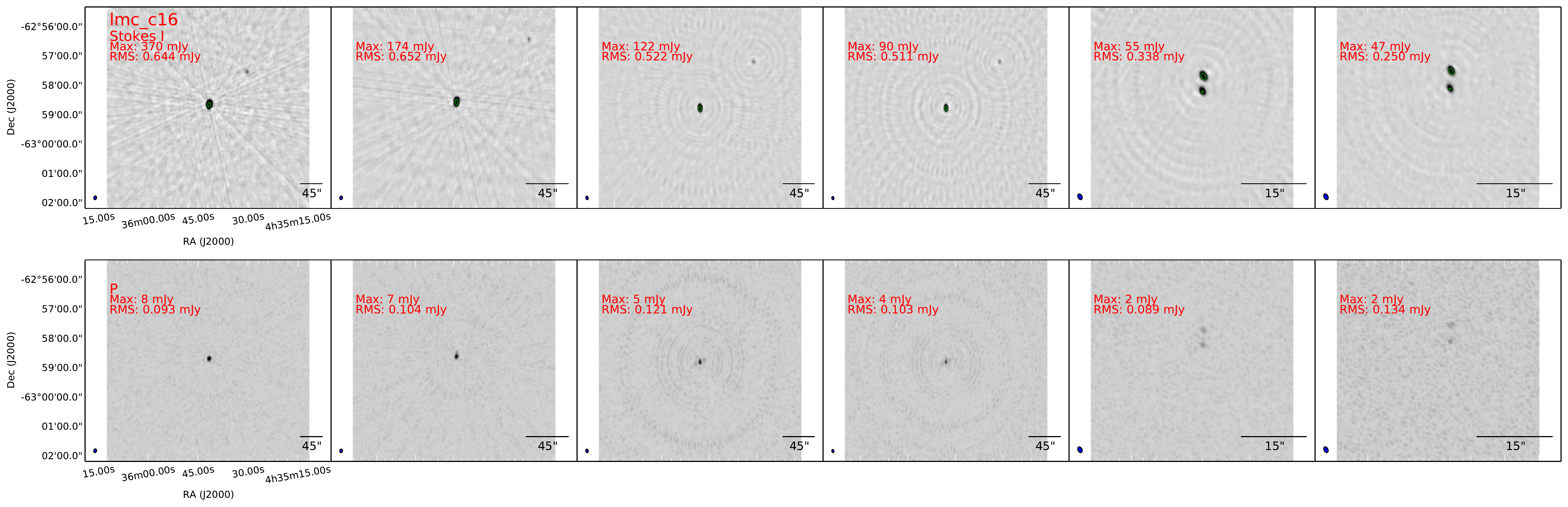}
\caption{Source: lmc\_c16. Central image frequencies: 1.4, 2.8, 4.8, 6.1, 8.3 and 9.6 GHz (left to right). The source is somewhat resolved at all frequencies from 1.3--10 GHz. The best-fit polarization model is of type TT.}
\label{fig:lmc_c16_showsrc}
\end{figure*}

\begin{figure*}[htpb]
\includegraphics[width=0.95\textwidth]{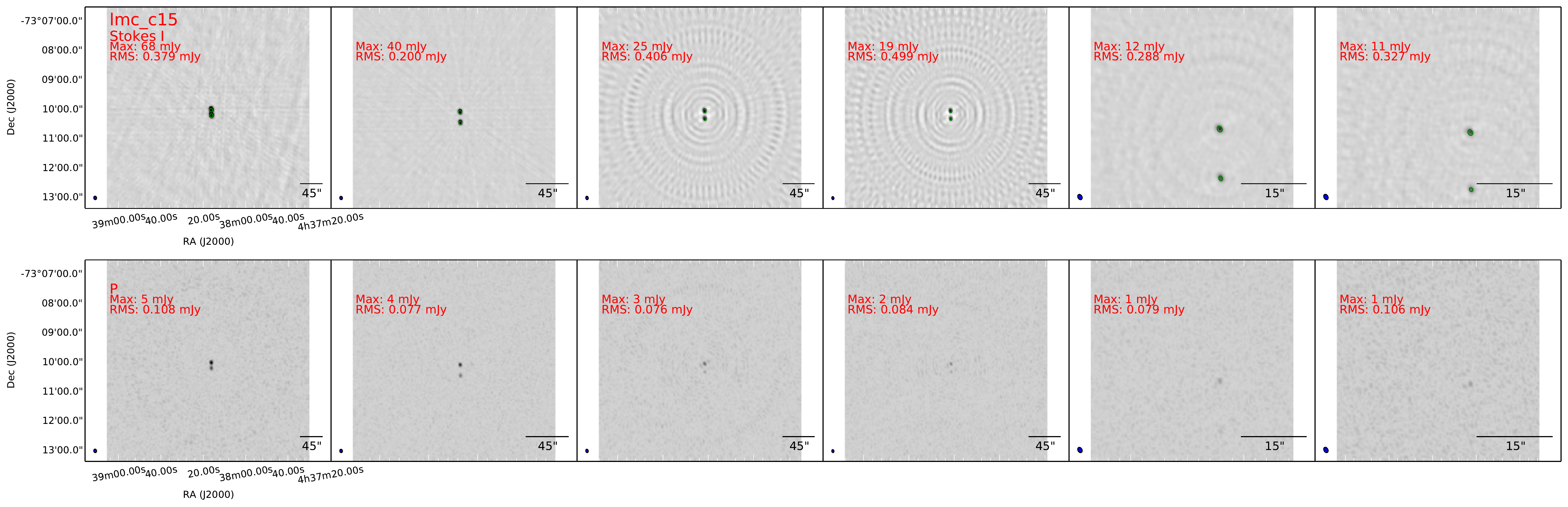}
\caption{Source: lmc\_c15. Central image frequencies: 1.4, 2.8, 4.8, 6.1, 8.3 and 9.6 GHz (left to right). The spectropolarimetric data represent the integrated Stokes I, Q, \& U flux densities over both components visible in the images above (which are themselves marginally resolved)), as described in Section \ref{sec-extract}. The best-fit polarization model is of type SSS.}
\label{fig:lmc_c15_showsrc}
\end{figure*}

\begin{figure*}[htpb]
\includegraphics[width=0.65\textwidth]{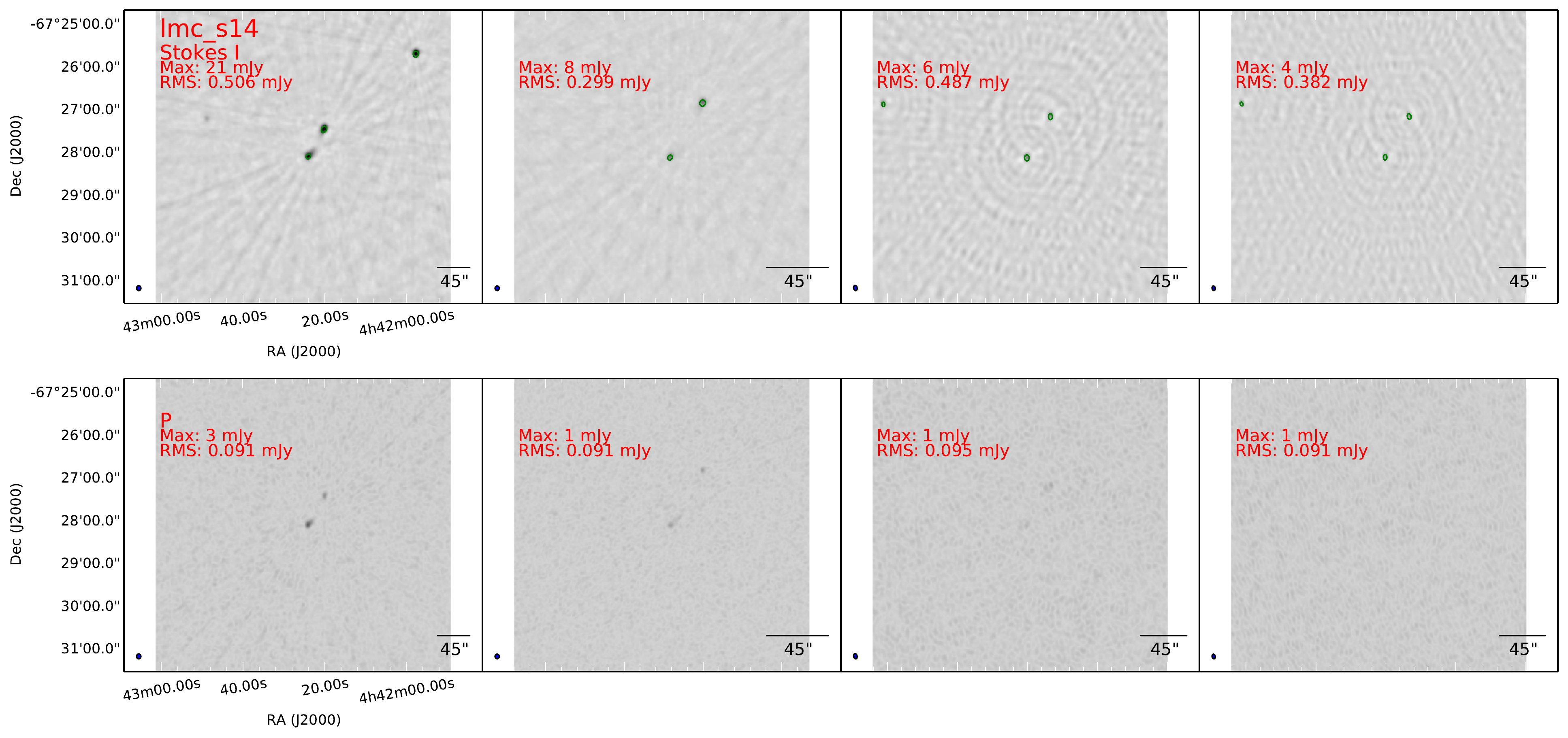}
\caption{Source: lmc\_s14. Central image frequencies: 1.4, 2.8, 4.8 and 6.1 GHz (left to right). The spectropolarimetric data represent the integrated Stokes I, Q, \& U flux densities over the entire extended source, as described in Section \ref{sec-extract}. The best-fit polarization model is of type TT.}
\label{fig:lmc_s14_showsrc}
\end{figure*}

\begin{figure*}[htpb]
\includegraphics[width=0.65\textwidth]{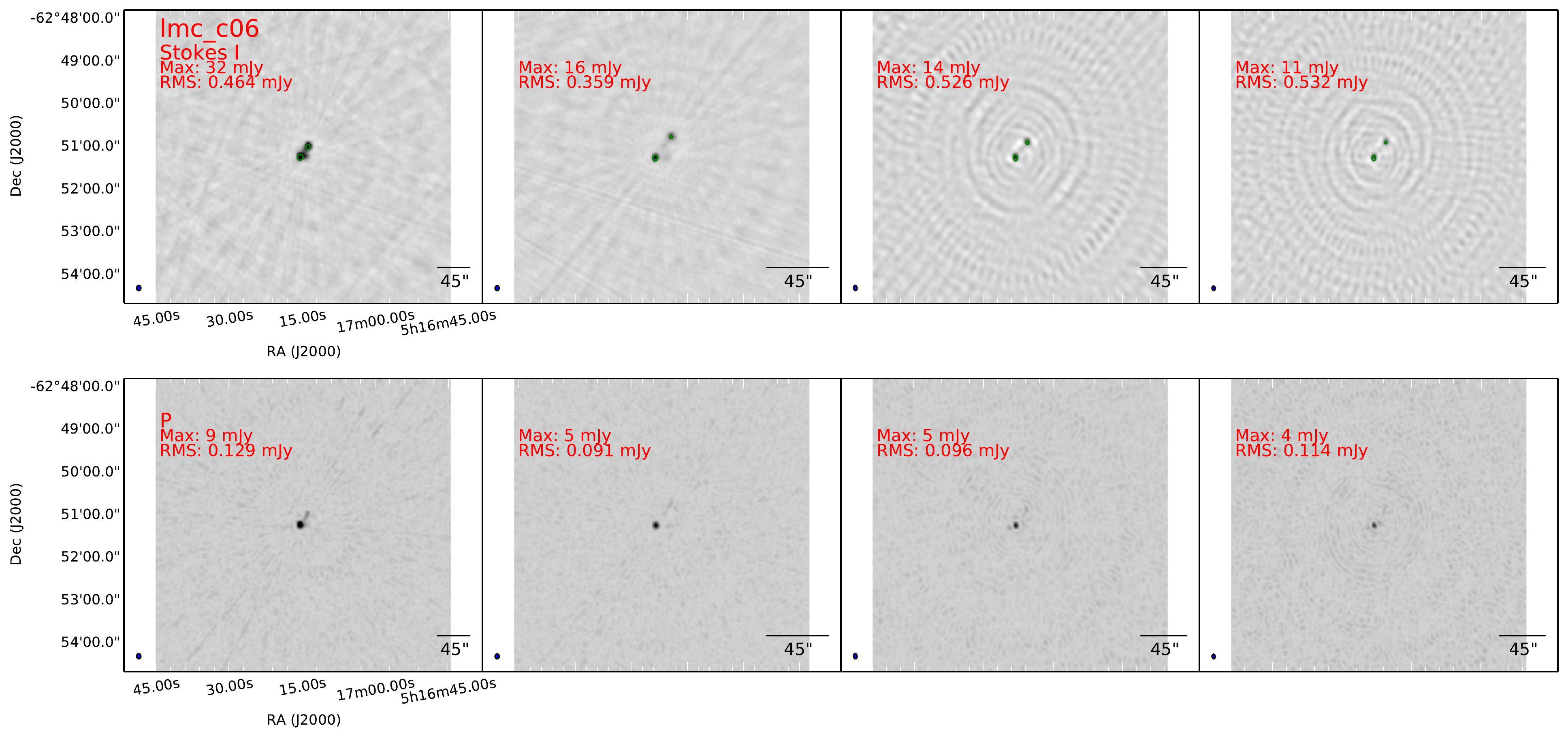}
\caption{Source: lmc\_c06. Central image frequencies: 1.4, 2.8, 4.8 and 6.1 GHz (left to right). The spectropolarimetric data represent the integrated Stokes I, Q, \& U flux densities over the entire extended source, as described in Section \ref{sec-extract}. The best-fit polarization model is of type TT.}
\label{fig:lmc_c06_showsrc}
\end{figure*}

\begin{figure*}[htpb]
\includegraphics[width=0.95\textwidth]{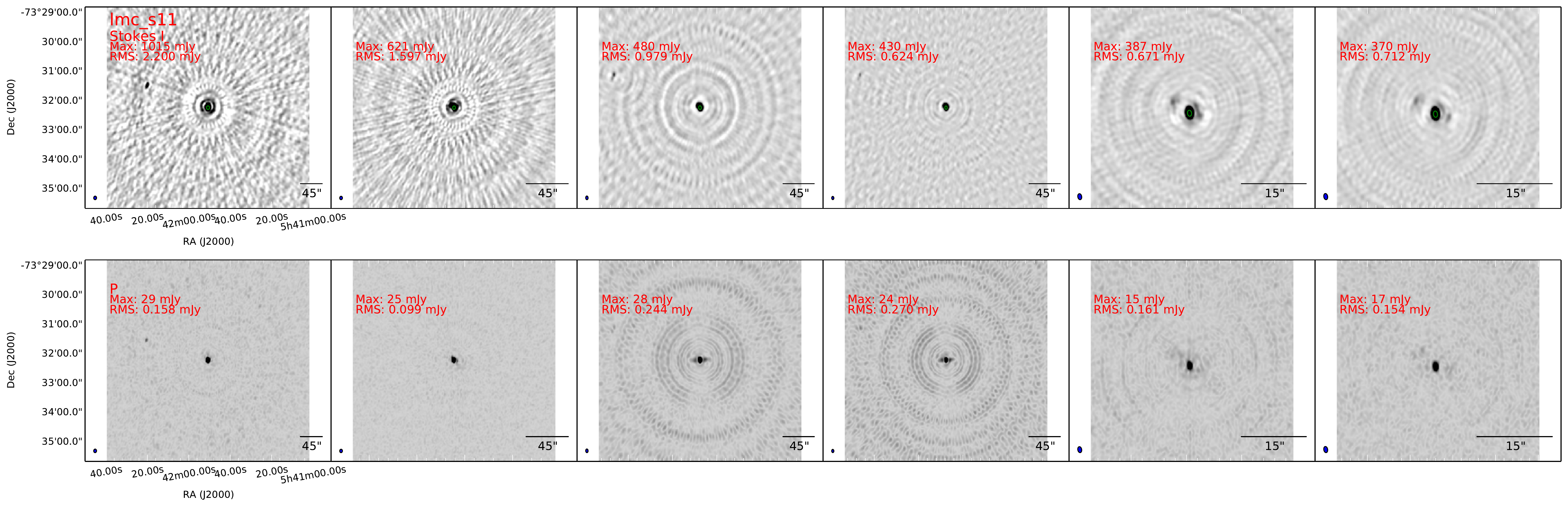}
\caption{Source: lmc\_s11. Central image frequencies: 1.4, 2.8, 4.8, 6.1, 8.3 and 9.6 GHz (left to right). The source is slightly resolved at all frequencies from 1.3--10 GHz, but resolves into three distinct components in the 3cm band. The best-fit polarization model is of type SSTT.}
\label{fig:lmc_s11_showsrc}
\end{figure*}

\begin{figure*}[htpb]
\includegraphics[width=0.95\textwidth]{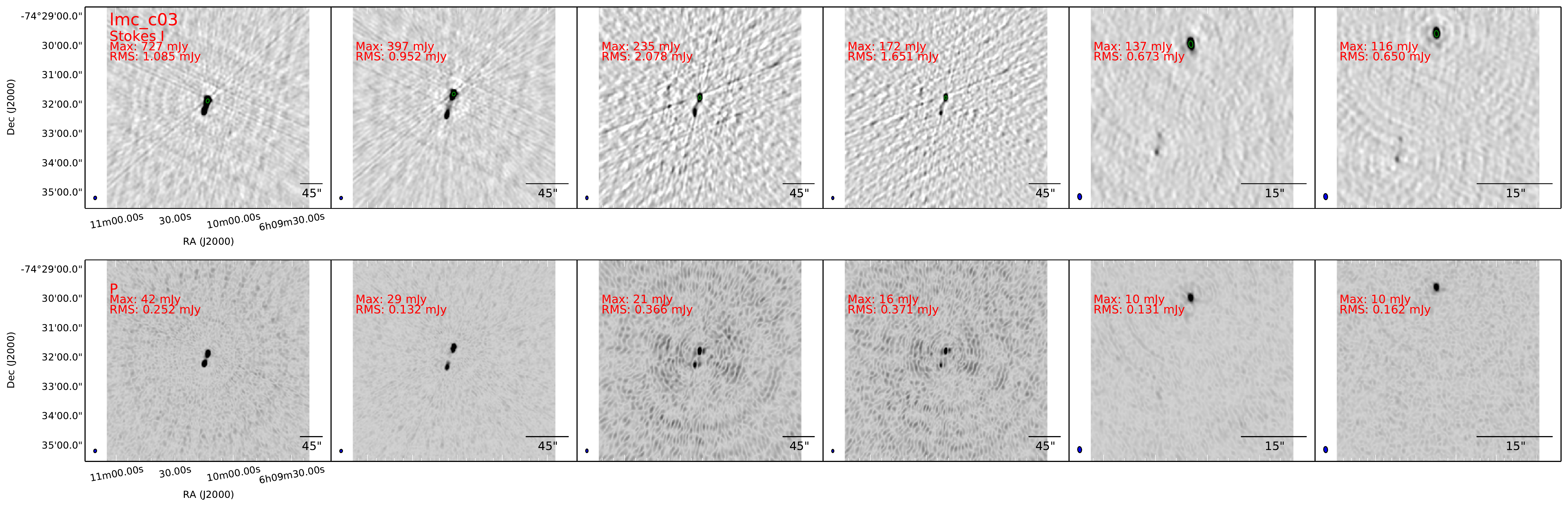}
\caption{Source: lmc\_c03. Central image frequencies: 1.4, 2.8, 4.8, 6.1, 8.3 and 9.6 GHz (left to right). The spectropolarimetric data represent the integrated Stokes I, Q, \& U flux densities over the entire extended source, as described in Section \ref{sec-extract}. The best-fit polarization model is of type SSTT.}
\label{fig:lmc_c03_showsrc}
\end{figure*}

\begin{figure*}[htpb]
\includegraphics[width=0.95\textwidth]{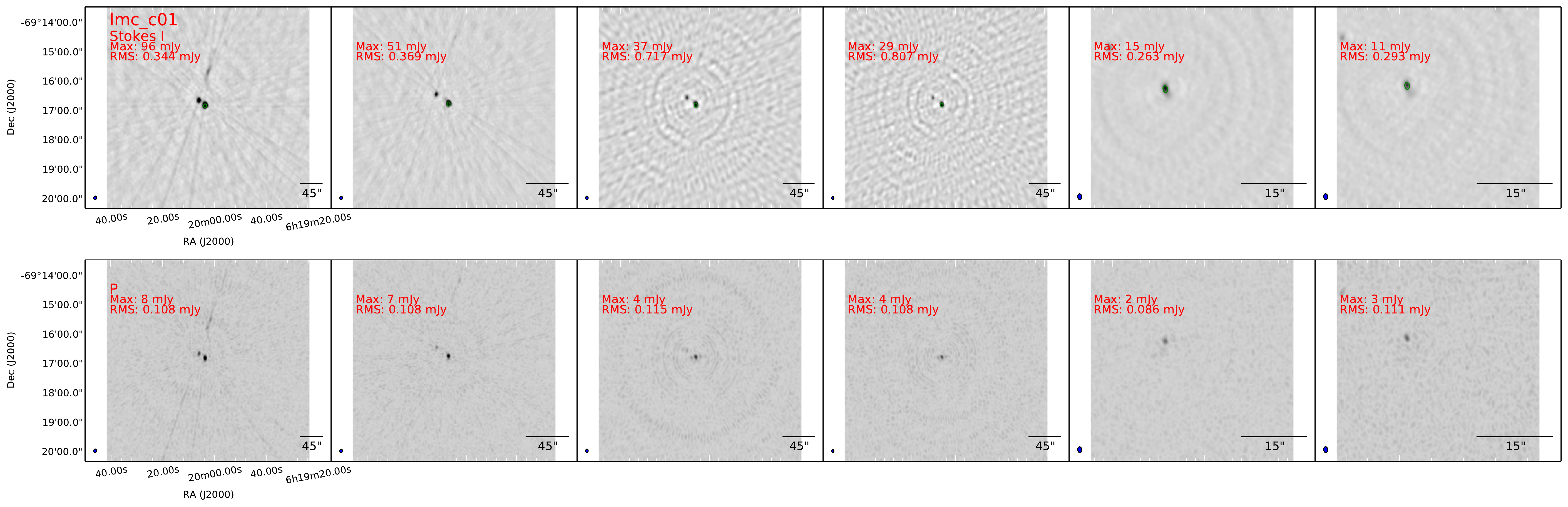}
\caption{Source: lmc\_c01. Central image frequencies: 1.4, 2.8, 4.8, 6.1, 8.3 and 9.6 GHz (left to right). The spectropolarimetric data for this source was extracted from the right-most component of the double source visible in the images above. This component is marginally resolved at all frequencies (see Table \ref{tab:SourceDat}). The best-fit polarization model is of type TT.}
\label{fig:lmc_c01_showsrc}
\end{figure*}

\begin{figure*}[htpb]
\includegraphics[width=0.65\textwidth]{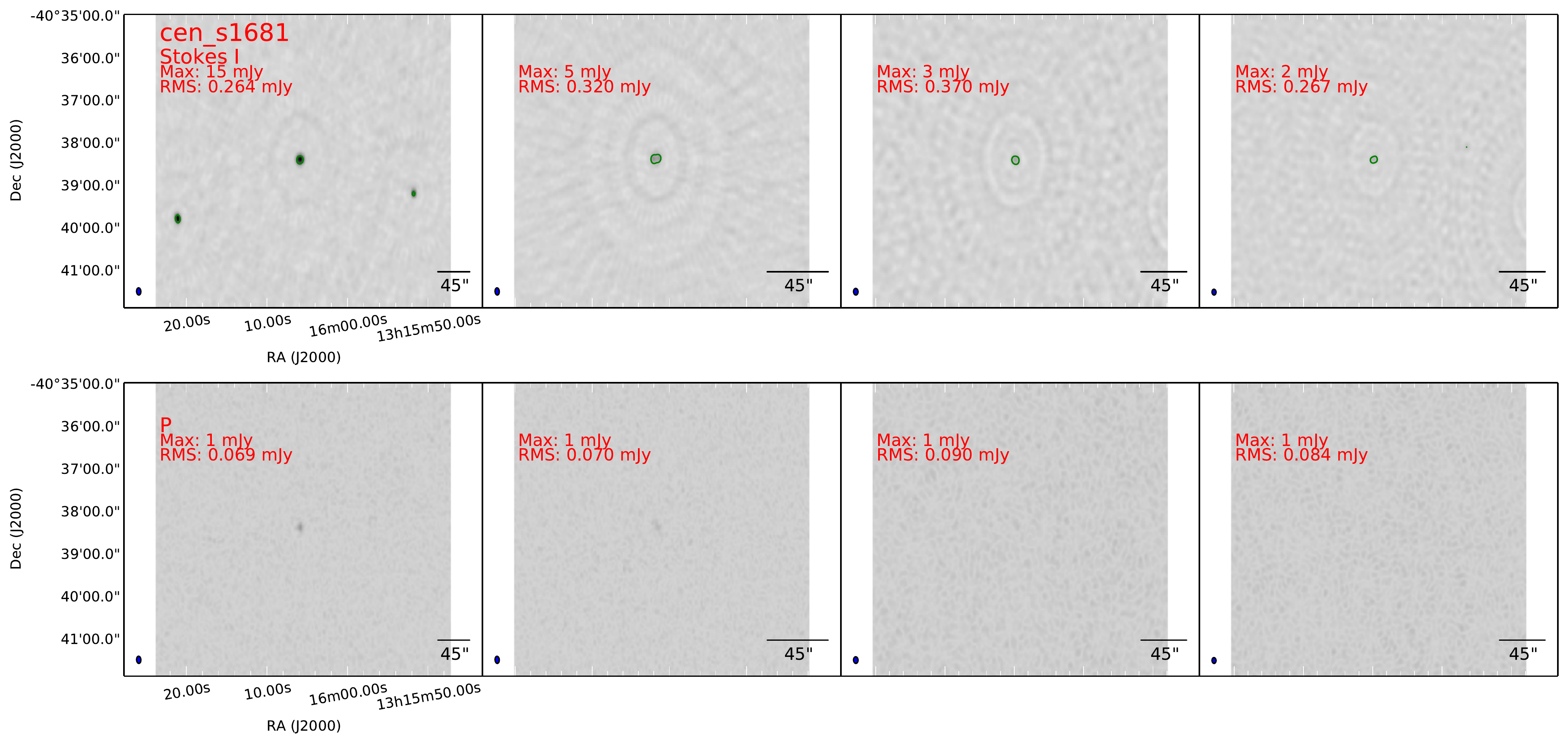}
\caption{Source: cen\_s1681. Central image frequencies: 1.4, 2.8, 4.8 and 6.1 GHz (left to right). The source is somewhat resolved at all frequencies from 1.3--6.4 GHz. The best-fit polarization model is of type TT.}
\label{fig:cena_s1681_showsrc}
\end{figure*}

\begin{figure*}[htpb]
\includegraphics[width=0.95\textwidth]{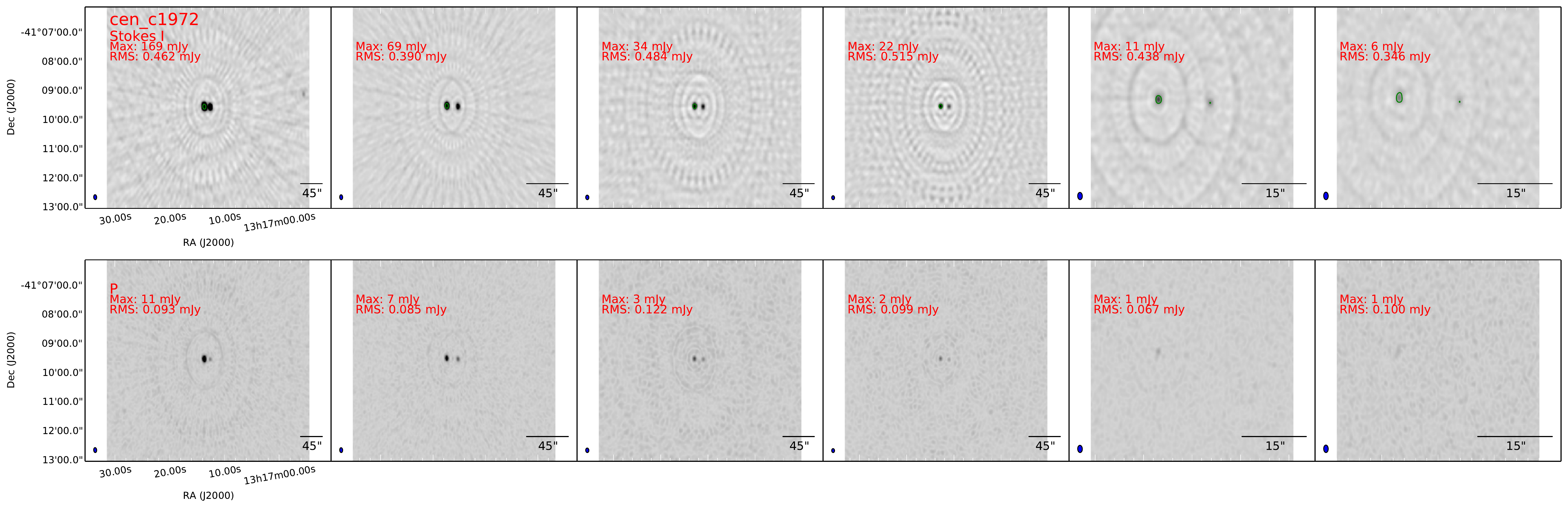}
\caption{Source: cen\_c1972. Central image frequencies: 1.4, 2.8, 4.8, 6.1, 8.3 and 9.6 GHz (left to right). The spectropolarimetric data for this source was extracted from the left-most component of the double source visible in the images above. This component becomes resolved itself at a frequency of $\sim$6 GHz (see Table \ref{tab:SourceDat}). The best-fit polarization model is of type TT.}
\label{fig:cena_c1972_showsrc}
\end{figure*}

\begin{figure*}[htpb]
\includegraphics[width=0.65\textwidth]{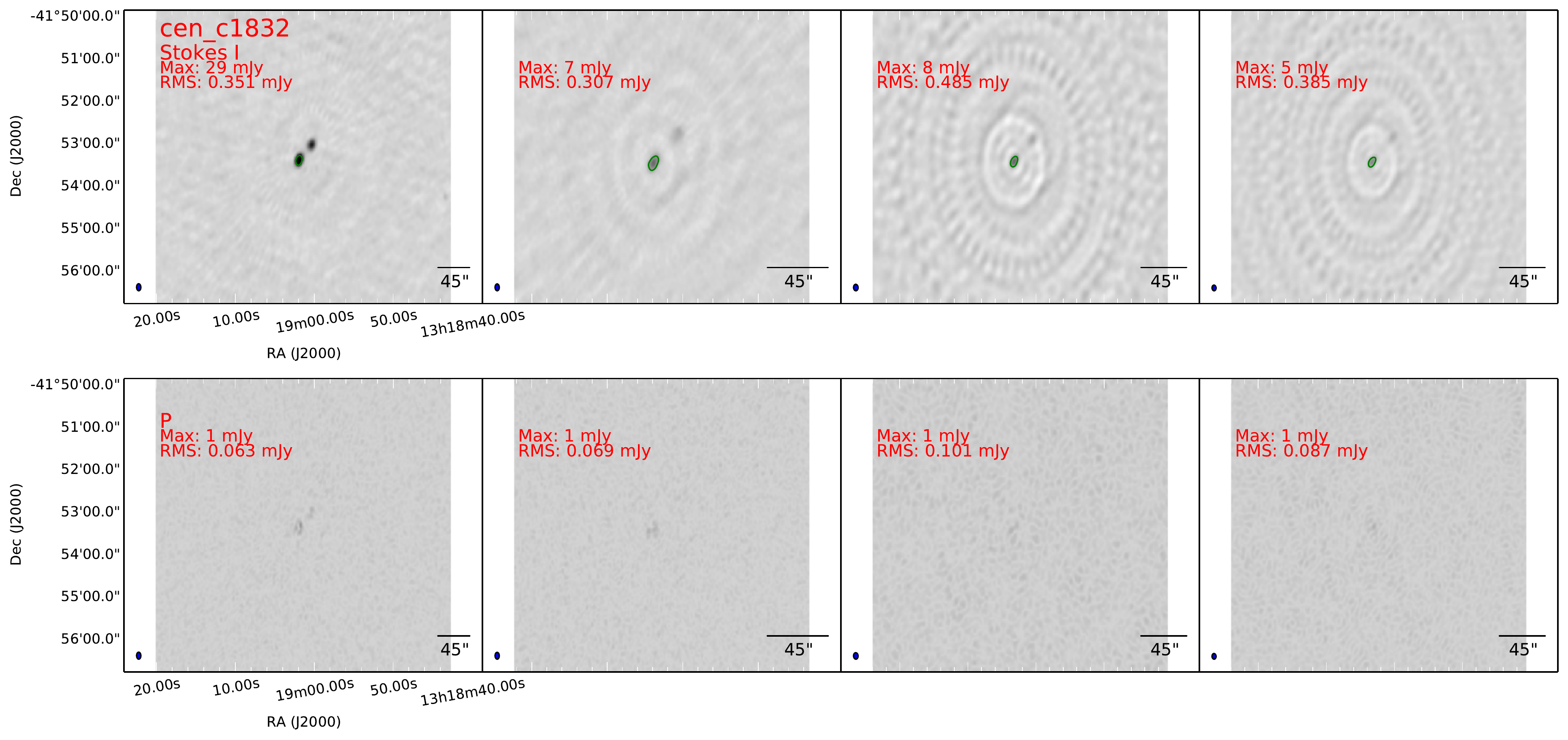}
\caption{Source: cen\_c1832. Central image frequencies: 1.4, 2.8, 4.8 and 6.1 GHz (left to right). The spectropolarimetric data represent the integrated Stokes I, Q, \& U flux densities over the entire extended source, as described in Section \ref{sec-extract}. The best-fit polarization model is of type ST.}
\label{fig:cena_c1832_showsrc}
\end{figure*}

\begin{figure*}[htpb]
\includegraphics[width=0.95\textwidth]{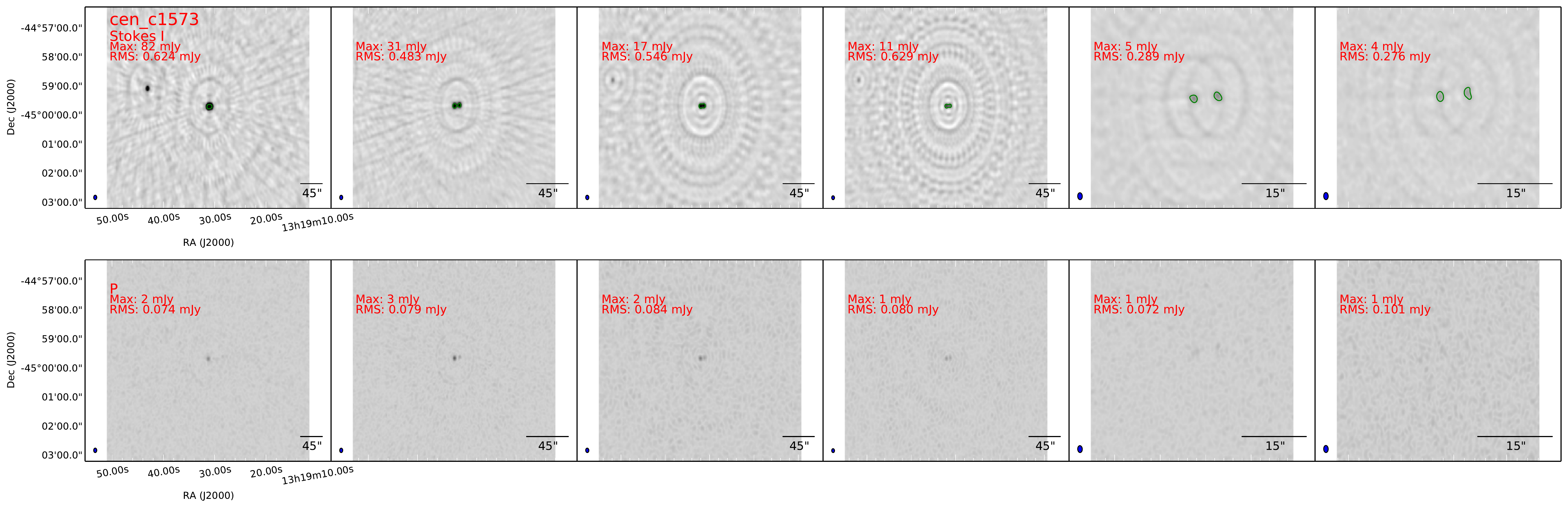}
\caption{Source: cen\_c1573. Central image frequencies: 1.4, 2.8, 4.8, 6.1, 8.3 and 9.6 GHz (left to right). The source is somewhat resolved at all frequencies from 1.3--10 GHz. The best-fit polarization model is of type ST.}
\label{fig:cena_c1573_showsrc}
\end{figure*}

\begin{figure*}[htpb]
\includegraphics[width=0.35\textwidth]{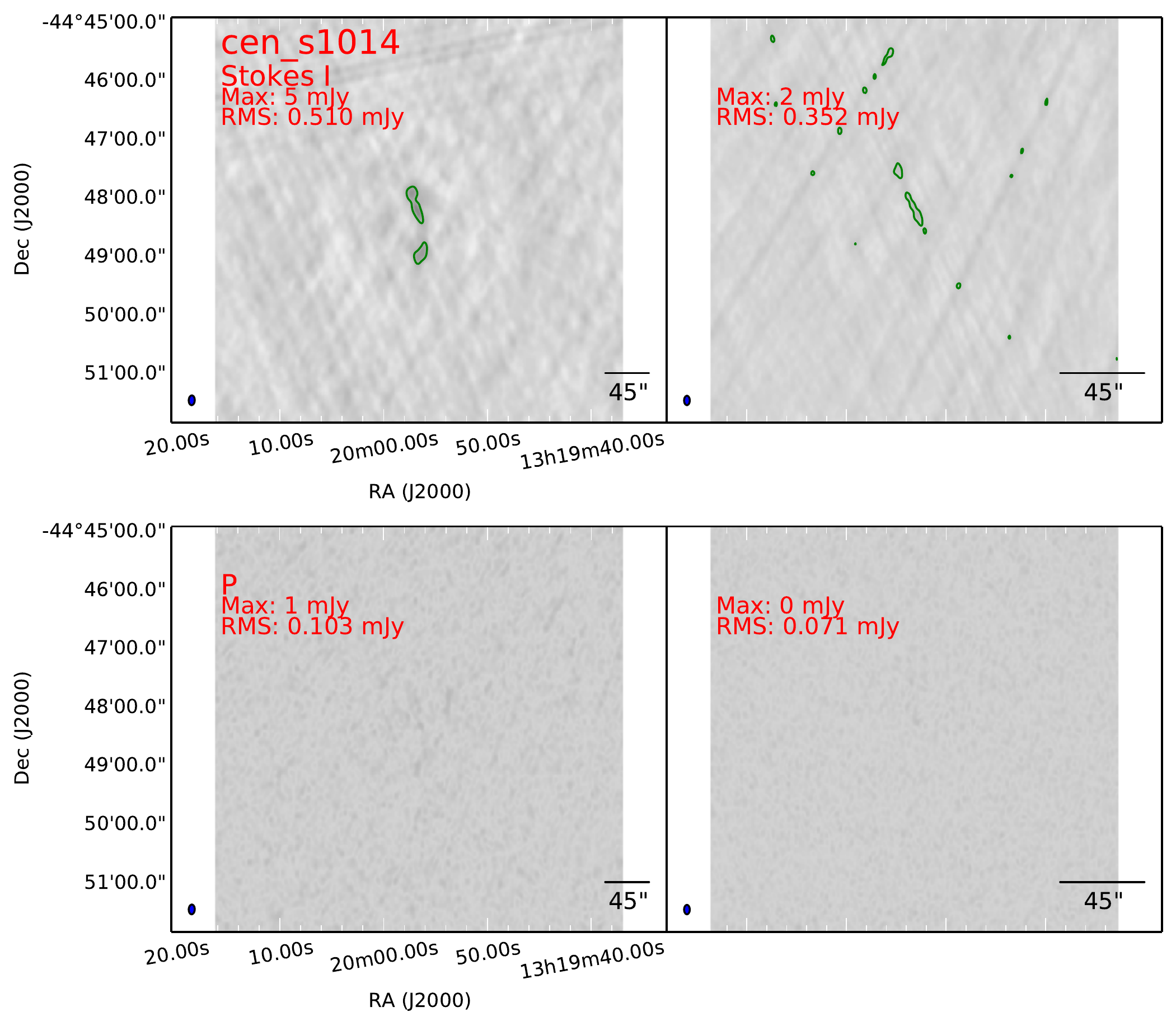}
\caption{Source: cen\_s1014. Central image frequencies: 1.4 and 2.8 GHz (left to right). The spectropolarimetric data represent the integrated Stokes I, Q, \& U flux densities over the entire extended source, as described in Section \ref{sec-extract}. The best-fit polarization model is of type SSS.}
\label{fig:cena_s1014_showsrc}
\end{figure*}

\begin{figure*}[htpb]
\includegraphics[width=0.95\textwidth]{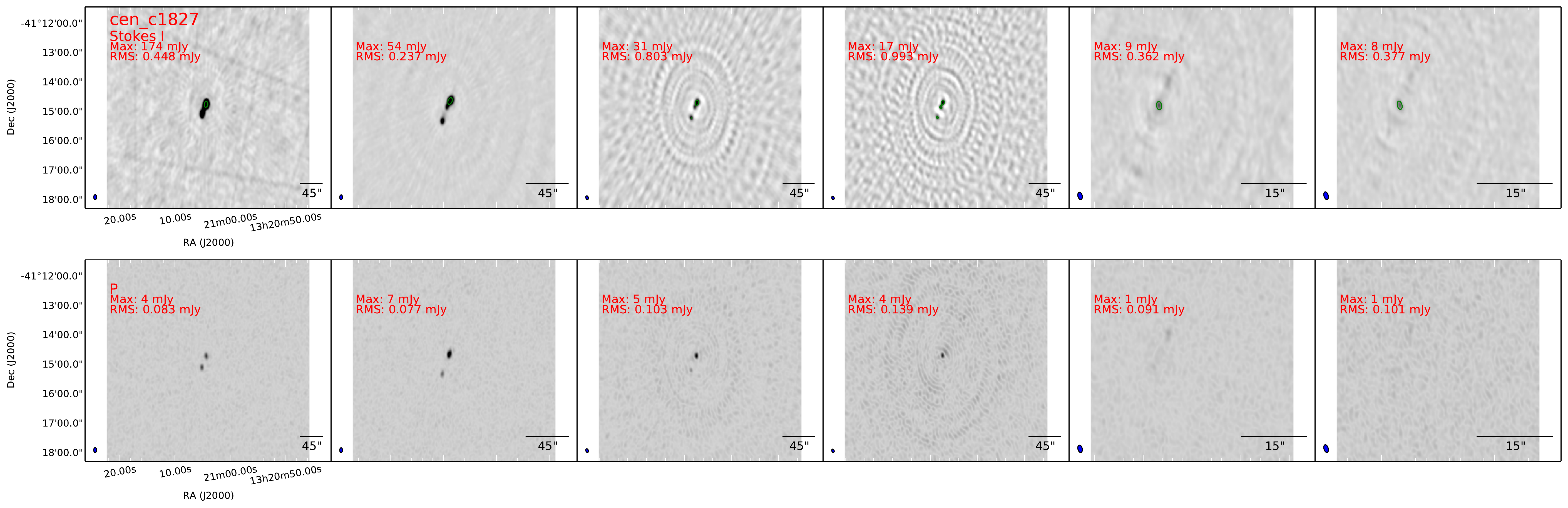}
\caption{Source: cen\_c1827. Central image frequencies: 1.4, 2.8, 4.8, 6.1, 8.3 and 9.6 GHz (left to right). The spectropolarimetric data represent the integrated Stokes I, Q, \& U flux densities over the entire extended source, as described in Section \ref{sec-extract}. The best-fit polarization model is of type TT.}
\label{fig:cena_c1827_showsrc}
\end{figure*}

\begin{figure*}[htpb]
\includegraphics[width=0.65\textwidth]{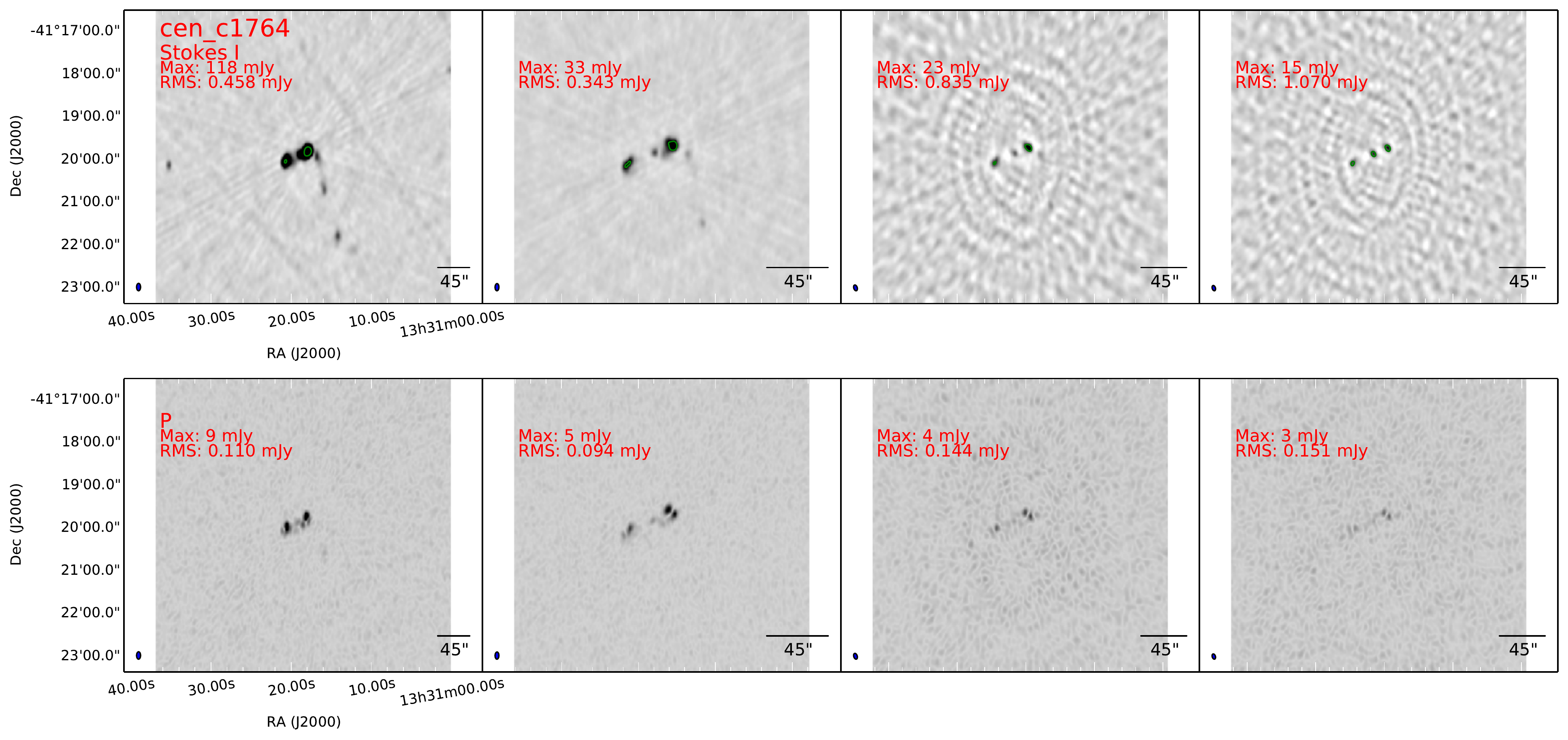}
\caption{Source: cen\_c1764. Central image frequencies: 1.4, 2.8, 4.8 and 6.1 GHz (left to right). The spectropolarimetric data represent the integrated Stokes I, Q, \& U flux densities over the entire extended source, as described in Section \ref{sec-extract}. Note that the data were extracted for the bright triple component source only --- the nearby double-component radio source that is visible at low frequencies was excluded. The best-fit polarization model is of type T.}
\label{fig:cena_c1764_showsrc}
\end{figure*}

\begin{figure*}[htpb]
\includegraphics[width=0.65\textwidth]{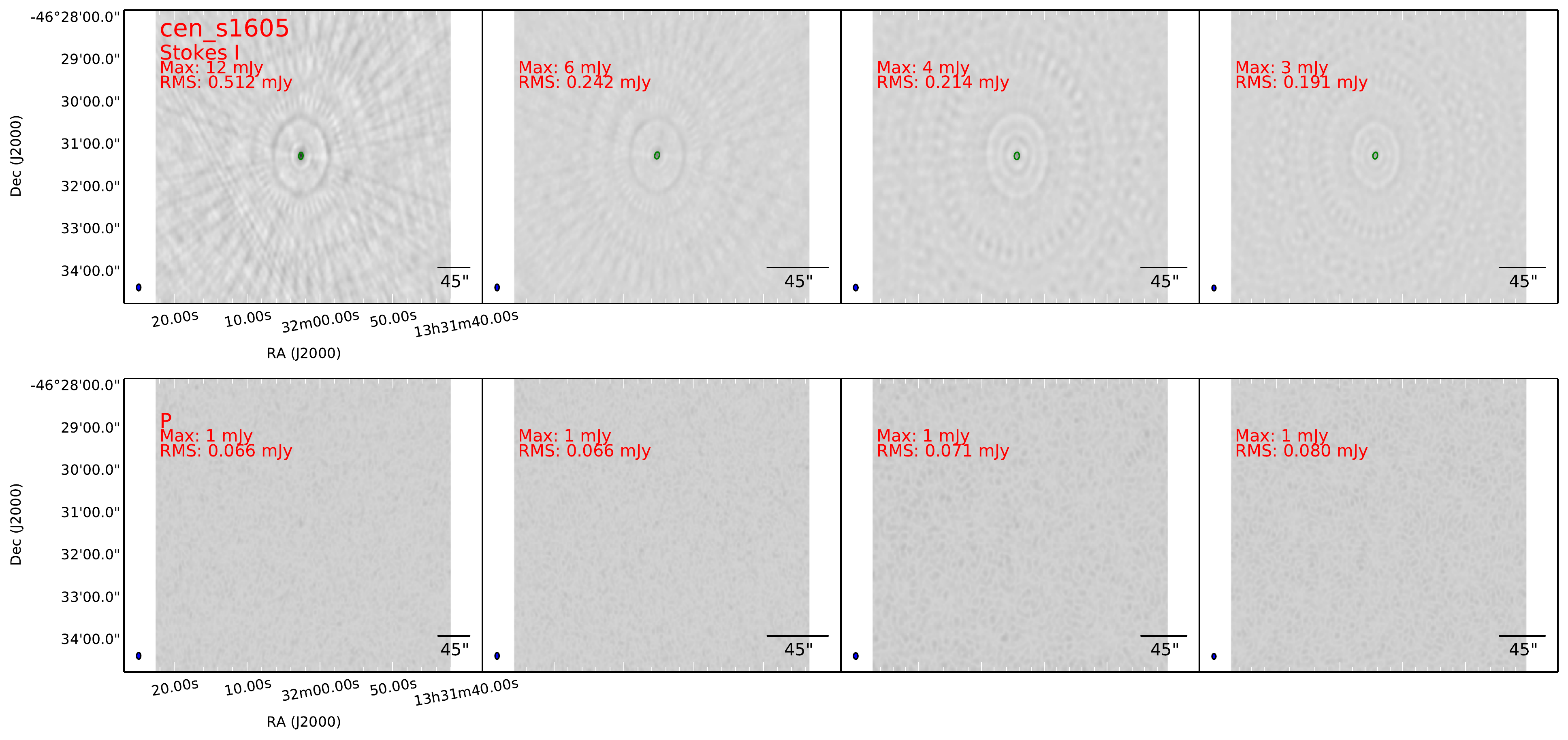}
\caption{Source: cen\_s1605. Central image frequencies: 1.4, 2.8, 4.8 and 6.1 GHz (left to right). The source is somewhat resolved at all frequencies from 1.3--6.4 GHz. The best-fit polarization model is of type TT.}
\label{fig:cena_s1605_showsrc}
\end{figure*}

\begin{figure*}[htpb]
\includegraphics[width=0.65\textwidth]{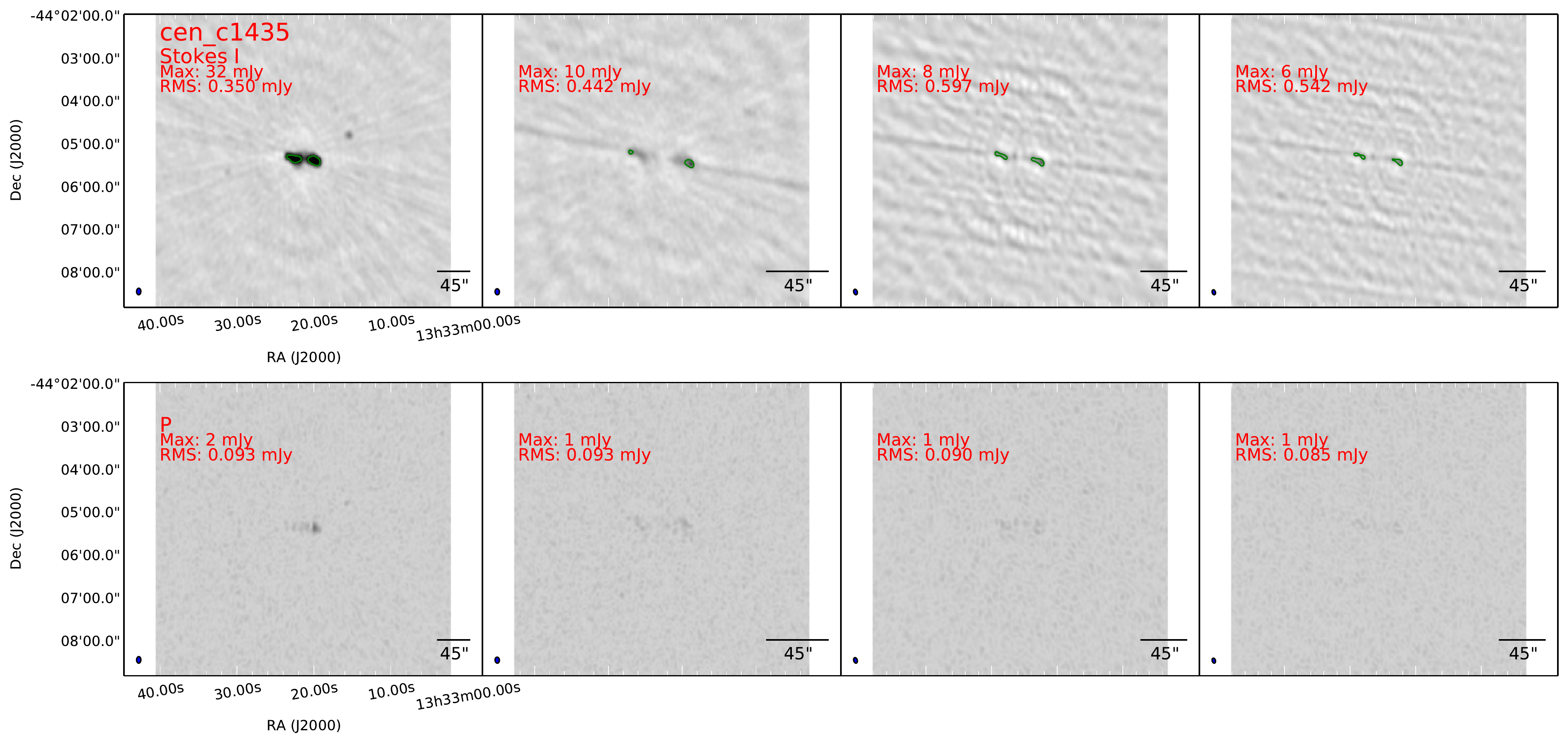}
\caption{Source: cen\_c1435. Central image frequencies: 1.4, 2.8, 4.8 and 6.1 GHz (left to right). The spectropolarimetric data represent the integrated Stokes I, Q, \& U flux densities over the entire extended source, as described in Section \ref{sec-extract}. The best-fit polarization model is of type SS.}
\label{fig:cena_c1435_showsrc}
\end{figure*}

\begin{figure*}[htpb]
\includegraphics[width=0.65\textwidth]{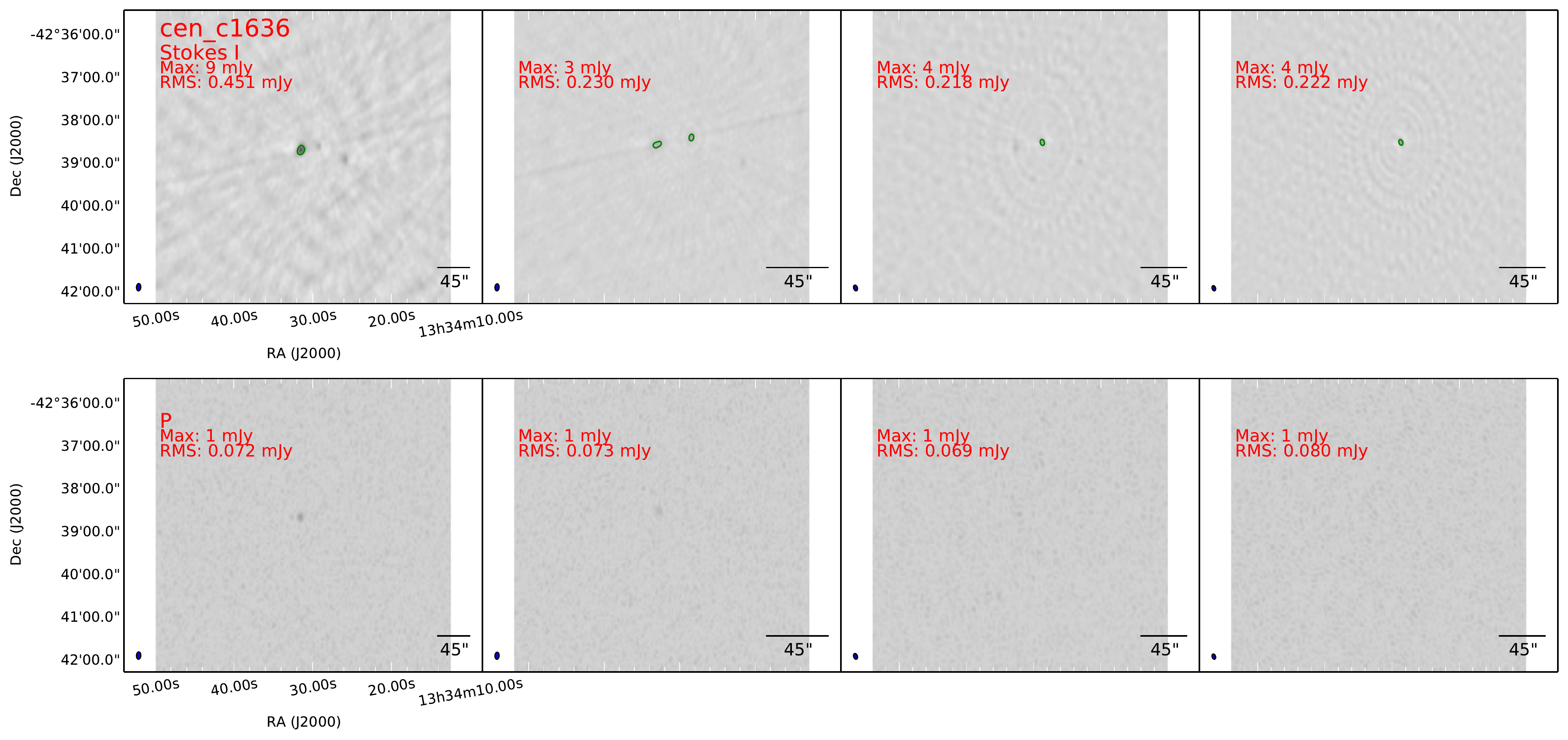}
\caption{Source: cen\_c1636. Central image frequencies: 1.4, 2.8, 4.8 and 6.1 GHz (left to right). The source is somewhat resolved at all frequencies from 1.3--6.4 GHz. The best-fit polarization model is of type TT.}
\label{fig:cena_c1636_showsrc}
\end{figure*}

\begin{figure*}[htpb]
\includegraphics[width=0.95\textwidth]{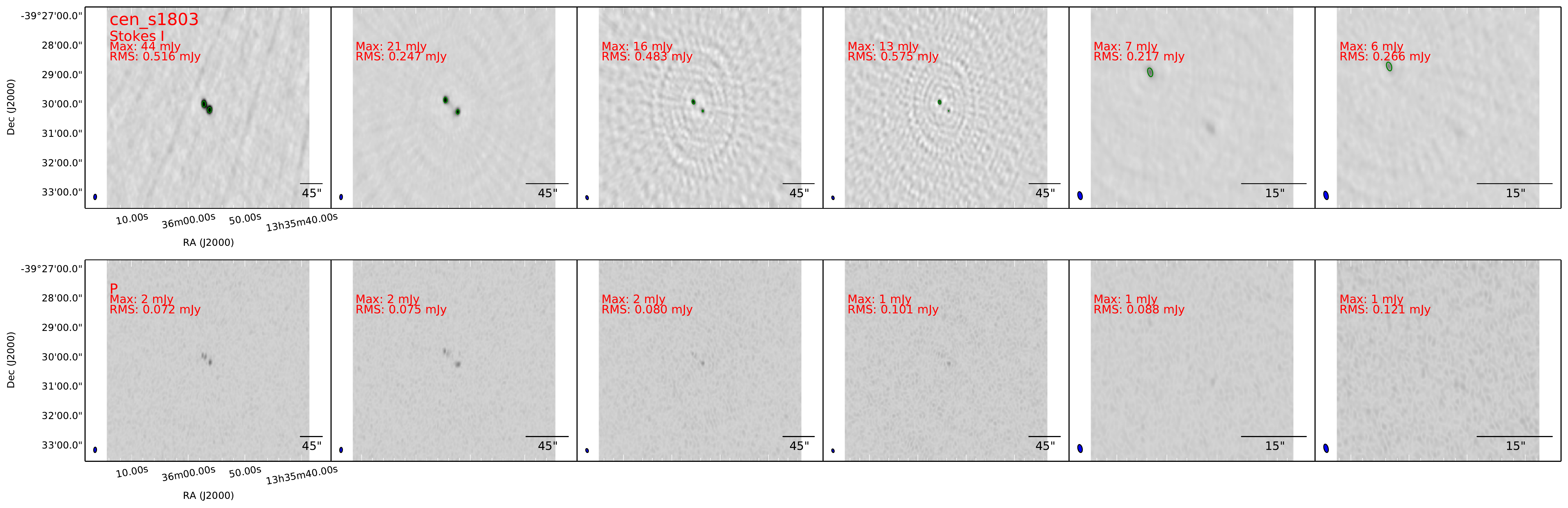}
\caption{Source: cen\_s1803. Central image frequencies: 1.4, 2.8, 4.8, 6.1, 8.3 and 9.6 GHz (left to right). The spectropolarimetric data represent the integrated Stokes I, Q, \& U flux densities over the entire extended source, as described in Section \ref{sec-extract}. The best-fit polarization model is of type SS.}
\label{fig:cena_s1803_showsrc}
\end{figure*}

\FloatBarrier

\section{Spectropolarimetric data}\label{sec-appendB}

We consider that our data might be useful for future research --- e.g. as standard cases for comparing and testing new polarimetric analysis algorithms, or for monitoring these sources for temporal changes in their broadband polarization properties. As such, we provide the raw frequency-dependent Stokes I, Q, U \& V data, along with their uncertainties, for each source in our sample (i.e. those sources listed in Table \ref{tab:SourceDat}) as named in the table caption. An example is presented below to provide an indication of format; the full set of tables can be found at VizieR. The data for each source have been calibrated, imaged, and processed as described in the main text.

\begin{table*} 
\caption{Excerpt of the spectropolarimetric data table for lmc\_s14}
\label{table:lmc_s14dat}
\scriptsize 
\tabcolsep=0.4cm 
\begin{tabular}{l l l l l l l l l} 
\hline 
\hline 
\multicolumn{1}{c}{(1)} & \multicolumn{1}{c}{(2)} & \multicolumn{1}{c}{(3)} & \multicolumn{1}{c}{(4)} & \multicolumn{1}{c}{(5)} & \multicolumn{1}{c}{(6)} & \multicolumn{1}{c}{(7)} & \multicolumn{1}{c}{(8)} & \multicolumn{1}{c}{(9)} 
\\ 
\multicolumn{1}{c}{$\nu$} & \multicolumn{1}{c}{$I$} & \multicolumn{1}{c}{$\Delta I$} & \multicolumn{1}{c}{$Q$} & \multicolumn{1}{c}{$\Delta Q$} & \multicolumn{1}{c}{$U$} & \multicolumn{1}{c}{$\Delta U$} & \multicolumn{1}{c}{$V$} & \multicolumn{1}{c}{$\Delta V$}
\\ 
\multicolumn{1}{c}{[GHz]} & \multicolumn{1}{c}{[Jy]} & \multicolumn{1}{c}{[Jy]} & \multicolumn{1}{c}{[Jy]} & \multicolumn{1}{c}{[Jy]} & \multicolumn{1}{c}{[Jy]} & \multicolumn{1}{c}{[Jy]} & \multicolumn{1}{c}{[Jy]} & \multicolumn{1}{c}{[Jy]} 
\\
\hline 
1.352807&0.0194&0.0005&0.0014&0.0002&0.0029&0.0003&-0.0001&0.0003\\
1.411809&0.0177&0.0004&0.0025&0.0002&0.0014&0.0002&0.0003&0.0002\\
1.453213&0.0157&0.0008&0.0023&0.0006&0.0008&0.0006&-0.0001&0.0007\\
1.476433&0.0177&0.0003&0.0030&0.0002&0.0003&0.0002&0.0001&0.0002\\
1.512018&0.0170&0.0005&0.0030&0.0004&-0.0002&0.0004&-0.0003&0.0004\\
1.641459&0.0149&0.0004&0.0022&0.0003&-0.0008&0.0003&0.0001&0.0003\\
1.667973&0.0150&0.0003&0.0020&0.0003&-0.0018&0.0002&-0.0003&0.0002\\
.&.&.&.&.&.&.&.&.\\
.&.&.&.&.&.&.&.&.\\
.&.&.&.&.&.&.&.&.\\
\hline 
\end{tabular} 
\end{table*}

\end{document}